# *Emergence and Synchronization in Chaotic Oscillators and in the Human Cortical Network*

Nir Lahav

Department of Physics

Ph.D. Thesis

Submitted to the Senate of Bar-Ilan University

Ramat-Gan, Israel March 2020

This work was carried out under the supervision of

Professor Shlomo Havlin

and Professor Reuven Cohen

From the Department of Physics and Department of Mathematics,

Bar-Ilan University

# Acknowledgments

I would like to thank Ashok Vaish for his continuous and generous support. I would also like to thank Kobi Flax, Dale Gillman, Eti Ben Simon and David Steinhart for their help.

# Table of Content

# Abstract

When we look at the world around us, we see complex physical systems and emergent phenomena. Emergence occurs when a system is observed to have properties that its parts do not have on their own. These properties or behaviors emerge only when the parts interact in a wider whole. Examples of emergence can vary from the synchronization of pendulum clocks hanging on the same wall to the phenomenon of life as an emergent property of chemistry. One of the most complex systems that exist in nature is the human brain. It contains on average 100 to 200 billion neurons and about 100 trillion synapses connecting them. From this vast neuronal dynamics, the ability to learn and store memory emerges as well as the ability to have complex cognitive skills, conscious experience and a sense of self.

In this work, we investigated how complex systems like the human brain create emergent properties. In order to do so, we used network theory (*paper 1*), chaos and synchronization theory (*paper 2 and 3*).

In recent years numerous attempts to understand the human brain were undertaken from a network point of view. A network framework takes into account the relationships between the different parts of the system and enables to examine how global and complex functions might emerge from network topology. Previous work revealed that the human brain features 'small world' characteristics and that cortical hubs tend to interconnect among themselves. However, in order to fully understand the topological structure of hubs, and how their profile reflects the brain's global functional organization, one needs to go beyond the properties of a specific hub and examine the various structural layers that make up the network. To address this topic further, in the first paper, we applied an analysis known in statistical physics and network theory as k-shell decomposition. The analysis was applied on a human cortical network, derived from MRI and DSI data of six participants. Such analysis enables to portray a detailed account of cortical connectivity focusing on different neighborhoods of inter-connected layers across the cortex. Our findings reveal that the human cortex is highly connected and efficient, and unlike the internet, network contains no isolated nodes.

The cortical network is composed of a nucleus alongside shells of increasing connectivity that formed one connected giant component, revealing the human brain's global functional organization. All these components were further categorized into three hierarchies in accordance with their connectivity profile, with each hierarchy reflecting different functional roles. Such a model may explain an efficient flow of information from the lowest hierarchy to the highest one, with each step enabling increased data integration and the emergence of new properties. At the top, the highest hierarchy (the nucleus) serves as a global interconnected collective and demonstrates a high correlation with consciousness-related regions, suggesting that the nucleus might serve as a platform for consciousness to emerge.

In order to investigate the emergence phenomenon, it is not enough to only analyze the structure of the network-- it is also necessary to analyze the dynamics of the network's nodes. Synchronization is a crucial emergent property, in which different nodes assimilate their dynamics until they become the same. As a result, different patterns and properties can emerge in the network. From fireflies and self-organized starling murmurations to neurons, synchronization has been reported in a diversity of systems. In order to examine the dynamics of emergence, we analyzed the synchronization of chaotic systems. In chaos theory, we can find a couple of emergent properties: the presence of strange attractors with their multifractal structure and the presence of chaotic synchronizations. By investigating the dynamics of strange attractors, we switched a point of view to the emergent domain of the fractal structures within strange attractors and how these structures change during the synchronization process.

The synchronization of coupled chaotic systems represents a fundamental example of self-organization and collective behavior. This well-studied phenomenon is classically characterized in terms of macroscopic parameters, such as Lyapunov exponents, that help predict the system's transitions into globally organized states. However, the local, microscopic, description of this emergent process continues to elude us. In the second paper we demonstrate that at the microscopic level, synchronization is captured through a gradual process of topological adjustment in phase space, in which the strange attractors of the two coupled systems continuously converge, taking similar forms, until complete topological synchronization ensues. We observed the local nucleation of topological synchronization in specific regions of the system's attractor, providing early signals of synchrony, that appear significantly before the onset of complete synchronization. This local synchronization initiates at regions of the attractor characterized by lower expansion rates, in which the chaotic trajectories are least sensitive to slight changes in initial conditions. Our findings offer a fresh and novel description of synchronization in chaotic systems, exposing its local embryonic stages that are overlooked by the currently established global analysis. Such local topological synchronization enables the identification of configurations where prediction of the state of one system is possible from measurements on that of the other, even in the absence of global synchronization.

In the third paper, we analyzed the relationship between the two emergent phenomena in chaos. The emergence of the multi-fractal structure of strange attractors and the emergence of chaotic synchronization. To capture the multi-fractal structure, we measured the general dimension of the system and measured how it evolves while increasing the coupling strength. We show that during the gradual process of topological adjustment in phase space the multifractal structures of each strange attractor of the two coupled systems continuously converge, taking similar form, until complete topological synchronization ensues.

Our analysis demonstrates that with this new approach we can expand our understanding of the synchronization process. Furthermore, according to our results chaotic synchronization has a similar property in different kinds of systems. Both in continuous systems and in discrete systems, with the right coupling, synchronization is initiated at the regions of the attractor characterized by lower density and creates what we termed a 'zipper effect'. The zipper- effect is a distinctive pattern in the multi-fractal structure of the system that acts as a signature of the microscopic buildup of the synchronization process. Topological synchronization offers a new perspective to chaotic synchronization and allows us to find new universal properties and expand our understanding of the synchronization process.

# 1. Introduction

When we look at the world around us, most of the time we will see complex physical systems and emergent phenomena. Emergence occurs when a system is observed to have properties its parts do not have on their own. These properties or behaviors emerge only when the parts interact in a wider whole. Examples of emergence can vary from the synchronization of pendulum clocks hanging on the same wall[1] to the phenomenon of life as an emergent property of chemistry[2,3]. One of the most complex systems that we find in nature is the human brain. It contains on average between 100- 200 billion neurons and about 100 trillion synapses between them[4]. From these vast neuronal dynamics emerge the ability to learn, to store memory, to have complex cognitive skills and the ability to have conscious experience and a sense of self.

In this work we investigated how complex systems like the human brain create emergent properties. In order to do so, we used network theory, chaos and synchronization theory. Applying network theory on real structural data of the human cortex revealed hierarchical organization of the cortical network in terms of data integration. As different information paths propagate along the hierarchies, they become more integrated and as a result emergence can occur (first paper). In order to investigate the emergence phenomenon, it's not enough to analyze only the structure of the network, it is necessary also to analyze the dynamics of the nodes of the network. One crucial emergent property is the synchronization phenomenon, in which different nodes assimilate their dynamics until they become the same. As a result, different patterns and properties can emerge in the network. From fireflies and self-organized starling murmurations to neurons, synchronization has been reported in a diversity of systems[5–9].

In order to examine the dynamics of emergence we analyzed synchronization of chaotic systems. In chaos theory we can find a couple of emergent properties, the presents of strange attractors with their multi fractal structure and chaotic synchronizations. By investigating the dynamics of strange attractors, we switched point of view from the time domain of the dynamics to the emergent domain of the fractal structures of strange attractors. Analyzing synchronization from the point of view of the emergent of the fractal structures of strange attractors revealed a new kind of synchronization that we named **topological synchronization.** This gave us new information about the process of the emergent of chaotic synchronization (second and third papers). Taking a larger perspective, I hope that these findings of the emergent of topological synchronization will deeper our understanding of the emergent properties we find in nature, particularly in the human neural network.

## *1.1 Network theory*

The scientific research in the domain of physics tries to explain the various phenomena in nature. In recent decades, these attempts have led physicists towards the complexity field, which is characterized by large systems that consist of multiple parts and have complex interactions between them. In order to solve these kind of problems physicists attempt to find suitable mathematical tools to analyze such large systems.

Some of these tools come from **Graph theory** or **network theory**. According to this approach the system would be examined as a *network*, the different elements of the systemwould be *vertices* or *nodes*, and the connections between the different elements of the system would be represented by *arcs, edges* or *links* of the network.

Indeed, in recent years numerous attempts to understand such complex systems were undertaken, from a network point of view[10–12]. A network framework takes into account the relationships between the different parts of the system and enables to examine how global and complex functions might emerge from network topology. In order to analyze a network, we can describe several network characteristics:

**Degree (k)** of a node is the number of edges that connect to the node. **Hub** is a node with degree above the average degree of the network. **Distance** between nodes is the shortest path between node i and node j. **Average diameter (L)** of the network is denoted by:

$$L = \frac{1}{N(N-1)} \sum_{i \neq j} d_{ij}$$

$d_{ij}$ – distance between node i and node j; N – total number of nodes in the network

**Local clustering coefficient ($c_i$)** of a node i reflects the probability that "myfriend's friend will also be my friend" (computed for each node). **Clustering coefficient (C)** is the average over all local $c_i$ and it provide an estimation of the amount of local structures in the network. Topologically it means that the network will have a large quantity of triangles:

$$C = \frac{1}{N} \sum_{i} c_i$$

**Small-world networks** are networks that are significantly more clustered than random networks, yet have approximately the same characteristic pathlength as random networks (high clustering coefficient and low average distance).

**Assortativity coefficient** is the Pearson correlation coefficient of the degreeof connected nodes. Positive values indicate a correlation between nodes of similar degree, while negative values indicate relationships between nodes of different degree. The assortativity coefficient lies between −1 and 1.

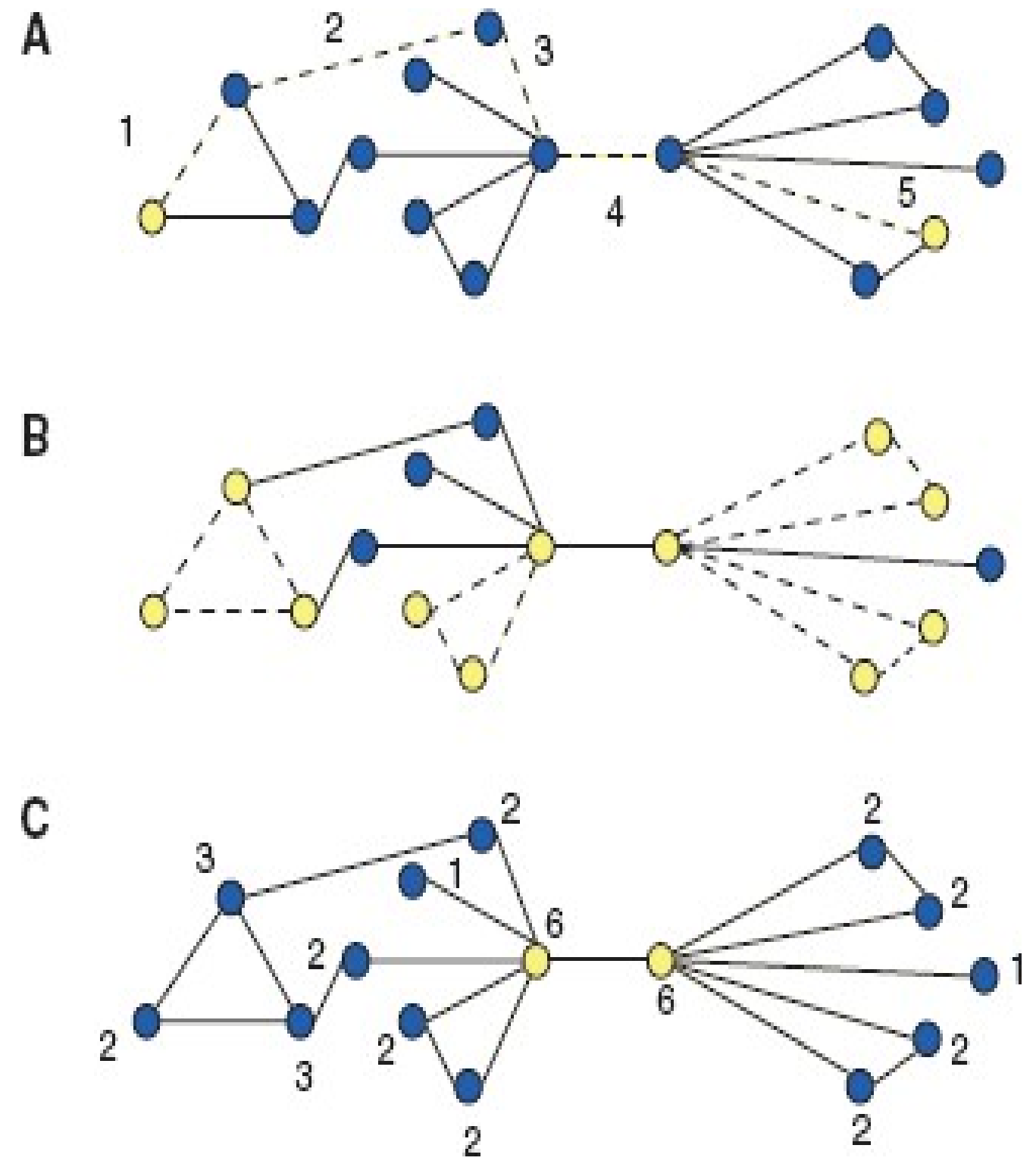

***Fig. 1**. Graph scheme: path length, clustering, average degree. Nodes are usually depicted by circular objects. Edges are the connections between these nodes. A, the path length between the two yellownodes is defined as the fewest number of edges that must be traversed to get from one to the other. In this case, five edges must be followed, and therefore the path length between these two nodes is five. B, a high clustering coefficient means that if two nodes are both connected to a third node, then they are probably also connected to each other. The calculation of the clustering coefficient takes into account the number of connected triangles (shown here with yellow nodes and dashed edges). C, the degree of a node is equal to the number of edges connected to it. A hub is defined as a node that has a degree larger than the average degree. The average degree in this network is 3.3, and therefore, both nodes with degree 6 are hubs (shown in yellow)*[13].

Previous work regarding the structural neuronal network revealed that the human brain features 'small world' characteristics (i.e. small average distance and large clustering coefficient which associated with a large number of local structures)[14–22]. They further found that cortical hubs tend to interconnect and interact among themselves[14,21,23,24]. For instance, van den Heuvel and Sporns demonstrated that hubs tend to be more densely connected among themselves than with nodes of lower degrees, creating a closed exclusive "rich club"[25–28]. These studies, however, mainly focused on the individual degree (i.e. the number of edges that connect to a specific node) of a given node, not taking into account how their neighbors' connectivity profile might also influence their role or importance. In order to better understand the topological structure of hubs, their relationship with other nodes, and how their connectivity profile might reflect the brain's global functional organization, one needs to go beyond the properties of a specific hub and examine the various structural layers that make up the network.

In order to explore the relations between network topology and its functional organization we applied a statistical physics analysis called k- shell decomposition[10,29–33] on a human cortical network derived from MRI and DSI data. Unlike regular degree analysis, k-shell decomposition does not only check a node's degree but it also considers the degree of the nodes connected to it. The k-shell of a node reveals how central this node is in the network with respect to its neighbors, meaning that a higher k-value signifies a more central node belonging to a more connected neighborhood in the network. Removing different degrees iteratively enabled us to uncover the most connected area of the network (i.e., the nucleus) as well as the connectivity shells that surround it. Therefore, every shell defines a neighborhood of nodes with similar connectivity.

The uniqueness of the k-shell decomposition method is that it takes into account both the degree of the node as well as the degree of the nodes connected to that node. Thus, we can examine groups of nodes, every group has its own unique connectivity pattern that can facilitate the emergent of new properties. In this way one can examine cortical anatomical regions according to their connectivity neighborhood. For each node in the network we determined its *shell level* (i.e. to which shell it belongs, or if it survived the whole process, it belongs to the highest level – the nucleus). We then calculated *shell levels* for every anatomical region, comprised of many nodes, according to the weighted average *shell level* of its nodes.

A few studies have already applied this analysis in a preliminary way, focusing mainly on the network's nucleus and its relevance to known functional networks[27,28,34]. For instance, Hagmann et al. revealed that the nucleus of the human cortical network is mostly comprised of default mode network regions[34]. However, in the first paper we show that when examined more carefully, k-shell decomposition analysis enables the creation of a topological model for the entire human cortex taking into account the nucleus as well as the different connectivity shells ultimately uncovering a reasonable picture of the global functional organization of the cortical network.

## 1.2 Chaos theory

Chaos theory is a branch of physics, part of nonlinear dynamics, which deals with disordered systems whose apparently-random states of disorder and irregularities are often governed by deterministic laws that are highly sensitive to initial conditions[35]. Small differences in initial conditions can yield widely diverging outcomes for such dynamical systems, rendering long-term prediction of their behavior impossible in general. This can happen even though these systems are deterministic, meaning that their future behavior follows a unique evolution and is fully determined by their initial conditions, with no random elements involved[36,37]. Mathematically, the sensitivity to initial conditions can be captures by Lyapunov exponents. A quantity that characterizes the rate of separation of infinitesimally close trajectories[35,37]. Quantitatively, two trajectories in state space with initial separation $\delta Z_0$ diverge at a rate given by:

$$|\delta Z(t)| \approx e^{\lambda t} |\delta Z_0|$$

where $\lambda$ is the Lyapunov exponent. The rate of separation can be different for different orientations of initial separation vector. Thus, there is a spectrum of Lyapunov exponents that is equal in number to the dimensionality of the state space. The largest one is called the Maximal Lyapunov exponent, and it determines a notion of predictability for the dynamical system. If it's positive, then the trajectories will diverge exponentially in time and thus the system is chaotic.

The difference between chaotic systems and real noise can be found not in the time series, that represents the behavior along the time axis, but in the state space of the systems, that represents all the states that a system can be in. If the chaotic system is dissipative, one can see that the chaotic system doesn't fill out all possible states in the state space, like noise does, instead it is bounded in a subset of states that creates distinct fractal shapes in the state space. This fractal shape represents all the states that the system can be in, and the system attracts to this particular subset. This complex attractor is known as strange attractor[35,37].

At a very early stage the system will convergeto its strange attractor and stay inside it until an external force will change the system (invariant set). Like every attractor, strange attractors have a basin of attraction, the set of initial states from where the trajectory will converge to the strange attractor. Strange attractor is an invariant minimal set of points that features mixing[35,38]. Which means that once the trajectory is in the strange attractor it will stay there and one cannot find inside the attractor a smaller attractor. Mixing describes an irreversible process like the one that occurs in thermodynamics as entropy increases. It means that with sufficient time, the trajectory from any subset on the attractor will reach anyother subset on the attractor.

Strange attractors differ from fix points and periodic attractors by the fact that its trajectory will never repeat itself. As a result, a strange attractor is 'home' for all aperiodic trajectories of the dynamical system[38]. Equivalently, strange attractors contain dense **u**nstable **p**eriodic **o**rbits[39–41]. Since all of the periodic orbits within the strange attractorare unstable, a trajectory will never settle down to any one of them. However, since the set of UPOs is dense within the chaotic set, a typical trajectory will wander incessantly in a sequence of close approaches to theseorbits. The more unstable an orbit, the less time that a trajectory will spend near it.

Fractal structures typically emerge in Strange attractors[35]. The fact that a strange attractor features fractal shapes means that all the states in it obey scale symmetry. Such symmetry is called self-similarity. Zooming in and out within the strange attractor will reveal statistically similar structures. Moreover, it means that the system will never visit a state twice, it will always jump to new states in all possible scales. This self-similarity can be captured mathematically with the property of fractal dimension (usually box-count dimension). The concept of a fractal dimension rests on the relationship between scaling and dimension[42]:

$$\lim_{r \to 0} N \sim r^{-D}$$

Where N is number of covers that are needed to cover a shape, r is the radius of the cover and D is the dimension. This scaling rule typifies conventional rules about geometry and dimensions. For lines, it quantifies that, because there is a linear relation between the number of covers and their radius, D=1. For squares, D=2, and so on. We can rearrange this relation to obtain:

$$D = \lim_{r \to 0} \frac{\ln(N)}{\ln(r^{-1})}$$

Now the dimension represents the scaling rule between the number of covers and their decreasing radius and in general it can be a non-integer value.

This equation captures the box counting dimension of a fractal, which is good approximation for its Hausdorff dimension[43].

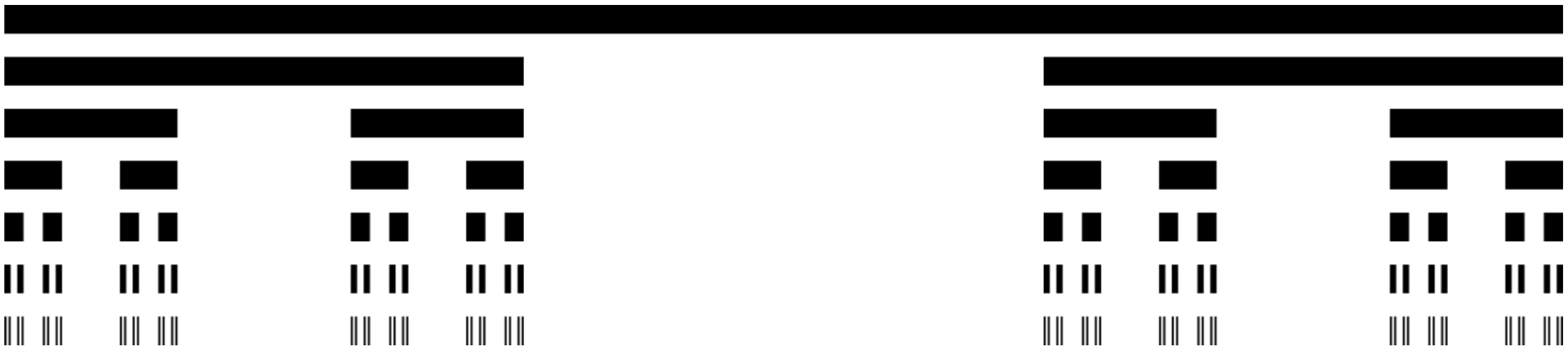

***Fig. 2**. Cantor set is created by iteratively deleting the middle third from a set of line segments. One starts by deleting the open middle third (1/3, 2/3) from the interval [0, 1], leaving two line segments. Next, the open middle third of each of these remaining segments is deleted, leaving four line segments.This process is continued ad infinitum. The number of covers that we need to cover all lines in iteration n is equal to* $2^n$. *The radius of the covers in iteration n is equal to* $3^{-n}$. *Hausdorff dimension of cantor set is:* $D_0 = \frac{ln\,2}{ln\,3}$

Most strange attractors have infinite scaling rules zooming in and out their structure, meaning that inside one strange attractor we can find infinite self-similarities and infinite fractal dimensions[44,45]. We say that these strange attractors are multi-fractals. Each fractal structure has a different probability for how long the trajectory will spend on it. The scaling exponent that dominates the attractor is revealed by the box counting dimension, but the attractor has many more types of self-similarities that are not revealed by this method. Every different self-similarity will represent mathematically asa different dimension (or a different scaling exponent) that depends on the probability of states to obey this scaling exponent. In order to investigate thewhole structure of a strange attractor one needs to measure not only the dominate dimension but all the dimensions inside the structure of theattractor.

In order to do so, we used general dimension estimation known as the Renyi dimension that can measure all the scaling rules of the strange attractor with respects to their probabilities[44–46]. Instead of one dimension, now the strange attractor will have a curve of dimensions from dimension $D_{-\infty}$ to dimension $D_{\infty}$ (the general term is Dq, with parameter q that can be any real number). The dominate dimension is the box counting dimension, which is a mixture of all the scaling exponents that will appear the most in the attractor. On the curve of the general dimension it will have the value of $D_0$. $D_{-\infty}$ represents a very rare scaling exponent on the strange attractor with asmall number of states obeying this exponent. $D_{\infty}$ represents another very rare scaling exponent of the strange attractor, but this time with high numberof states.

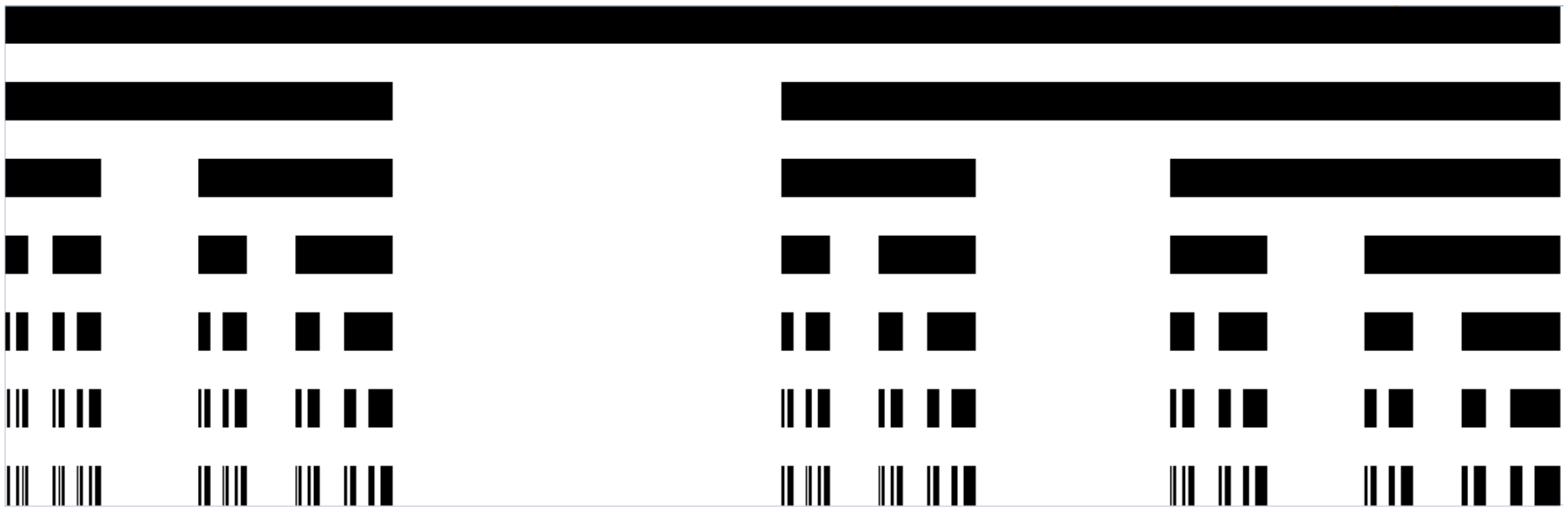

***Fig. 3**. Asymmetric Cantor set, built by removing the second quarter at each iteration and giving a value for the left and right remaining sections. The result is a multi-fractal with different fractal dimensions. For example, if we will choose only the left side sections, we'll see that the size of a sectionis one quarter of the previous sections' size. It is a rare scaling rule that appears only once. If we will assign to the left section a high value, it will be represented by the fractal dimension of $D_{\infty}$ (which represents rare scaling rule with high number of states). If we will choose only the right-side sections, we'll see that the size of a section is two quarters the size of the previous section. It is another rare scaling rule and if we will assign to the right section low value, it will be represented by the fractal dimension of $D_{-\infty}$ (which represents rare scaling rule with small number of states). $D_0$ is the box counting dimension, which is a mixture of all the scaling exponents that will appear most of the time in the set. Here, $D_0$=0.63[45].*

Until now we were talking about the emergent of multi-fractal structures inside strange attractors. Another property that can emerge in chaotic systems is chaotic synchronization. Although chaotic systems have high sensitivity for initial conditions, still, surprisingly, if two chaotic systems are coupled, they could synchronize their dynamics and create new and synchronized chaotic dynamics[39,47,48]. Every oscillation can be assigned with a different coupling strength, which corresponds to the level of interaction one oscillation has with another oscillation. For example, we can use a unidirectional couplingin which only one oscillation interacts with the other. This kind of couplingwill cause a master slave system in which the "slave" oscillation follows the "master" oscillation. If both the oscillations have the same coupling strengththan both of them will follow each other and a new synchronize chaotic dynamics will appear in both of the systems. This kind of systems is known as bidirectional coupled systems.

Typically, there are three kinds of synchronizations, phase synchronization, lag synchronization and complete synchronization[39,48]. In Phase synchronization the two oscillators follow the same phases and frequencies but not the same power (or amplitude). In complete synchronization the oscillators have the same frequencies and the same amplitudes. In other words, the dynamic activity of both oscillators will be the same over time. Lag synchronization represents a case of intermediate synchronization between phase and complete synchronizations. In this case the oscillations have the same frequencies and amplitudes, but they will have a time lag between them, e.g. one system will be behind the other. Typically, with no coupling strength the two oscillators will not be synchronized. As the coupling strength will increase the systems will gain first phase synchronization and then lag synchronization Until eventually, with enough amount of coupling strength, the systems will reach complete sync. In chaotic systems one can obtain phase, lag and approximate complete synchronizations[39].

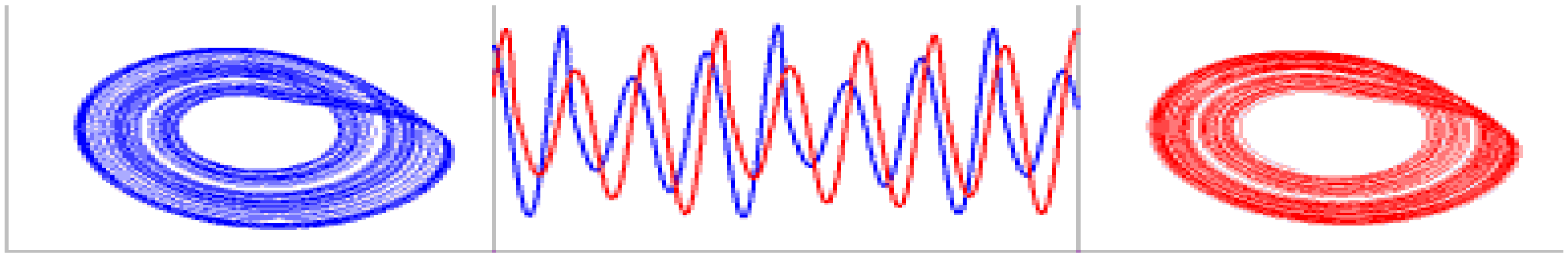

***Fig. 4***. *Example of Phase synchronization between two coupled chaotic systems (red and blue). In this case the chaotic system is a Rossler system (left and right figures are 2D trajectories of the strange attractors of the two oscillations in state space). The middle figure depicts the time series of the two oscillation. Note that the oscillations oscillate in the same frequencies but their power\amplitudes are different (photo: Arkady Pikovsky).*

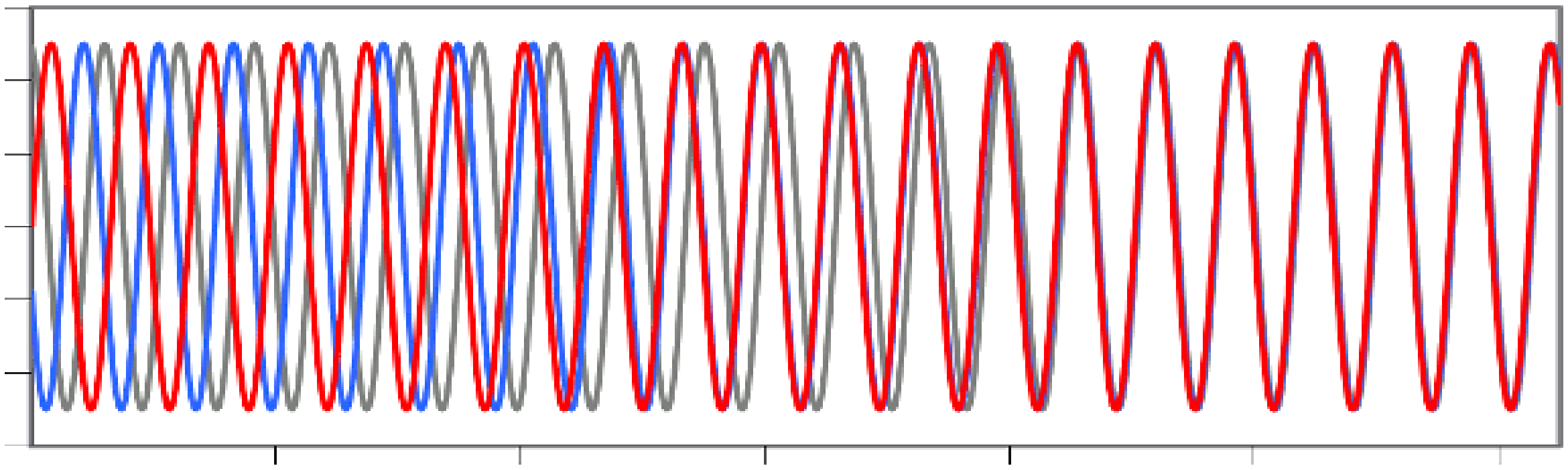

***Fig. 5***. *Example of lag and complete synchronizations of three oscillations (blue, gray and red). In the left side of this time series the oscillations are in lag synchronization. Both have the same frequency and the same power but with a time lag between them. In the right side of the time series all the oscillations are in complete sync (photo: Dev Gualtieri, Tikalon LLC).*

The analysis of chaotic synchronization is based on the analysis of the time series of the oscillators in order to distinguish between phase, lag and complete synchronization. Rosenblum et al[48] showed in their paper that an analysis of the Lyapunov spectrum can also indicate the transitions between different kinds of synchronization. Lyapunov spectrum indicates the transitions between nonsynchronous to phase synchronization and also indicates the transition between phase synchronization to lag synchronization in chaotic systems.

Typically, in a continuous 3D chaotic system (like Rossler or Lorentz systems) the Lyapunov spectrum will consist of one negative, one positive and one zero Lyapunov exponent. When two such oscillators are coupled, the system will now have six Lyapunov exponents. In low coupling, phase synchronization appears. by this transition, one of the zero Lyapunov exponent becomes negative. Further increase of coupling leads to the occurrence of the relationship between the chaotic amplitudes. As a result, the states of two interacting systems coincide (if shifted in time) to lag synchronization. in the Lyapunov spectrum this transition corresponds to the zero crossing by one of the positive Lyapunov exponents that now becomes negative. The motion in the originally six-dimensional phase space is now confined to a nearly three-dimensional manifold, thus corresponding to characterization of a synchronous regime via attractor dimensions. Further increase of coupling decreases the time shift, so the systems tend to be completely synchronized[48].

Lyapunov spectrum captures the average behavior of the system and offers a macroscopic description of the synchronization process. Unstable periodic orbits can offer mesoscopic description of synchronization by means of changes to the UPO's as a result of synchronization[48–52]. These measurements shed a light on the synchronization process, but they don't give a full description of the emergence process of chaotic synchronization. Inorder to do so we need a microscopic description of the process. To achieve that, we described the chaotic synchronization process from the emergent point of view of the multi fractal structure of the attractors. In the second and third papers we analyzed how the multi fractal structure of strange attractors assimilate along the process of chaotic synchronization and describe this Topological synchronization.

## 2. References

1. Huygens C. *Oeuvres Complètes de Christiaan Huygens. Publiées Par La SociétéHollandaise Des Sciences.* . Vol 1. La Haye: M. Nijhoff; 1888. https://www.biodiversitylibrary.org/item/61160.

2. Corning PA. The re-emergence of "emergence": A venerable concept in search of a theory. *Complexity*. 2002;7(6):18-30.

3. Luisi PL. *The Emergence of Life: From Chemical Origins to Synthetic Biology*. Cambridge University Press; 2016.

4. Azevedo FAC, Carvalho LRB, Grinberg LT, et al. Equal numbers of neuronal and nonneuronal cells make the human brain an isometrically scaled-up primate brain. *J Comp Neurol*. 2009;513(5):532-541. doi:10.1002/cne.21974

5. Néda Z, Ravasz E, Brechet Y, Vicsek T, Barabási A-L. Self-organizing processes: The sound of many hands clapping. *Nature*. 2000;403(6772):849.

6. Schmidt RC, Carello C, Turvey MT. Phase transitions and critical fluctuations in the visual coordination of rhythmic movements between people. *J Exp Psychol Hum Percept Perform*. 1990;16(2):227.

7. Varela F, Lachaux J-P, Rodriguez E, Martinerie J. The brainweb: phase synchronization and large-scale integration. *Nat Rev Neurosci*. 2001;2(4):229.

8. Agladze NN, Halaidych O V, Tsvelaya VA, et al. Synchronization of excitable cardiac cultures of different origin. *Biomater Sci*. 2017;5(9):1777-1785.

9. Hildenbrandt H, Carere C, Hemelrijk CK. Self-organized aerial displays ofthousands of starlings: a model. *Behav Ecol*. 2010;21(6):1349-1359. doi:10.1093/beheco/arq149

10. Carmi S. A model of internet topology using K-shell decomposition. *PNAS*. 2007;104(27):11150-11154.

11. Cohen R, Havlin S. *Complex Networks: Structure, Robustness and Function*. Cambridge University Press; 2010.

12. Newman MEJ. The structure and function of complex networks. *SIAM Rev.* 2003;45:167-256.

13. Bassett DS. Small World Brain Networks. *Neurosci*. 2006;12(512).

14. Achard S, Salvador R, Whitcher B, Suckling J, Bullmore ED. A resilient, low- frequency, small-world human brain functional network with highly connected association cortical hubs. *J Neurosci*. 2006;26(1):63-72.

15. Bullmore E, Sporns O. Complex brain networks: graph theoretical analysis ofstructural and functional systems. *Nat Rev Neurosci*. 2009;10(3):186-198.

16. He Y, Chen ZJ, Evans AC. Small-world anatomical networks in the human brainrevealed by cortical thickness from MRI. *Cereb Cortex*. 2007;17(10):2407-2419.

17. Sporns O, Chialvo DR, Kaiser M, Hilgetag CC. Organization, development and function of complex brain networks. *Trends Cogn Sci*. 2004;8(9):418-425.

doi:10.1016/j.tics.2004.07.008S1364-6613(04)00190-1 [pii]

18. Sporns O, Zwi JD. The small world of the cerebral cortex. *Neuroinformatics*. 2004;2(2):145-162. doi:NI:2:2:145 [pii]10.1385/NI:2:2:145

19. Stam CJ, Jones BF, Nolte G, Breakspear M, Scheltens P. Small-world networks and functional connectivity in Alzheimer's disease. *Cereb Cortex*. 2007;17(1):92-99. doi:bhj127 [pii]10.1093/cercor/bhj127

20. Stam CJ, Reijneveld JC. Graph theoretical analysis of complex networks in the brain. *Nonlinear Biomed Phys*. 2007;1(1):3. doi:1753-4631-1-3 [pii]10.1186/1753-4631-1-3

21. van den Heuvel MP, Stam CJ, Boersma M, Pol HEH. Small-world and scale-freeorganization of voxel-based resting-state functional connectivity in the human brain. *Neuroimage*. 2008;43(3):528-539. http://www.sciencedirect.com/science/article/pii/S1053811908009130.

22. Ponten SC, Bartolomei F, Stam CJ. Small-world networks and epilepsy: graph theoretical analysis of intracerebrally recorded mesial temporal lobe seizures. *Clin Neurophysiol*. 2007;118(4):918-927. doi:S1388-2457(06)01843-8 [pii]10.1016/j.clinph.2006.12.002

23. Buckner RL, Sepulcre J, Talukdar T, et al. Cortical hubs revealed by intrinsic functional connectivity: mapping, assessment of stability, and relation to Alzheimer's disease. *J Neurosci*. 2009;29(6):1860-1873.

24. Eguiluz VM, Chialvo DR, Cecchi GA, Baliki M, Apkarian A V. Scale-free brain functional networks. *Phys Rev Lett*. 2005;94(1):18102. http://www.ncbi.nlm.nih.gov/entrez/query.fcgi?cmd=Retrieve&db=PubMed&dopt=Citation&list_uids=15698136.

25. Collin G, Sporns O, Mandl RCW, van den Heuvel MP. Structural and functional aspects relating to cost and benefit of rich club organization in the human cerebral cortex. *Cereb Cortex*. 2014;24(9):2258-2267.

26. Harriger L, Van Den Heuvel MP, Sporns O. Rich club organization of macaquecerebral cortex and its role in network communication. *PLoS One*. 2012;7(9):e46497.

27. van den Heuvel MP, Sporns O. Rich-club organization of the human connectome. *J Neurosci*. 2011;31(44):15775-15786.

28. van den Heuvel MP, Sporns O, Collin G, et al. Abnormal rich club organizationand functional brain dynamics in schizophrenia. *JAMA psychiatry*. 2013;70(8):783-792.

29. Adler J. Bootstrap percolation. *Phys A Stat Mech its Appl*. 1991;171(3):453-470.

30. Alvarez-Hamelin JI, Dall'Asta L, Barrat A, Vespignani A. k-core decomposition of Internet graphs: hierarchies, self-similarity and measurement biases. *arXiv Preprcs/0511007*. 2005.

31. Alvarez-Hamelin JI, Dall'Asta L, Barrat A, Vespignani A. Large scale networks fingerprinting and visualization using the k-core decomposition. In: *Advances inNeural Information Processing Systems*. ; 2005:41-50.

32. Modha DS, Singh R. Network architecture of the long-distance pathways in themacaque brain. *Proc Natl Acad Sci*. 2010;107(30):13485-13490.

33. Pittel B, Spencer J, Wormald N. Sudden emergence of a giantk-core in a randomgraph. *J Comb Theory, Ser B*. 1996;67(1):111-151.

34. Hagmann P, Cammoun L, Gigandet X, et al. Mapping the structural core of human cerebral cortex. *PLoS Biol*. 2008;6(7):e159. doi:07-PLBI-RA-4028 [pii]10.1371/journal.pbio.0060159

35. Strogatz SH. *Nonlinear Dynamics and Chaos: With Applications to Physics,Biology, Chemistry, and Engineering*. Westview press; 2014.

36. Kellert SH. *In the Wake of Chaos: Unpredictable Order in Dynamical Systems*. University of Chicago press; 1993.

37. Boeing G. Visual analysis of nonlinear dynamical systems: chaos, fractals, self-similarity and the limits of prediction. *Systems*. 2016;4(4):37.

38. Kravtsov YA. *Limits of Predictability*. Vol 60. Springer Science & Business Media;2012.

39. Boccaletti S, Kurths J, Osipov G, Valladares DL, Zhou CS. The synchronization ofchaotic systems. *Phys Rep*. 2002;366(1-2):1-101.

40. Auerbach D, Cvitanović P, Eckmann J-P, Gunaratne G, Procaccia I. Exploringchaotic motion through periodic orbits. *Phys Rev Lett*. 1987;58(23):2387.

41. Cvitanović P. Periodic orbits as the skeleton of classical and quantum chaos. *Phys D Nonlinear Phenom*. 1991;51(1-3):138-151.

42. Iannaccone PM, Khokha M. *Fractal Geometry in Biological Systems: An Analytical Approach*. CRC Press; 1996.

43. Falconer K. *Fractal Geometry: Mathematical Foundations and Applications*. JohnWiley & Sons; 2004.

44. Hentschel HGE, Procaccia I. The infinite number of generalized dimensions offractals and strange attractors. *Phys D Nonlinear Phenom*. 1983;8(3):435-444.

45. Halsey TC, Jensen MH, Kadanoff LP, Procaccia I, Shraiman BI. Fractal measures and their singularities: the characterization of strange sets. *Phys Rev A*. 1986;33(2):1141.

46. Martinez VJ, Jones BJT, Dominguez-Tenreiro R, Weygaert R. Clusteringparadigms and multifractal measures. *Astrophys J*. 1990;357:50.

47. Pecora LM, Carroll TL. Synchronization in chaotic systems. *Phys Rev Lett*.1990;64(8):821.

48. Rosenblum MG, Pikovsky AS, Kurths J. From phase to lag synchronization in coupled chaotic oscillators. *Phys Rev Lett*. 1997;78(22):4193.

49. Heagy JF, Carroll TL, Pecora LM. Desynchronization by periodic orbits. *Phys Rev E*. 1995;52(2):R1253.

50. Yanchuk S, Maistrenko Y, Mosekilde E. Synchronization of time-continuous chaotic oscillators. *Chaos An Interdiscip J Nonlinear Sci*. 2003;13(1):388-400.

51. Pazó D, Zaks MA, Kurths J. Role of unstable periodic orbits in phase and lag synchronization between coupled chaotic oscillators. *Chaos An Interdiscip J Nonlinear Sci*. 2003;13(1):309-318.

52. Pikovsky A, Zaks M, Rosenblum M, Osipov G, Kurths J. Phase synchronization of chaotic oscillations in terms of periodic orbits. *Chaos An Interdiscip J Nonlinear Sci*. 1997;7(4):680-687.

# 3. Appendix Papers:

## 3.1. K-shell decomposition reveals hierarchical cortical organization of the human brain

**Nir Lahav*[1], Baruch Ksherim*[1], Eti Ben-Simon[2,3], Adi Maron-Katz[2,3], Reuven Cohen[4], Shlomo Havlin[1].**

*1. Dept. of Physics, Bar-Ilan University, Ramat Gan , Israel*
*2.Sackler Faculty of Medicine, Tel Aviv University, Tel-Aviv , Israel*
*3.Functional Brain Center, Wohl Institute for Advanced Imaging, Tel-Aviv Sourasky Medical Center, Tel- Aviv, Israel*
*4.Dept. of Mathematics, Bar-Ilan University, Ramat Gan, Israel*

PAPER

# *K*-shell decomposition reveals hierarchical cortical organization of the human brain

**Nir Lahav[1,5], Baruch Ksherim[1,5], Eti Ben-Simon[2,3], Adi Maron-Katz[2,3], Reuven Cohen[4] and Shlomo Havlin[1]**

1 Department of Physics, Bar-Ilan University, Ramat Gan, Israel
2 Sackler Faculty of Medicine, Tel Aviv University, Tel-Aviv, Israel
3 Functional Brain Center, Wohl Institute for Advanced Imaging, Tel-Aviv Sourasky Medical Center, Tel- Aviv, Israel
4 Department of Mathematics, Bar-Ilan University, Ramat Gan, Israel
5 Authors contributed equally to this work.

**E-mail:** freenl@gmail.com

**Keywords:** network theory, *k*-shell decomposition, cortical hubs, graph theory, modular hierarchies in the brain, human brain, consciousness

## Abstract

In recent years numerous attempts to understand the human brain were undertaken from a network point of view. A network framework takes into account the relationships between the different parts of the system and enables to examine how global and complex functions might emerge from network topology. Previous work revealed that the human brain features 'small world' characteristics and that cortical hubs tend to interconnect among themselves. However, in order to fully understand the topological structure of hubs, and how their profile reflect the brain's global functional organization, one needs to go beyond the properties of a specific hub and examine the various structural layers that make up the network. To address this topic further, we applied an analysis known in statistical physics and network theory as *k-shell decomposition analysis.* The analysis was applied on a human cortical network, derived from MRI\DSI data of six participants. Such analysis enables us to portray a detailed account of cortical connectivity focusing on different neighborhoods of inter-connected layers across the cortex. Our findings reveal that the human cortex is highly connected and efficient, and unlike the internet network contains no isolated nodes. The cortical network is comprised of a nucleus alongside shells of increasing connectivity that formed one connected giant component, revealing the human brain's global functional organization. All these components were further categorized into three hierarchies in accordance with their connectivity profile, with each hierarchy reflecting different functional roles. Such a model may explain an efficient flow of information from the lowest hierarchy to the highest one, with each step enabling increased data integration. At the top, the highest hierarchy (the nucleus) serves as a global interconnected collective and demonstrates high correlation with consciousness related regions, suggesting that the nucleus might serve as a platform for consciousness to emerge.

'..And you ask yourself, where is my mind?' The pixies (Where is my mind)

## Introduction

The human brain is one of the most complex systems in nature. In recent years numerous attempts to understand such complex systems were undertaken, in physics, from a network point of view (Newman 2003, Carmi 2007, Colizza and Vespignani 2007, Goh *et al* 2007, Cohen and Havlin 2010). A network framework takes into account the relationships between the different parts of the system and enables to examine how global and complex functions might emerge from network topology. Previous work revealed that the human brain features 'small world' characteristics (i.e. small average distance and large clustering coefficient associated with a large number of local structures (Sporns and Zwi 2004, Sporns *et al* 2004, Achard *et al* 2006, He *et al* 2007, Ponten *et al* 2007, Reijneveld *et al* 2007, Stam and Reijneveld 2007, Stam *et al* 2007, van den Heuvel *et al* 2008, Bullmore and Sporns 2009), and that cortical hubs tend to interconnect and interact among themselves (Eguiluz *et al* 2005, Achard *et al* 2006, van den Heuvel *et al* 2008, Buckner *et al* 2009). For instance, van den Heuvel and Sporns demonstrated that hubs tend to be more densely connected among themselves than with nodes of lower degrees, creating a closed exclusive '*rich club*' (van den Heuvel and Sporns 2011, Harriger *et al* 2012, van den Heuvel *et al* 2013, Collin *et al* 2014). These studies, however, mainly focused on the individual degree (i.e. the number of edges that connect to a specific node) of a given node, not taking into account how their neighbors' connectivity profile might also influence their role or importance. In order to better understand the topological structure of hubs, their relationship with other nodes, and how their connectivity profile might reflect the brain's global functional organization, one needs to go beyond the properties of a specific hub and examine the various structural layers that make up the network.

In order to explore the relations between network topology and its functional organization we applied a statistical physics analysis called *k-shell decomposition* (Adler 1991, Pittel *et al* 1996, Alvarez-Hamelin *et al* 2005a, 2005b, Carmi 2007, Garas *et al* 2010, Modha and Singh 2010) on a human cortical network derived from MRI and DSI data. Unlike regular degree analysis, *k*-shell decomposition does not only check a node's degree but also considers the degree of the nodes connected to it. The *k*-shell of a node reveals how central this node is in the network with respect to its neighbors, meaning that a higher *k*-value signifies a more central node belonging to a more connected neighborhood in the network. By removing different degrees iteratively, the process enables to uncover the most connected area of the network (i.e., the *nucleus*) as well as the connectivity *shells* that surround it. Therefore, every shell defines a neighborhood of nodes with similar connectivity (see figure 1). A few studies have already applied this analysis in a preliminary way, focusing mainly on the network's nucleus and its relevance to known functional networks (Hagmann *et al* 2008, van den Heuvel and Sporns 2011). For instance, Hagmann *et al* revealed that the nucleus of the human cortical network is mostly comprised of *default mode network* (DMN) regions (Hagmann *et al* 2008). However, when examined more carefully, *k*-shell decomposition analysis, as shown here, enables the creation of a topology model for the entire human cortex taking into account the nucleus as well as the different connectivity shells ultimately uncovering a reasonable picture of the global functional organization of the cortical network. Furthermore, using previously published *k*-shell analysis of internet network topology (Carmi 2007) we were able to compare cortical network topology with other types of networks.

We hypothesize that using *k*-shell decomposition would reveal that the human cortical network exhibits a *hierarchical structure* reflected by shells of higher connectivity, representing increasing levels of data processing and integration all the way up to the nucleus. We further assume that different groups of shells would reflect various cortical functions, with high order functions associated with higher shells. In this way we aim to connect the structural level with the functional level and to uncover how complex behaviors might emerge from the network.

## Materials and methods

### Imaging

The networks for our analysis were derived from two combined brain imaging methods, MRI/DSI recorded by *Patric Hagmann's group* from University of Lausanne (for all the functions and data sets, please refer to : http://brain-connectivity-toolbox.net/). This imaging data only covers the cortex and does not include the Insula or other sub cortical structures. Using this data, clusters of gray matter formed the nodes while fibers of white matter formed the edges of the cortical network. In this technique, 998 cortical ROIs were used to construct the nodes of each network and 14 865 edges were derived from white matter fibers (for more specific details please see Hagmann *et al* 2008). Six structural human cortical networks were transformed into six connection matrices by Patric Hagmann's group, derived from five right handed subjects (first two networks were derived from the same subject in different times). These connection matrices were utilized to calculate the network's properties and to apply the *k*-shell decomposition analysis. We used binary connection matrices ('1'—connected, '0'—disconnected) and not weighted connection matrices because of known difficulties in determining the appropriate weights and how to normalize them (Hagmann *et al* 2003, 2007, van den Heuvel and Pol 2010, van den Heuvel and Sporns 2011). In order to create single subject binary networks we assigned every weighted link that was different from zero as '1'. In order to create an average network from all 6 networks a 50% threshold was used, i.e. a link should appear in more than half of the networks in order to be included in the average network.

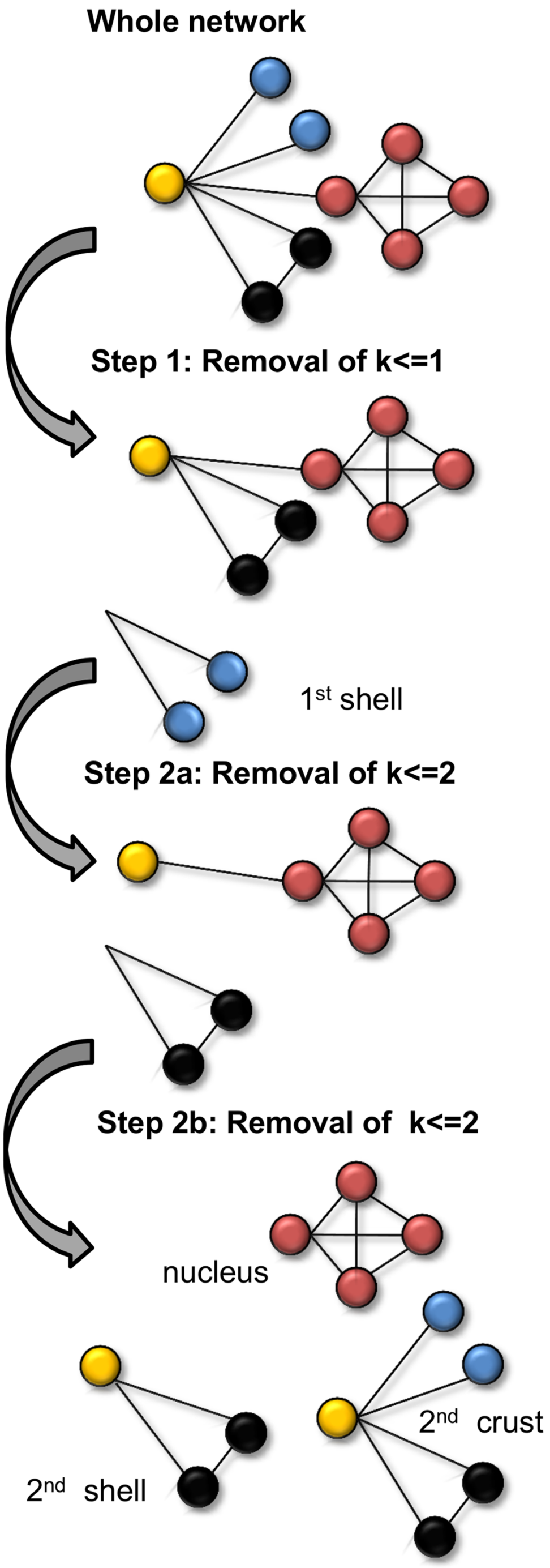

**Figure 1.** *K*-shell decomposition process. *K*-shell decomposition takes into account the degree of the node as well as the degree of the nodes connected to it. This example shows the difference of the *K*-shell method compared with regular degree count. Top panel: the whole network. The yellow node is a hub ($k = 5$) and thus one might think that it would be in the nucleus. But on the first step of the process ($k = 1$), two of its neighbors will be removed to the first shell (blue). When re-computing the degree of the remaining nodes we notice there are no more nodes with only one link. The remaining network is the 1st core. On step 2a ($k = 2$), another two of its neighbors will be removed (black). Then, when re-computing the degree of each node (step 2b), the yellow node has a low degree ($k = 1$) and will be removed to the second shell. The process stops in $k = 3$ when the remaining nodes will be removed and no node will remain in the network. *K*-core is composed of the remaining network in a given *k* step and the *nucleus* is defined as the final *k*-core in the process. The nucleus of this network is thus the 2nd-core, the group of the last remaining nodes (red). *K*-crust includes the nodes that have been removed until step *k* of the process. This network has 5 nodes in its 2nd-crust (blue, black and yellow. for more details see methods).

In order to connect between our structural network and known functional networks the 998 nodes were clustered into 66 known anatomical regions in accordance with Hagmann *et al* (2008).

## Network theory

Several network characteristics were used in our analysis:

*Degree* ($k$) of a node is the number of edges that connect to the node.

*Hub* is a node with degree above the average degree of the network.

*Distance* between nodes is the shortest path between node $i$ and node $j$.

*Average diameter* ($L$) of the network is denoted by:

$$L = \frac{1}{N(N-1)}\sum_{i \neq j} d_{ij}.$$

$d_{ij}$ is the distance between node $i$ and node $j$; $N$ is the total number of nodes in the network

*Local clustering coefficient* ($c_i$) of a node $i$ reflects the probability that 'my friend's friend will also be my friend' (computed for each node). *Clustering coefficient* is the average over all local $c_i$ and it provides estimation of the amount of local structures in the network. Topologically it means that the network will have a large quantity of triangles: $C = \frac{1}{N}\sum_i c_i$.

*Small-world networks* are networks that are significantly more clustered than random networks, yet have approximately the same characteristic path length as random networks (high clustering coefficient and low average distance).

*Assortativity coefficient* is the Pearson correlation coefficient of degree between pairs of linked nodes. Positive values indicate a correlation between nodes of similar degree, while negative values indicate relationships between nodes of different degree. Assortativity coefficient lies between $-1$ and 1.

We also examined whether the cortical network exhibits a *hierarchal structure* (not to be confused with the hierarchies derived from $k$-shell decomposition analysis) in which hubs connect nodes which are otherwise not directly connected. Networks with a hierarchal structure have a power law clustering coefficient distribution- $C \sim K^{-\beta}$ which means that as the node degree increases ($k$) the clustering coefficient ($C$) decreases. The presence of hubs with low clustering coefficient means that the network has a hierarchal structure (since hubs connect nodes which are not directly connected, triangles with hubs are not frequent) (see supplementary material 6 for further details).

## *K*-shell decomposition method

In the $k$-shell decomposition method we revealed the network's nucleus as well as the shells that surround it. The $k$-shell of a node indicates the centrality of this node in the network with respect to its neighbors. The method is an iterative process, starting from degree $k = 1$ and in every step raising the degree to remove nodes with lower or similar degree, until the network's nucleus is revealed, along the following steps:

**Step 1**. Start with connectivity matrix $M$ and degree $k = 1$.

**Step 2**. Remove all nodes with degree $\leqslant k$, resulting in a pruned connectivity matrix $M'$.

**Step 3**. From the remaining set of nodes, compute the degree of each node. If nodes have degree $\leqslant k$, step 2 is repeated to obtain a new $M'$; otherwise, go back to step 1 with degree $k = k + 1$ and $M = M'$.

**Stop** when there are no more nodes in $M'$ ($M' = 0$).

The $k$-shell is composed of all the new removed nodes (along with their edges) in a given $k$ step. Accumulating the removed nodes of all previous steps (i.e. all previous $k$-shells) is termed the $k$-crust. The $k$-core is composed of the remaining network in a given $k$ step and the *nucleus* is defined as the final $k$-core in the process. In the end of every step a new $k$-shell, $k$-crust and $k$-core are produced of the corresponding $k$ degree. In the end of the process the nucleus is revealed with the most central nodes of the network, and the rest of the nodes are removed to the different *shells* (see figure 1). Typically, in the process of revealing the nucleus, all removed nodes in the $k$-crust eventually connect to each other forming one *giant component*.

The uniqueness of $k$-shell decomposition method is that it takes into account both the degree of the node as well as the degree of the nodes connected to that node. Thus, we can examine groups of nodes, every group with its own unique connectivity pattern. In this way one can examine cortical anatomical regions according to their connectivity neighborhood. For each node in the network we determined its *shell level* (i.e. to which shell it belongs, or if it survived the whole process, it belongs to the highest level—the nucleus). We then calculated *shell levels* for every anatomical region, comprised of many nodes, according to the weighted average *shell level* of its nodes.

## Statistics and random networks

In order to evaluate the significant of the properties of the cortical network each result was compared to that of a randomized network. The network was randomized by keeping the degree distribution and sequence of the matrix intact and only randomizing the edges between the nodes (Rubinov and Sporns 2010). For each cortical network several random networks were computed with different amount of randomized edges (from 1% until 100% of the edges). This process was repeated several times iteratively. $K$-shell decomposition was applied for each of the randomized networks. Since the results of the cortical network were resilient to small perturbations (1% of the edges randomized) we raise the amount of randomization. For greater amount of randomization the results were fixed around an average value after 5 iterations (or more) using 100% random edges. Thus we took the random networks to be with 100% randomized edges and 5 iterations.

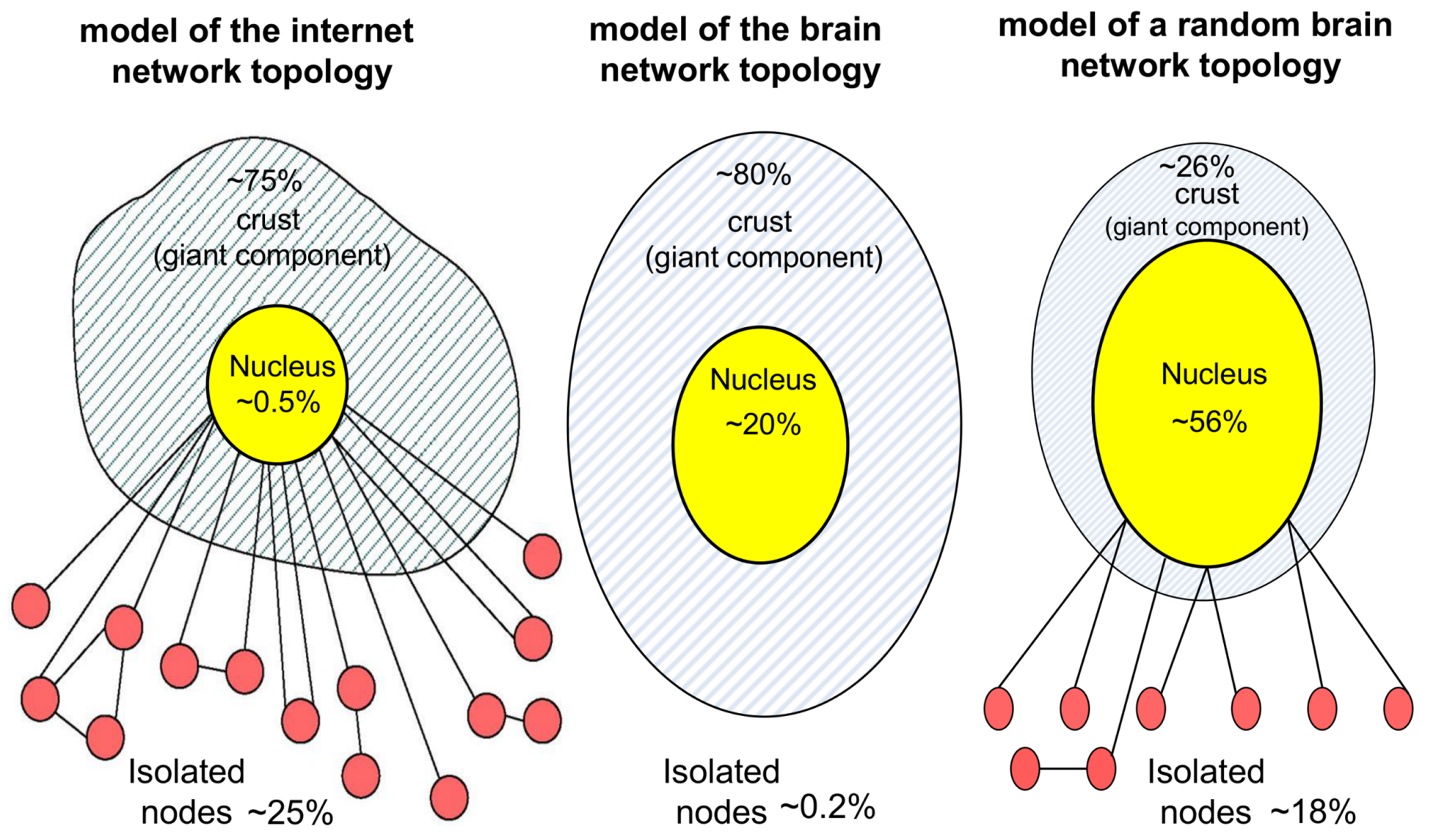

**Figure 2.** Topology of the cortical network. Topology of the cortical network (middle) compared with the internet topology, after Carmi (2007) (left) and random cortex network (right). In the cortical network the nucleus consists of 20% of the nodes while the remaining 80% compose a one giant component from all the removed nodes in the different shells. Note, a much bigger nucleus in the random cortical network and contrary to the cortical network larger amount of isolated nodes in both random and internet topologies.

To assess statistical significance of our results across networks, permutation testing was used (van den Heuvel and Pol 2010). Matrix correlations across 6 networks were computed and compared with correlations obtained from 1000 random networks. These random network correlations yielded a null distribution comprised of correlations between any two networks obtained from the random topologies. Next, we tested whether the real correlations significantly exceeded the random correlations, validated by a $p$-value $< 0.01$. Moreover, the significance of the observed connectivity within and between hierarchies was evaluated using a random permutation test. In this test, each node was randomly assigned with a hierarchy, while preserving the connectivity structure of the graph as well as hierarchy sizes. This process was repeated 10,000 times (creating a null model), and in each repetition, the number of connections within each hierarchy and between each pair of hierarchies was calculated. For each pair of hierarchies, a connectivity $p$-value was calculated using the fraction of the permutations in which the number of connections linking them was equal or higher than this number in the real data. Resulting $p$-values were corrected for multiple comparisons using the false discovery rate (FDR) procedure thresholded at 0.05.

## Results

### Cortex network topology

The results of the $K$-shell decomposition process revealed that the human cortex topology model has an 'egg-like' shape (see figure 2). In the 'middle', 22% ($\pm$12%) of the networks' nodes formed the nucleus ('the yolk' in the egg analogy) and 'surrounding' the nucleus about 77% ($\pm$12%) of the removed nodes formed the shells. These removed nodes did not reach the nucleus and connected to each other to form one *giant component*. The nucleus has on average 217 nodes ($\pm$117) and the giant component has on average 770 nodes ($\pm$121). The rest of the nodes are *isolated nodes*. These removed nodes did not connect to the giant component, and essentially connect to the rest of the network solely through the nucleus (some nodes are not connect to any other node in the network and thus were removed; on average $9 \pm 6$ nodes per cortical network).

Over all 6 networks, the average $k$-core of the nucleus was 19($\pm$1), which means that during the iterative process the nucleus was revealed after the removal of 19($\pm$1) shells. Thus, the minimum degree in the nucleus is 20 and the average degree of the nodes in the nucleus is 45 ($\pm$4). In comparison, the average degree across the entire cortical network is 29 ($\pm$1), demonstrating that the nucleus contains hubs with significantly higher degree than that of the average network. In addition, the nucleus had considerably lower average distance compared to the average distance of the entire cortical network (2 $\pm$ 0.2 versus 3 $\pm$ 0.1, respectively). This finding means that it takes 2 steps, on average, to get from one node to any other node in the 217 nodes of the nucleus.

The giant component is formed in a process similar to a first order phase transition with several critical points, as for the internet (Pittel *et al* 1996, Carmi 2007). In the beginning of the process islands of removed nodes were forming and growing, but at some stage all of these islands connect together to form the giant component (see figure S1 for more details). This abrupt phase transition occurred, on average, in $k$-crust 15 $\pm$ 1 (i.e. big islands of removed nodes were formed in crust 14, comprised of all previous shells including shell 14, but in crust 15 all of these islands disappear and a single giant component is formed). There is no significant difference between the number of removed nodes that were added to crust 15 compared to crust 14, yet a phase transition had occurred, suggesting that the difference is in the amount of the removed hubs. In crust 15, for the first time, enough hubs (which connect to lower degree nodes) were removed at once and connect all the islands to form the giant component. Later, another critical point is observed. On average in crust 18 ($\pm$1), a very large amount of nodes are removed at once to join the giant component (on average 282 nodes comprising 37% $\pm$ 10% of all the nodes in the giant component (see figure S1). This may suggest that the process reached yet another group of higher hubs which have been removed along with their connections. These hubs connect to significantly more nodes than the previous hubs leading to a massive removal of nodes. We also note that the giant component features small world characteristics similar to the entire network ($C = 0.4$ for both giant component and the whole network, average distance is 3.6 $\pm$ 0.5 for the giant component, slightly higher than that of the whole network (3 $\pm$ 0.1), see figure S2).

### Cortex network topology in comparison to other networks

The cortical network topology is found to be very different from the topologies of a randomized cortex or the internet network (at the autonomous systems level) which displayed a 'medusa-like' shape (Carmi 2007) (see figure 2). In addition to the nucleus and the giant component both random and internet topologies have a large amount of isolated nodes, forming the 'medusa legs' in the medusa shape (on average 17% in the randomized cortical networks and 25% in the internet network, unlike close to 0.3% $\pm$ 0.3% in the cortical network).

In addition, the average nucleus size of the randomized cortex is nearly three times bigger than the average nucleus of the human cortex (56% versus 20%). The cortical nucleus contains only 50% of the hubs, the rest fall on average in the last 4–5 shells before the nucleus, while in the random cortex 100% of all hubs reached the nucleus (see figure S3). A network that displays a significant amount of hubs on several levels and not just in the nucleus could support a hierarchical structure that enables modular integration, as evident in cortical function (Christoff and Gabrieli 2000, Gray *et al* 2002, Northoff and Bermpohl 2004, Northoff *et al* 2006, Bassett *et al* 2008). Note that in the cortical network the hubs outside the nucleus start on average at shell 14–15 which supports the hypothesis that the first phase transition (shell 15 $\pm$ 1) is due to the removal of those hubs (as mentioned above).

### Correlation between topology and known brain functions

In the $k$-shell decomposition analysis the connections of a node as well as its neighborhood determine at which shell that node will be removed. Neighborhood of high degree will be removed in a higher shell, or might survive the entire process and be part of the nucleus. Therefore, the giant component is comprised of different shells which represent different neighborhood densities of connectivity. These shells, corresponding to known cortical networks, enable an effective examination of cortical hierarchical organization.

We, therefore, examined the functional attributes of the nodes found in the nucleus and in all shells, by checking the shell level of every anatomical region (mapping how many nodes from the anatomical region have been removed to the different shells). Subsequently, we were able to score each anatomical region in accordance with its place in the network's hierarchy represented by its shell level. This characterization is demonstrated to be more accurate than just analyzing the average degree of each anatomical region (see figure S4 and supplementary material 1 for further details).

Furthermore, we examined the nucleus and revealed known functional areas that are *always* found in the nucleus across all 6 networks (see figure 3). These areas comprise the entire *bilateral midline region* and overlap with five major functional networks: motor and motor planning, the default network, executive control network, high order visual areas and the salience network (see table 1 for full details). In contrast, several known functional areas were *never* in the nucleus across all 6 networks. These areas include most of the *right temporal lobe* (e.g. the fusiform gyrus, A1, V5), right Broca and Wernicke homologues and right inferior parietal cortex. Interestingly, all the areas that never appear in the nucleus are from the right hemisphere. Furthermore, 70% of all the lowest shells are from the right hemisphere while 60% of the areas that are *always* in the nucleus belong to the left hemisphere (see supplementary material 2 for more details ).

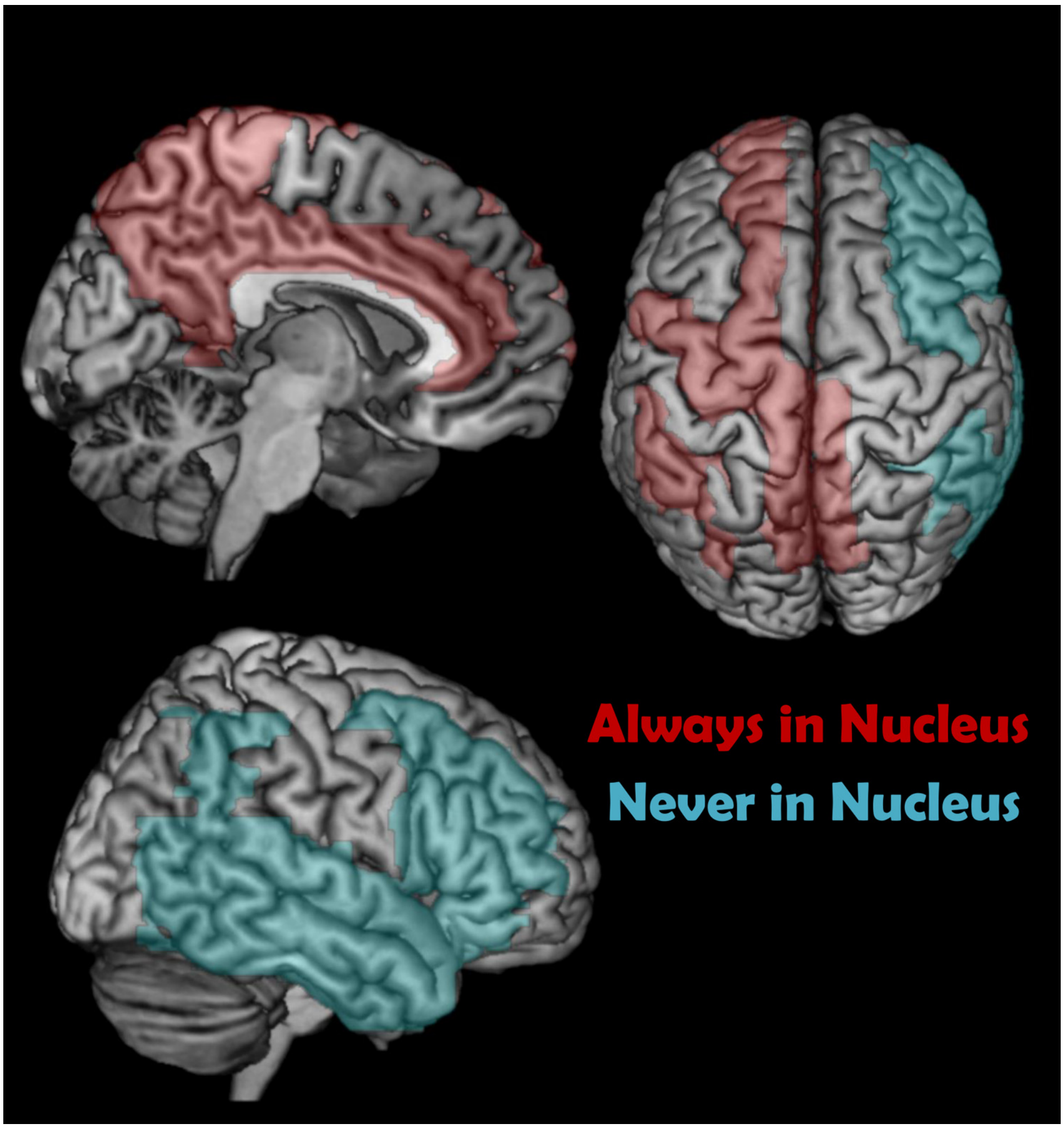

**Figure 3.** Anatomical regions and the network nucleus. Brain maps displaying anatomical regions that are always in the nucleus (red) and never in the nucleus (blue). Note that all the regions that never reach the nucleus are from the right hemisphere.

Next, we used the critical points that were observed during the giant component formation (see supplementary material 3 for more details) in order to detect and establish different hierarchies of shells. Briefly, the creation of the giant component corresponded to the shell threshold of a middle hierarchy and the creation of the nucleus corresponded to the threshold of a high hierarchy. This analysis resulted in three major hierarchal groups (low, middle and high) as portrayed in figure 4.

The first hierarchal group consists of regions found in the lowest shells (average shell level 8.8, number of nodes/edges: 99/730 respectively). The removed nodes of this group are distributed across the shells with relatively high standard deviation (4.42, e.g. fusiform gyrus, entorhinal cortex, parahippocampal cortex. See table 1 and figure 5 for full details). Notably, in this hierarchal group 75% of the regions were bilateral and 50% of the regions were never in the nucleus. The second hierarchy is a middle group which includes nodes found in the highest shells, but still not in the nucleus (number of nodes/edges: 335/4377 respectively). This group can be further subdivided to two subgroups, *distributed middle* and *localized middle* according to their average shell level and standard deviation. The average shell level of the distributed middle group is 14.5 ($\pm$3.07). This subgroup includes regions like right A1, right V5 and right Broca's homologue (for full details see table 1 and figure 5(d)). The average shell level of the localized middle group is 16.67 ($\pm$1.13). This subgroup includes regions like right wernicke homologue and right middle frontal gyrus. In the middle hierarchy 56% of the regions are bilateral and 40% of the regions are from the right hemisphere (in localized middle 88% right). 48% of the regions in this hierarchy were never found in the nucleus (for full details see table 1 and figure 5(c)).

The third group is the highest hierarchy which contains regions predominantly found in the nucleus (number of nodes/edges: 561/8430 respectively). This group can also be subdivided to two subgroups, *distributed high* and *localized high* according to their average shell level and standard deviation. Average shell level of distributed high is 16.92 ($\pm$2.82). This subgroup includes the superior frontal gyrus, left Wernicke, left Broca and left V5. The average shell level of the localized high group is 19.30 (±0.97) and includes the precuneus and the cingulate cortex (for full details see table 1 and figure 5). In this hierarchal group 69% of the regions were bilateral while 28% of the regions belonged to the left hemisphere. 44% of the regions in this hierarchy were always in the nucleus (66% in localized high). Altogether, all the regions that are always in the nucleus are from the high hierarchy while the regions that never reached the nucleus are from lower hierarchies.

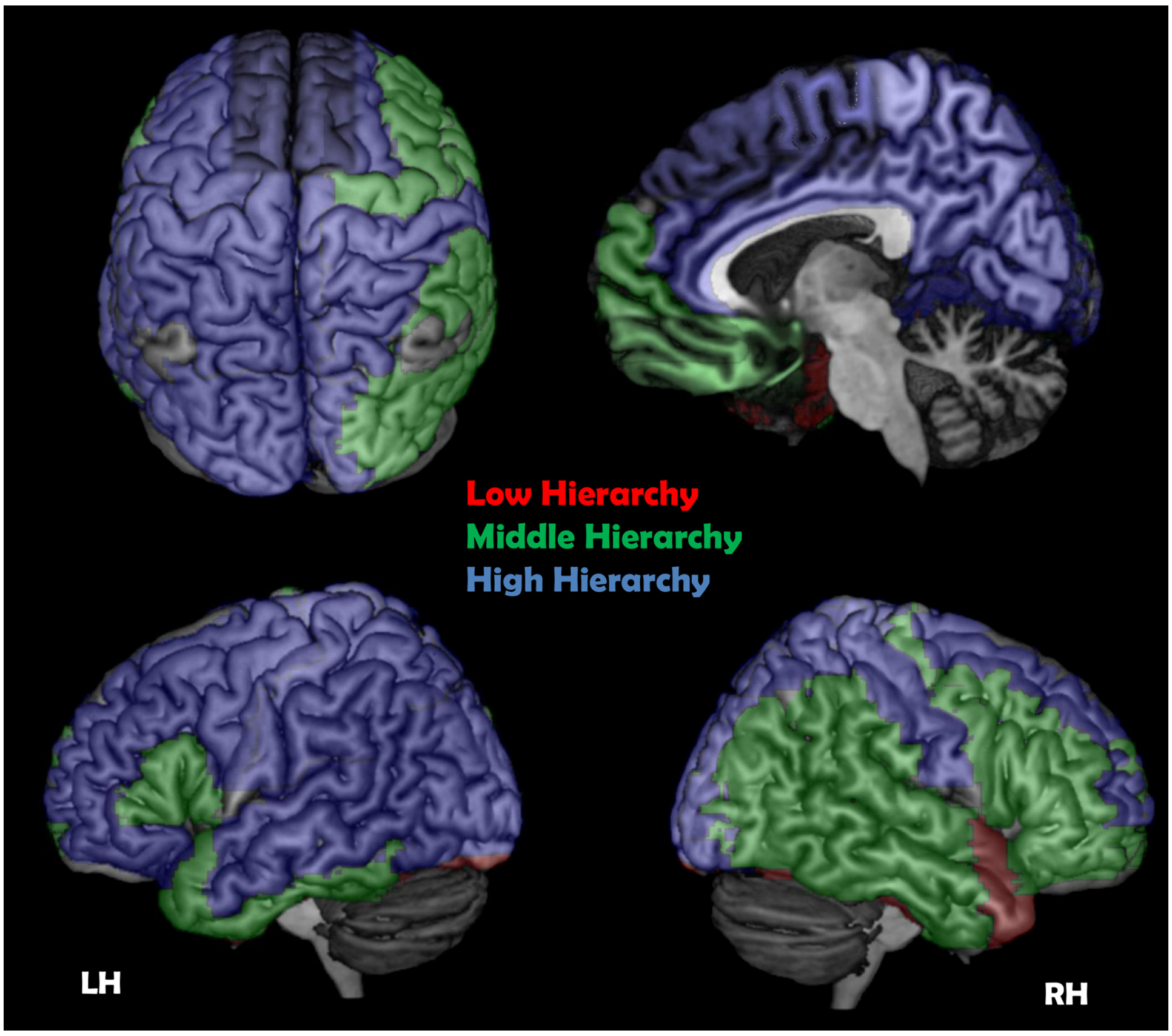

**Figure 4.** Anatomical regions according to their hierarchies. Brain maps displaying cortical anatomical regions according to their hierarchies. Red—low hierarchy, green—middle hierarchy, blue—high hierarchy. One can divide the cortex to low hierarchy regions found in the lateral bottom part of the cortex, middle hierarchy in the right lateral middle part of the cortex, and high hierarchy in the left lateral middle, lateral top and mid-line part of the cortex. RH—right hemisphere, LH—left hemisphere.

Using the shell score we could further estimate the average shell level of known functional regions or networks (see table S2). Interestingly, average shell level often reflected known functional lateralization as detailed in table 2. For instance, Wernicke's area is found in the high hierarchy (average shell level 18.3) and its right homologue in the middle hierarchy (average shell level 17), never reaching the nucleus. In a similar way, the average shell level of Broca's area is 14.5 while its right homologue's average is 14.1. Both of these regions are found in the middle hierarchy but the right homologue never reaches the nucleus. In addition, right primary motor region and right TPJ are found in the middle hierarchy (and also never reach the nucleus) whereas their left counterparts are found in the high hierarchy (and left primary motor region always reaches the nucleus). Functional lateralization was also evident when looking at the network level. For instance, the salience, executive control and sensorimotor networks (average shell level 17.3, 16.8 and 17.5, respectfully) reveal leftward dominance in terms of the amount of regions that reach the nucleus (more than 50%, see table 2). This lateralization effect was especially evident in the middle hierarchy; for instance 88% of the regions in localized middle belong to the right hemisphere while most of their left homologues found in the high hierarchy. These findings can be explained by hemispheric dominance given that all of the subjects were right handed and also by well-known language lateralization (Gazzaniga and Sperry 1967). $K$-shell decomposition analysis managed to recover these known functional attributes which were not detected in regular methods using degree count.

The functional network with the highest average shell level was the DMN with a score of 18.1. 81% of its regions were found in the high hierarchy with 70% always reaching the nucleus. Following the DMN, the salience and the sensorimotor networks also demonstrate high average shell level (17.3 and 17.5, respectfully) reflecting their high functional relevance. The visual ventral stream (i.e. the 'what' stream Goodale and Milner 1992) has a very low shell level of 12 comprising 75% of the low hierarchy. 40% of its regions never reach

**Table 1.** Cortical anatomical regions according to hierarchies.

| Anatomical region | Side | Function |
|---|---|---|
| **Localized high** | | |
| Paracentral lobule | Mid | SMA—sensorimotor network (always) |
| Caudal anterior cingulate cortex | L | Salience network (always) |
| Caudal anterior cingulate cortex | R | Salience\executive control network (always) |
| Inferior parietal cortex | L | DMN, sensorimotor network, visual dorsal stream (always) |
| Posterior cingulate cortex | Mid | DMN (always) |
| Rostral anterior cingulate cortex | Mid | Salience\executive control network, DMN (always) |
| Precuneus | Mid | DMN (always) |
| Isthmus of the cingulate cortex | R | DMN (always) |
| Pericalcarine cortex | R | Primary visual area |
| Postcentral gyrus | L | Primary somatosensory cortex—sensorimotor network |
| Superior parietal cortex | L | Executive control, sensory integration, sensorimotor network, visual dorsal stream |
| Supramarginal gyrus | L | Wernicke area, TPJ |
| Bank of the superior temporal sulcus | L | Visual dorsal stream |
| Cuneus | R | Visual |
| **Distributed high** | | |
| Superior frontal cortex | L | DMN\executive\salience, sensorimotor network (always) |
| Precentral gyrus | L | Primary motor cortex—sensorimotor network (always) |
| Superior temporal cortex | L | Wernicke ,TPJ, visual dorsal stream |
| Pericalcarine cortex | L | Primary visual |
| Pars orbitalis | L | Executive control network |
| Middle temporal cortex | L | V5 (visual dorsal stream), DMN |
| Lateral occipital cortex | L | Primary visual, visual ventral stream |
| Isthmus of the cingulate cortex | L | DMN |
| Cuneus | L | Visual |
| Rostral middle frontal cortex | L | Executive control network, DMN |
| Superior parietal cortex | R | Executive, sensory integration, sensorimotor network, visual dorsal stream |
| Superior frontal cortex | R | DMN\executive\salience\ sensorimotor network |
| Postcentral gyrus | R | Primary somatosensory cortex—sensorimotor network |
| Lingual gyrus | R | Visual |
| **Localized middle** | | |
| Inferior parietal cortex | R | DMN, sensorimotor network, visual dorsal stream (never) |
| Caudal middle frontal cortex | R | Executive control network, sensorimotor network (never) |
| Bank of the superior temporal sulcus | R | Visual dorsal stream (never) |
| Supramarginal gyrus | R | Wernicke homologue, TPJ (never) |
| Superior temporal cortex | R | Wernicke homologue, TPJ, visual dorsal stream (never) |
| Frontal pole | R | Executive control network |
| Frontal pole | L | Salience and executive control networks |
| Medial orbitofrontal cortex | R | Stimulus-reward associations |
| **Distributed middle** | | |
| Pars triangularis | R | Broca homologue (never) |
| Pars triangularis | L | Broca |
| Middle temporal cortex | R | V5 (visual dorsal stream), DMN (never) |

**Table 1.** (Continued.)

| Anatomical region | Side | Function |
|---|---|---|
| Pars opercularis | R | Broca homologue (never) |
| Pars opercularis | L | Broca |
| Inferior temporal cortex | R | Visual association, visual ventral stream (never) |
| Inferior temporal cortex | L | Visual association, visual ventral stream |
| Rostral middle frontal cortex | R | Salience and executive control networks (never) |
| Pars orbitalis | R | Salience and executive control networks (never) |
| Transverse temporal cortex | R | Primary auditory cortex (never) |
| Temporal pole | L | Salience network |
| Lateral orbitofrontal cortex | L + R | Stimulus-reward associations |
| Medial orbitofrontal cortex | L | Stimulus-reward associations |
| Precentral gyrus | R | Primary motor cortex—sensorimotor network |
| Caudal middle frontal cortex | L | Executive control network, DMN, sensorimotor network |
| Lateral occipital cortex | R | Primary visual, visual ventral stream |
| **Low** | | |
| Temporal pole | R | Salience network (never) |
| Parahippocampal cortex | R | Hippocampal support, visual ventral stream (never) |
| Parahippocampal cortex | L | Hippocampal support, visual ventral stream |
| Fusiform gyrus | R | Face recognition, visual ventral stream (never) |
| Fusiform gyrus | L | Face recognition, visual ventral stream |
| Entorhinal cortex | R | Hippocampal support, visual ventral stream (never) |
| Entorhinal cortex | L | Hippocampal support, visual ventral stream |
| Lingual gyrus | L | Visual association |

DMN = default mode network, TPJ = temporal parietal junction. Always = region that always reaches the nucleus for all networks, never = region that never reaches the nucleus for all networks.

the nucleus. In contrast, the visual dorsal stream (where\how stream), has one of the highest average shell level, 17.7%, and 60% of its regions found within the high hierarchy. Interestingly, both streams reveal left dominance in terms of average shell level, with most of their right regions never reaching the nucleus. These results are detailed in table 2 and in supplementary material 2.

## Connections between hierarchies

In order to examine the connections between the different hierarchies, we compared the number of connections within each hierarchy to the number of connections with other hierarchies (calculated as a percentage of its total connections). Within the lowest hierarchy it was found that only 22% ± 6.33% were self-connections and the rest were distributed between the middle group (30% ± 3.36%) and the highest group (48% ± 4.24%). In the middle hierarchy approximately half of the connections (52% ± 2.6%) were self-connections and 41.5% ± 2.6% were linked to the highest group. Interestingly, only 7% ± 0.77% of the connections from the middle hierarchy were linked to the lowest hierarchy. The highest hierarchy exhibited the highest levels of self-connections (72% ± 1.6%). Only 22.5% ± 1.5% of its connections were linked to the middle hierarchy and 6% ± 0.6% to the lowest hierarchy (for more details see table S1). These findings suggest a flow of information from the lowest to the highest hierarchy with each step enabling greater local processing, possibly supporting increased data integration.

We further tried to distinguish the differences between localized and distributed hierarchies. Distributed hierarchies have high standard deviation of the shell distribution and localized hierarchies have small standard deviation of the shell distribution (see figure 5). Notably, while most of the edges of the localized hierarchies were mainly self-connections or connections to their distributed partner in the same hierarchy (e.g. distributed to localized middle), the distributed hierarchies displayed more connections to other hierarchies (~15% in distributed subgroups compared to only ~8% in localized subgroups) supporting their role in cross-hierarchy data integration. Moreover, many of these connections were also across similar categories (e.g. distribute middle with distribute high, app. 25%). Furthermore, the distributed and localized subgroups within the same hierarchy displayed a large amount of connections between themselves (~33% of their connections), supporting the fact that they originate from the same hierarchy. The significance of the observed connectivity within and between hierarchies was evaluated using a random permutation test (see 'statistics and random networks' in materials and methods section). The results showed that connectivity within each hierarchy is significantly higher than a null-model (FDR $q < 0.0005$) and that connectivity between all hierarchies was significantly lower (FDR $q < 0.0005$) than expected according the size of the hierarchies (null-model; see figure 6).

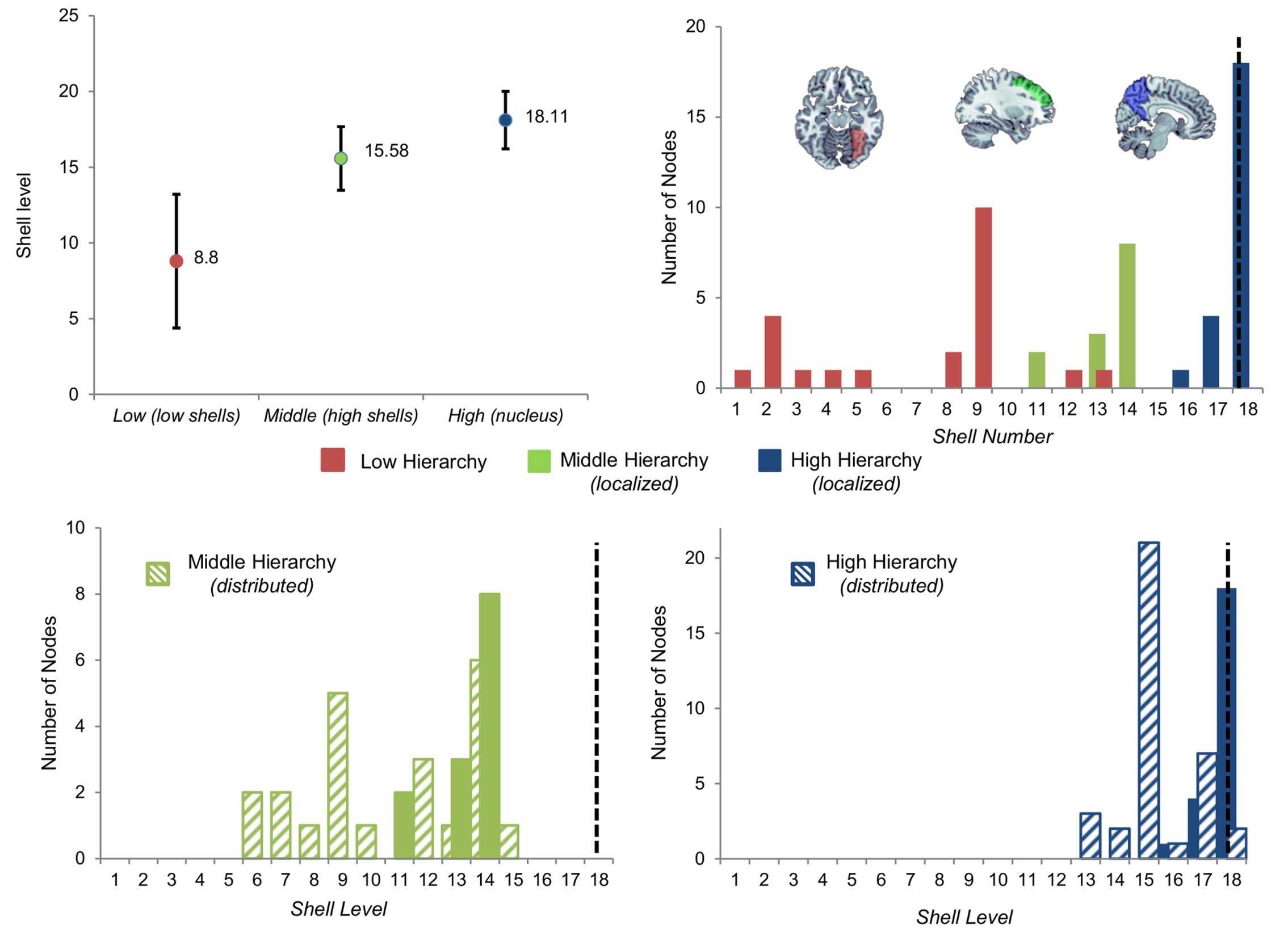

**Figure 5.** Hierarchies of the cortical network. *Top left panel:* average shell level of the hierarchies. *X*-axis: hierarchy, *Y*-axis: shell level. *Top right panel:* an example of a single anatomical region representing each hierarchy (derived from average cortical network over all 6 networks. For exact data see supplementary data 1 and figure S5). Right precuneus as an example of localized high hierarchy regions (blue). Notably, this area always reached the nucleus. Right caudal middle frontal as an example of localized middle hierarchy regions (green). Notably, this area never reached the nucleus. Right fusiform gyrus as an example of low hierarchy regions (red). Note the high standard deviation of the shell distribution. This region never reached the nucleus. *X*-axis: *k*-shell number; *Y*-axis: number of nodes. Dashed line: nucleus. *Bottom left:* right lateral occipital cortex as an example of distributed middle hierarchy regions (striped green, localized middle hierarchy as above). *X*-axis: *k*-shell number; *Y*-axis: number of nodes. Dashed line: nucleus. *Bottom right:* Left precentral gyrus as an example of distributed high hierarchy regions (striped blue, localized high hierarchy as above). This area always reached the nucleus. *X*-axis: *k*-shell number; *Y*-axis: number of nodes, dashed line: nucleus.

## Discussion

In the current study we applied the *k*-shell decomposition analysis to reveal the global functional organization of the human cortical network. Using this analysis we managed to build a model of cortex topology and connect the structural with the functional level. Our findings indicate that the human cortex is highly connected and efficient, compared to other networks, comprised of a nucleus and a giant component with virtually no isolated nodes. The giant component consists of different degree shells which represent different neighborhoods of connectivity, revealing the global properties of the cortical network. Together with the nucleus, these connectivity shells were categorized into three hierarchies representing an increasing number of regional connections, possibly supporting an increase in data processing and integration within each hierarchy. In accordance, the highest hierarchy was predominantly comprised of left and midline cortical regions (including regions of the DMN) known to be associated with high-order functions (Northoff *et al* 2006). Lastly, this collective of interconnected regions, integrating information throughout the cortex, might allow global properties such as consciousness to emerge.

### Network properties

Although we had only data of the cortex (and no subcortical regions) our findings demonstrate, in accordance with previous works (Achard *et al* 2006, Cohen and Havlin 2010, Ekman *et al* 2012) that the cortical network is resilient to small perturbations, highly organized, interconnected and much more efficient compared with a random cortical network or the internet network. *K*-shell decomposition analysis further proved to be more accurate and provide better resolution of network properties compared to standard methods (e.g. counting degrees, for full details see supplementary material 1).

**Table 2.** Laterality effects.

| Anatomical region | Left | Right |
|---|---|---|
| Precentral gyrus (primary motor cortex) | High (always) | Middle |
| Inferior parietal | High (always) | Middle (never) |
| Supra marginal gyrus (Wernicke area,TPJ) | High | Middle (never) |
| Superior temporal (Wernicke area ,TPJ) | High | Middle (never) |
| Lateral occipital cortex (primary visual) | High | Middle |
| Lingual gyrus (visual association) | Low | High |
| Bank of the superior temporal sulcus (vision) | High | Middle (never) |
| Pars orbitalis (executive control network) | High | Middle (never) |
| Middle temporal (V5, DMN) | High | Middle (never) |
| Rostral middle frontal cortex (executive control network, DMN) | High | Middle (never) |
| Superior frontal cortex | High (always) | High |
| Caudal middle frontal cortex (executive control network, DMN) | Middle | Middle (never) |
| Inferior temporal cortex (visual association) | Middle | Middle (never) |
| Pars triangularis (Broca homologue) | Middle | Middle (never) |
| Pars opercularis (Broca homologue) | Middle | Middle (never) |
| Temporal pole (salience network) | Middle | Low (never) |
| Parahippocampal cortex | Low | Low (never) |
| Fusiform gyrus | Low | Low (never) |
| Entorhinal cortex | Low | Low (never) |
| Functional networks | | |
| Dorsal stream (where stream) | 100% high | 80% middle (80% never) |
| Ventral stream (what stream) | 60% low | 60% low (80% never) |
| Auditory network | 100% high | 100% middle (100% never) |
| Executive control network | 77% high | 55% middle |
| Default mode network | 89% high (55% always) | 71% high (57% always) |
| Salience network | 60% (always) | 40% (always) |
| Sensorimotor network | 83% (always) | 17% (always) |

DMN = default mode network, TPJ = temporal parietal junction. Always = region that always reaches the nucleus for all networks, never = region that never reaches the nucleus for all networks.

The two main components of the cortical network, the nucleus and the giant component, both have small world properties though they might serve different roles. A higher clustering coefficient of the giant component alongside short average distance of the nucleus suggest that the majority of local processing takes place within the giant component while the nucleus mainly adds shortcuts and global structures to the network. Indeed, although the nucleus is highly connected, it includes only 50% of all hubs unlike the random nucleus which includes all network hubs (see figures S2 and S3). These 'peripheral' hubs were located in the giant component and, as previously suggested (Achard *et al* 2006), might enable efficient data integration and local information processing. Hubs outside the nucleus might therefore, serve as local processors integrating information from lower shells and transfer it forward to a higher hierarchy, eventually reaching the nucleus (for more information see supplementary material 4).

## Network hierarchies and data integration

*K*-shell decomposition analysis reveals that the creation of the giant component entails several critical points. From these critical points we could characterize three major neighborhoods of connectivity or three hierarchies (for more details see supplementary material 3). The regions in the lowest hierarchy appeared to be mostly involved in localized sensory perception (e.g. the fusiform face area and visual 'what' stream Goodale and Milner 1992). The different nodes within this hierarchy broadly distributed along the shells which might enable efficient data transfer and processing before sending it to higher hierarchies.

The middle hierarchy is found to be composed of high shells with high degree nodes, though half of them never reached the nucleus, a property that separates these regions from the high hierarchy. Functional regions found in this hierarchy appeared to be involved in high cognitive functions and data integration. For instance, most of the auditory network and regions involved in the integration of audio and visual perception were found in the middle hierarchy. In addition 40% of the executive control network (including right dorsolateral PFC, a crucial region in executive control and working memory Raz and Buhle 2006) and the right dorsal visual stream (where stream, Goodale and Milner 1992) are found in this hierarchy. Broca's area was also located in the middle hierarchy as well as other homologue regions related to language such as Broca and Wernicke homologues.

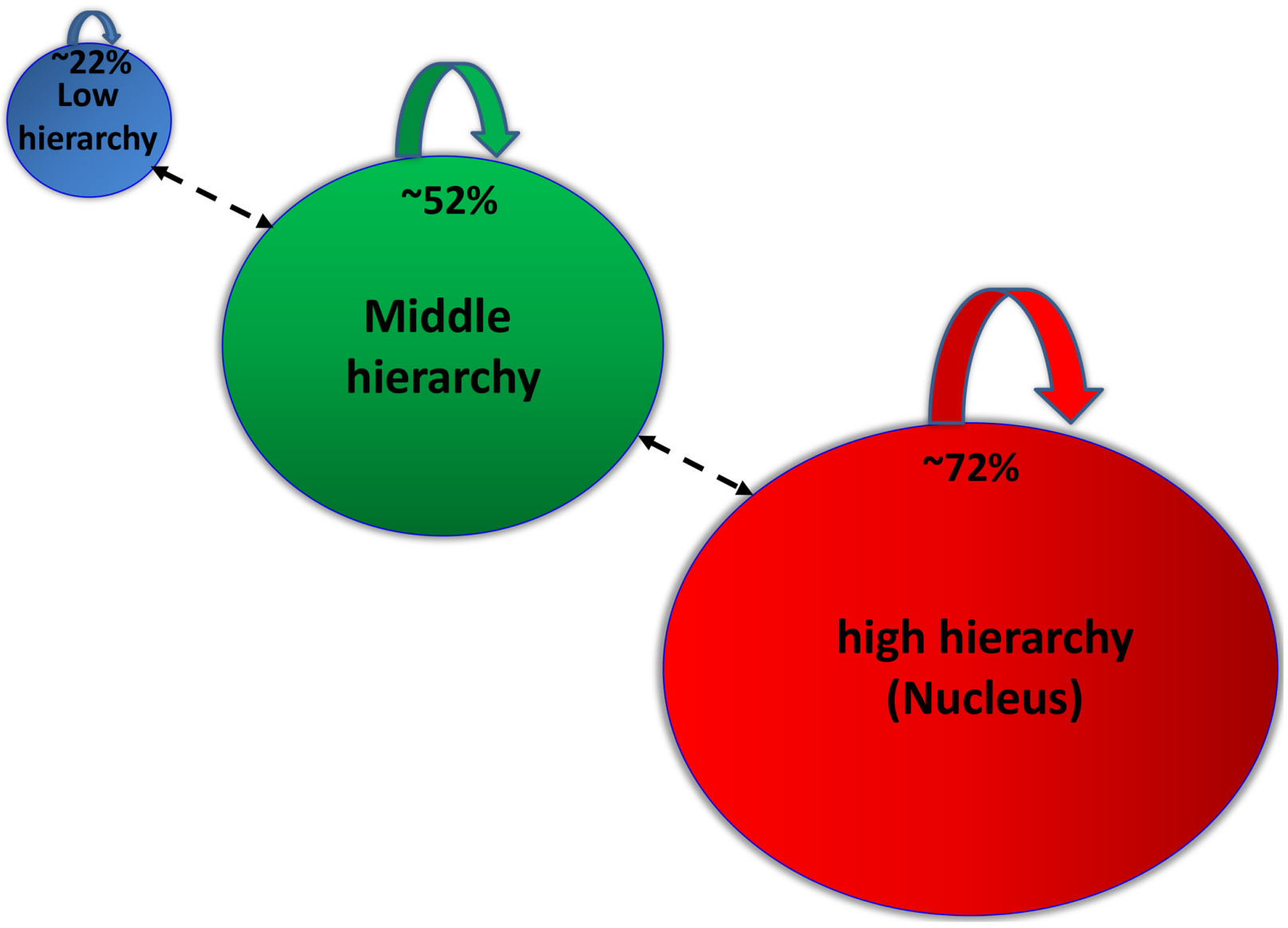

**Figure 6.** Connections between hierarchies. The size of the hierarchies represents total amount of intra hierarchy connections. Connections within any hierarchy is found to be significantly higher (arrows) and connections between hierarchies was significantly smaller (dash arrows) than expected when taking into account the size of the hierarchies, supporting the modularity nature of every hierarchy. Note the increased self-connections as the hierarchies increase (percent connections are normalized by the total amount of connections in each hierarchy).

The high hierarchy contained regions predominantly found in the nucleus. All regions that reached the nucleus across all cortical networks are found in this hierarchy. Unlike other hierarchies, this unique hierarchy is a single, highly interconnected component, which enables high levels of data integration and processing, probably involved in the highest cognitive functions. In accordance, the high hierarchy exhibited the highest amount of self-connections across hierarchies suggesting that it processes data mostly within itself (see figure 6). The nucleus (represented by the high hierarchy) has a very strong overlap with the DMN (81%), in accordance with the result of Hagmann *et al* (2008), and also with the visual cortex (75%), sensorimotor network (75%) and salience network (71%). The visual dorsal stream and the executive control network also display 60% overlap with the nucleus. Interestingly, all the regions that never appear in the nucleus (across all 6 networks) belong to the right hemisphere, while a strong tendency towards the left hemisphere appeared when examining the nucleus. As mentioned above, all the regions that reached the nucleus are mostly midline or left hemisphere regions. Roughly speaking, the left hemisphere is comprised of high hierarchy regions and the right hemisphere is comprised of middle hierarchy regions (see figure 4 and supplementary material 3 and 2).

Looking across hierarchies it's evident that the lowest hierarchy has the smallest amount of connections to other hierarchies and within itself; the middle hierarchy has more connections, almost equally distributed between itself and others; and the high hierarchy has the largest amount of connections, most of them within itself (see figure 6). Interestingly, self-connections within each hierarchy are significantly higher (and between hierarchies significantly smaller) than expected in a null model which takes into account the size of the hierarchies. This finding suggests that every hierarchy can be seen as a different module mostly involved in self-processing and only then in the transfer of information to other hierarchies (Hagmann *et al* 2008, Bullmore and Sporns 2009, van den Heuvel and Sporns 2011). Regarding cross hierarchy connections, it is important to note that most of the connections between middle and high hierarchies occur in their distributed subgroups. This finding suggests that in every hierarchy distributed regions are more involved in data transfer and integration across hierarchies, while localized regions deal more with data processing.

Assuming that data integration requires cross hierarchy connections (the amount of data that a hierarchy receives from other hierarchies—the centrality of the hierarchy Rubinov and Sporns 2010) and data processing depend on interconnected regions (the amount of calculations taking place inside the hierarchy—specialized processing within densely interconnected module Rubinov and Sporns 2010), then data integration and processing seem to increase as we step up in the hierarchies. These findings could therefore suggest a flow of information from the lowest to the highest hierarchy with every hierarchy integrating more data and executing further processing, in line with previous studies and theoretical works (Christoff and Gabrieli 2000, Damasio 1999, Gray *et al* 2002, Northoff *et al* 2006). The low hierarchy receives information, performs specific calculations with its small amount of intra connections and passes the information to the higher hierarchies. The middle hierarchy is further able to integrate more data and locally process more information. At the top, the nucleus receives the most information from all other hierarchies and executes further processing using its dense interconnections, suggesting its vital involvement in data integration within the cortical network.

## The nucleus as a platform for consciousness

The regions in the nucleus form one component and constitute the most connected neighborhoods in the cortical network with the highest degrees. In contrast to the giant component, which mostly exhibits local structures (i.e. high clustering coefficient), all the regions in the nucleus form global structures (see supplementary material 4) and densely connect within themselves creating a unique *interconnected collective* all over the brain that demonstrates the properties of a module. The regions and profile of this collective are consistent with previous works (Hagmann *et al* 2008, van den Heuvel and Sporns 2011, Collin *et al* 2014), mostly comprised of posterior medial and parietal regions. Furthermore, in Hagmann *et al*'s structural cortical core, 70% of the core's edges were self-connections, similar to our findings within the high hierarchy (72%). In addition, this structural core forms one module and connected with *connector hubs* to all other modules in the network, reflecting our results that the nucleus is a single interconnected module with increased global structures. These findings further suggest that the distributed high hierarchy is composed of such connector hubs, in charge of connecting other hierarchies with the nucleus.

A strong inter-connected nucleus has also been demonstrated by Sporns *et al* suggesting a rich club organization of the human connectome (van den Heuvel and Sporns 2011, van den Heuvel *et al* 2013, Collin *et al* 2014). Their results revealed a group of '12 strongly interconnected bihemispheric hub regions, comprising, in the cortex, the precuneus, superior frontal and superior parietal cortex'. These six cortical regions were part of our more detailed interconnected nucleus which further includes more regions of the high hierarchy (see table 1). This interconnected collective module creates one global structure, involving regions from all over the cortex, which may create one global function. Given recent theories that explain consciousness as a complex process of global data integration (Tononi and Edelman 1998, Damasio 1999, Dehaene and Naccache 2001, Balduzzi and Tononi 2008, Godwin *et al* 2015), in particular Global Work space Theory and integrated information theory (Tononi and Edelman 1998, Dehaene and Naccache 2001, Balduzzi and Tononi 2008), one can postulate that such global function could be related to conscious abilities. We therefore suggest that *the global interconnected collective module of the nucleus can serve as a platform for consciousness to emerge*. Integrated information theory suggests that, 'to generate consciousness, a physical system must have a large repertoire of available states (information) and it must be unified, i.e. it should not be decomposable into a collection of causally independent subsystems (integration)' (Tononi and Edelman 1998, Balduzzi and Tononi 2008). The nucleus can satisfy both of these requirements, receiving the most information from all other hierarchies, choosing relevant information from all different types of information and integrating it to a unified function using its global interconnected collective.

Indeed, all of the regions in the nucleus have been previously correlated to consciousness activities (Goodale and Milner 1992, Gray *et al* 2002, Northoff and Bermpohl 2004, Achard *et al* 2006, Northoff *et al* 2006, Christoff *et al* 2009, Godwin *et al* 2015), especially midline and fronto-parietal regions. The nucleus, receiving the most information from all other hierarchies and integrating it to a unified global function, is therefore a perfect candidate to be the high integrative, global work space region in which consciousness can emerge (for more information see supplementary material 5).

## Study limitations

Some limitation issues have to be taken into account when interpreting the current results. First, our network is limited only to the cortex. The 'real' brain goes beyond the data being analyzed here. Future studies should examine the entire brain network and include the insula and subcortical regions in order to determine the exact profile of the hierarchies and the nucleus. It is possible, for instance, that regions within the low hierarchy (e.g. the fusiform gyrus) might belong to higher hierarchies and are affected by lack of subcortical regions (such as the hippocampus). Another possibility is that some subcortical regions such as the thalamus would be part of the nucleus (van den Heuvel and Sporns 2011). Lastly, the structural connections of our network were mapped with DSI followed by computational tractography (Hagmann *et al* 2003, 2007, 2008, Schmahmann *et al* 2007). Although DSI has been shown to be especially sensitive with regard to detecting fiber crossings (Hagmann *et al* 2003, 2007, 2008, Schmahmann *et al* 2007), it must be noted that this method may be influenced by errors in fiber reconstruction, and systematic detection biases. Reveley *et al* demonstrated another limitation of the DTI and DSI techniques (Reveley *et al* 2015), in which the local association fibers near the cortex gray matter impede tractography by acting as barriers that prevent communication of track lines between the cortex and deeper white matter; thus limiting the detectability of long cortical connections throughout the brain.

## Conclusions

The current study used *k*-shell decomposition analysis in order to reveal the global functional organization of the human cortical network. Consequently, we built a model of human cortex topology and revealed the hierarchical structure of the cortical network. In addition, this analysis proved to be more accurate than standard methods in the characterization of cortical regions and hierarchies. Our findings indicate that the human cortex is highly connected and efficient, compared to other networks, comprised of a nucleus and a giant component with virtually no isolated nodes. The giant component consists of different connectivity shells, which we categorized into three hierarchies representing an increasing number of regional connections. Such a topological model could support an efficient flow of information from the lowest hierarchy to the highest one, with each step enabling more data integration and data processing. At the top, the highest hierarchy (the *global interconnected collective module*) receives information from all previous hierarchies, integrates it into one global function and thus might serve as a platform for consciousness to emerge.

## Acknowledgments

We would like to thank Mr Kobi Flax for his crucial role in data analysis, without him this work would not be accomplished! We would also like to thank Dr Itay Hurvitz for his invaluable comments. We thank the European MULTIPLEX (EU-FET project 317532) project, the Israel Science Foundation, ONR and DTRA for financial support.

## References

Achard S, Salvador R, Whitcher B, Suckling J and Bullmore E 2006 A resilient, low-frequency, small-world human brain functional network with highly connected association cortical hubs *J. Neurosci.* **26** 63–72
Adler J 1991 Bootstrap percolation *Physica* A **171** 453–70
Alvarez-Hamelin J I, Dall'Asta L, Barrat A and Vespignani A 2005a K-core decomposition of Internet graphs: hierarchies, self-similarity and measurement biases (arXiv:cs/0511007)
Alvarez-Hamelin J I, Dall'Asta L, Barrat A and Vespignani A 2005b Large scale networks fingerprinting and visualization using the k-core decomposition *Conf. on Advances in Neural Information Processing Systems 18* pp 41–50
Balduzzi D and Tononi G 2008 Integrated information in discrete dynamical systems: motivation and theoretical framework *PLoS Comput. Biol.* **4** e1000091
Bassett D S, Bullmore E, Verchinski B A, Mattay V S, Weinberger D R and Meyer-Lindenberg A 2008 Hierarchical organization of human cortical networks in health and schizophrenia *J. Neurosci.* **28** 9239–48
Buckner R L, Sepulcre J, Talukdar T, Krienen F M, Liu H, Hedden T, Andrews-Hanna J R, Sperling R A and Johnson K A 2009 Cortical hubs revealed by intrinsic functional connectivity: mapping, assessment of stability, and relation to Alzheimer's disease *J. Neurosci.* **29** 1860–73
Bullmore E and Sporns O 2009 Complex brain networks: graph theoretical analysis of structural and functional systems *Nat. Rev. Neurosci.* **10** 186–98
Carmi S 2007 A model of internet topology using K-shell decomposition *Proc. Natl Acad. Sci.* **104** 11150–4
Christoff K and Gabrieli J D 2000 The frontopolar cortex and human cognition: evidence for a rostrocaudal hierarchical organization within the human prefrontal cortex *Psychobiology* **28** 168–86
Christoff K, Gordon A M, Smallwood J, Smith R and Schooler J W 2009 Experience sampling during fMRI reveals default network and executive system contributions to mind wandering *Proc. Natl Acad. Sci.* **106** 8719–24
Cohen R and Havlin S 2010 *Complex Networks: Structure, Robustness and Function* (Cambridge: Cambridge University Press)
Colizza V and Vespignani A 2007 Invasion threshold in heterogeneous metapopulation networks *Phys. Rev. Lett.* **99** 148701
Collin G, Sporns O, Mandl R C and van den Heuvel M P 2014 Structural and functional aspects relating to cost and benefit of rich club organization in the human cerebral cortex *Cereb. Cortex* **24** 2258–67
Damasio A R 1999 *The Feeling of What Happens: Body and Emotion in the Making of Consciousness* (Boston, MA: Houghton Mifflin Harcourt)
Dehaene S and Naccache L 2001 Towards a cognitive neuroscience of consciousness: basic evidence and a workspace framework *Cognition* **79** 1–37
Eguiluz V M, Chialvo D R, Cecchi G A, Baliki M and Apkarian A V 2005 Scale-free brain functional networks *Phys. Rev. Lett.* **94** 018102
Ekman M, Derrfuss J, Tittgemeyer M and Fiebach C J 2012 Predicting errors from reconfiguration patterns in human brain networks *Proc. Natl Acad. Sci.* **109** 16714–9
Garas A, Argyrakis P, Rozenblat C, Tomassini M and Havlin S 2010 Worldwide spreading of economic crisis *New J. Phys.* **12** 113043
Gazzaniga M S and Sperry R W 1967 Language after section of the cerebral commissures *Brain* **90** 131–48
Godwin D, Barry R L and Marois R 2015 Breakdown of the brain's functional network modularity with awareness *Proc. Natl Acad. Sci.* 201414466
Goh K-I, Cusick M E, Valle D, Childs B, Vidal M and Barabási A-L 2007 The human disease network *Proc. Natl Acad. Sci.* **104** 8685–90
Goodale M A and Milner A D 1992 Separate visual pathways for perception and action *Trends Neurosci.* **15** 20–5
Gray J R, Braver T S and Raichle M E 2002 Integration of emotion and cognition in the lateral prefrontal cortex *Proc. Natl Acad. Sci.* **99** 4115–20
Hagmann P, Cammoun L, Gigandet X, Meuli R, Honey C J, Wedeen V J and Sporns O 2008 Mapping the structural core of human cerebral cortex *PLoS Biol.* **6** e159
Hagmann P, Kurant M, Gigandet X, Thiran P, Wedeen V J, Meuli R and Thiran J P 2007 Mapping human whole-brain structural networks with diffusion MRI *PLoS One* **2** e597

Hagmann P, Thiran J P, Jonasson L, Vandergheynst P, Clarke S, Maeder P and Meuli R 2003 DTI mapping of human brain connectivity: statistical fibre tracking and virtual dissection *Neuroimage* **19** 545–54

Harriger L, van den Heuvel M P and Sporns O 2012 Rich club organization of macaque cerebral cortex and its role in network communication *PloS One* **7** e46497

He Y, Chen Z J and Evans A C 2007 Small-world anatomical networks in the human brain revealed by cortical thickness from MRI *Cereb. Cortex* **17** 2407–19

Modha D S and Singh R 2010 Network architecture of the long-distance pathways in the macaque brain *Proc. Natl Acad. Sci.* **107** 13485–90

Newman M E J 2003 The structure and function of complex networks *SIAM Rev.* **45** 167–256

Northoff G and Bermpohl F 2004 Cortical midline structures and the self *Trends Cogn. Sci.* **8** 102–7

Northoff G, Heinzel A, de Greck M, Bermpohl F, Dobrowolny H and Panksepp J 2006 Self-referential processing in our brain—a meta-analysis of imaging studies on the self *Neuroimage* **31** 440–57

Pittel B, Spencer J and Wormald N 1996 Sudden emergence of a giantk-core in a random graph *J. Comb. Theory* B **67** 111–51

Ponten S C, Bartolomei F and Stam C J 2007 Small-world networks and epilepsy: graph theoretical analysis of intracerebrally recorded mesial temporal lobe seizures *Clin. Neurophysiol.* **118** 918–27

Raz A and Buhle J 2006 Typologies of attentional networks *Nat. Rev. Neurosci.* **7** 367–79

Reijneveld J C, Ponten S C, Berendse H W and Stam C J 2007 The application of graph theoretical analysis to complex networks in the brain *Clin. Neurophysiol.* **118** 2317–31

Reveley C, Seth A K, Pierpaoli C, Silva A C, Yu D, Saunders R C, Leopold D A and Frank Q Y 2015 Superficial white matter fiber systems impede detection of long-range cortical connections in diffusion MR tractography *Proc. Natl Acad. Sci.* **112** E2820–8

Rubinov M and Sporns O 2010 Complex network measures of brain connectivity: uses and interpretations *Neuroimage* **52** 1059–69

Schmahmann J D, Pandya D N, Wang R, Dai G, D'Arceuil H E, de Crespigny A J and Wedeen V J 2007 Association fibre pathways of the brain: parallel observations from diffusion spectrum imaging and autoradiography *Brain* **130** 630–53

Sporns O, Chialvo D R, Kaiser M and Hilgetag C C 2004 Organization, development and function of complex brain networks *Trends Cogn. Sci.* **8** 418–25

Sporns O and Zwi J D 2004 The small world of the cerebral cortex *Neuroinformatics* **2** 145–62

Stam C J, Jones B F, Nolte G, Breakspear M and Scheltens P 2007 Small-world networks and functional connectivity in Alzheimer's disease *Cereb. Cortex* **17** 92–9

Stam C J and Reijneveld J C 2007 Graph theoretical analysis of complex networks in the brain *Nonlinear Biomed. Phys.* **1** 3

Tononi G and Edelman G M 1998 Consciousness and complexity *Science* **282** 1846–51

van den Heuvel M P and Pol H E H 2010 Exploring the brain network: a review on resting-state fMRI functional connectivity *Eur. Neuropsychopharmacology* **20** 519–34

van den Heuvel M P and Sporns O 2011 Rich-club organization of the human connectome *J. Neurosci.* **31** 15775–86

van den Heuvel M P, Sporns O, Collin G, Scheewe T, Mandl R C, Cahn W, Goñi J, Pol H E H and Kahn R S 2013 Abnormal rich club organization and functional brain dynamics in schizophrenia *JAMA Psychiatry* **70** 783–92

van den Heuvel M P, Stam C J, Boersma M and Pol H H 2008 Small-world and scale-free organization of voxel-based resting-state functional connectivity in the human brain *Neuroimage* **43** 528–39

### 3.1.1. K-shell decomposition reveals hierarchical cortical organization of the human brain

## *Supporting information*

# K-shell decomposition reveals hierarchical cortical organization of the human brain

## Supporting information

### Supplementary material 1: Comparison to standard methods

In order to examine whether the hierarchical layers of the cortical network, as revealed by k shell decomposition method, could be detected by a simple degree count we compared our results with standard methods that take into account only node degree. First we compared cortical assortativity using the regular method (calculating the likelihood hubs will connect to other hubs) compared with the likelihood nodes from high shells will connect other nodes from similar shells. Results revealed higher assortativity (0.46) for high-shell nodes to connect together (0.06 in random cortices) than for hubs to connect together (0.16; -0.01 in random cortices) suggesting that shell characterization provides more accurate description of network properties.

Second, we compared the correlation between connectivity matrices across networks by node degrees and by examining node shells. While the correlation by node degree across all networks was on average $r^2= 0.49$, the correlation according to node shell was significantly higher, on average $r^2=0.63$ ($p< 2E-7$). Third, we compared the average degree of different nodes comprising an anatomical region with their shell distribution. In order to compare the different distributions, we normalized the degree distribution and the shell distribution to be between 0 and 1(see Fig. S4). On average, shell distribution was narrow and more localized compared to the degree distribution of each anatomical region. In accordance, average variance of the degree distribution was significantly higher than shell distribution variance (0.026 vs. 0.015, respectively, $p<0.007$). This finding suggests that shell distribution represent the different anatomical regions more accurately compared to degree distribution.

Next, we examined whether the shell profile of different regions is similar to that detected by a simple degree count. When examining anatomical regions with large number of hubs or with high hub density (see Fig. S8 and Fig. S5), similarities can be found between these areas and the nucleus (i.e. precuneus, anterior and posterior cingulate). However, when examined more carefully, significant differences appear. For example, regions like the left pericalcarine cortex, postcentral gyrus, right lingual gyrus and more, were always in the nucleus or in the highest hierarchy though their amount or density of hubs is relatively low and thus could not be predicted by counting degrees. Laterality effects might also be overlooked when simply counting degrees. For instance, examining primary motor cortex and Wernicke's area by counting their degrees display a similar degree in both hemispheres. In contrast, right homologue of Wernicke's area and right primary motor region were found in the middle hierarchy whereas their left counterparts were found in the high hierarchy (and left primary motor region always reached the nucleus). Therefore, K shell decomposition analysis could prove more useful when trying to determine language lateralization and motor dominance.

Other laterality effects that might be overlooked when simply counting degrees include regions like the superior temporal cortex and inferior parietal cortex. These regions display high amount or density of hubs in both hemispheres, but looking at average shell only left regions were found in the highest hierarchy (and always in the nucleus) while right hemisphere regions never reach the nucleus (or the highest hierarchy). The opposite case was also found where regions like the rostral middle frontal gyrus and pars orbitalis reach the highest hierarchy for the left hemisphere although they don't have large amount or high density of hubs in either hemisphere. To summarize, comparing standard methods that take into account only the degree of the nodes with k-shell decomposition revealed that k-shell analysis is more accurate and could provide better resolution of network properties.

**Supplementary material 2: Network based shell level and functional lateralization**

Interestingly, average shell level often reflected known functional lateralization as detailed in Table 2. Moreover, examining network based shell level revealed high left dominance across all examined networks. The salience and the sensorimotor networks (average shell level 17.3 and 17.5, respectfully) revealed leftward dominance in terms of the amount of regions that reach the nucleus (more than 50%, see Table 2). These networks had 73% of their regions in the high hierarchy with approximately 55% of them always reaching the nucleus. The executive control network had an average shell level of 16.8 with similar left dominance (64% left) 60% of its regions were found in the high hierarchy. The other 40% belong to the middle hierarchy (71% right and most of them never reach the nucleus). The functional network with the highest average shell level was the default mode network (DMN) with a score of 18.1. 81% of its regions were found in the high hierarchy with 70% always reaching the nucleus. Accordingly, there was a 56% overlap between the regions that always reach the nucleus and the DMN (42% overlap in the salience network; 40% in the sensorimotor network and only 22% overlap in the executive control network).

The highest Overlap with regions that never reach the nucleus was found in the auditory cortex, 75% of its regions never reach the nucleus. Next, the dorsal and the ventral visual streams with 40% overlap. Executive control network and salience network display 20% overlap and the DMN \sensorimotor both display 12% overlap with the regions that never reach the nucleus. Interestingly all the regions that never appear in the nucleus (across all 6 networks) belong to the right hemisphere. On the other hand, a strong tendency towards the left hemisphere was evident in the nucleus. Nearly all the regions in the nucleus were midline or left hemisphere regions. Roughly speaking, most of the left hemisphere was comprised of high hierarchy regions and most of the right hemisphere was comprised of middle hierarchy regions (see Fig. 4). This lateralization effect was found in most functional networks (executive, dorsal stream, sensorimotor and salience) except midline networks (DMN and the visual cortex). Left lateralization was also evident in many areas and networks across hierarchies. This effect was especially evident in the middle hierarchy; for instance 88% of the regions in localized middle belong to the right hemisphere while most of their left homologues were found in the high hierarchy. For instance, the right homologue of Wernicke's area, right primary motor region and right TPJ were found in the middle hierarchy (and also never reach the nucleus) whereas their left counterparts were found in the high hierarchy (and left primary motor region always reached the nucleus). These findings can be explained by hemispheric dominance given that all of our subjects were right handed and also by well-known language lateralization (Gazzaniga and Sperry, 1967). K-shell decomposition analysis managed to recover these known functional attributes which were not detected in regular methods using degree count.

**Supplementary material 3: The creation of the giant component**

The formation of the giant component was similar to a first order phase transition. When looking at this formation process two critical points can be detected: the first critical point can be detected at shells 14-15 when the first hubs were removed and enabled the creation of the giant component from islands of removed nodes. A second critical point occurred when a large number of nodes (37% of all the nodes in the giant component, see Fig. S1) were removed at once and joined the giant component (on average shell 18, one shell before the nucleus was revealed). We used these critical points in order to detect and distinguish between different shell hierarchies. The first critical point distinguishes between the low hierarchy and the middle hierarchy and the formation of the nucleus distinguished between the middle and the high hierarchy. More specifically, at first, nodes from the low hierarchy were removed, but only when nodes from the **distributed middle** hierarchy were being removed (shell average ~14.5 – first critical point) enough hubs were added to connect all removed nodes and create one giant component. Later, when nodes from the **localized middle** hierarchy (shell average of 16.7) and nodes from the **distributed high** hierarchy (shell average of 17 – second critical point) were removed, an additional large amount of hubs were removed along with their neighborhoods. Third critical point in the cortical network occurred on average in shell 19 when the nucleus was revealed (corresponding with the average shell level of distributed high- 16.92 (±2.82) and localized high- 19.30 (±0.97) hierarchies) distinguishing the high hierarchy.

**Supplementary material 4: comparison between the whole network and the giant component**

The cortical network and the giant component that comprise it, both have small world properties. The fact that the giant component alone has small world characteristics (high clustering coefficient and low average distance) (Newman, 2003) suggests that it's a very organized and efficient sub network similar to the highly connected nucleus. Furthermore, the cortical network exhibits a hierarchical structure (not to be confused with the hierarchies derived from k-shell decomposition analysis). In which, as the node degree increases the clustering coefficient decreases. Thus, the nucleus, which has high average degree, does not increase the clustering coefficient of the network (see Supplementary material 6). The average clustering coefficient of the whole network is just like the average clustering coefficient of the giant component, making the nucleus negligible in that perspective. It seems that most of the clustering coefficients (local structures) of the entire network were due to the giant component. On the other hand, as the degree of a node increases its average distance decreases (see Fig. S7) meaning that hubs decrease the average distance of the network (Newman, 2003). In accordance, the nucleus reduced the average distance of the network making it shorter than that of the giant component (see Fig. S2). These results suggest that the majority of local processing is conducted in the giant component while the nucleus mainly adds shortcuts and global structures to the network.

**Supplementary material 5: correlations between the nucleus and consciousness activity**

When correlating disorders of consciousness with brain activity, it has been shown that consciousness activity correlates with midline and fronto-parietal regional activities and with high connectivity levels of the precuneus and posterior cingulate (Vanhaudenhuyse *et al.*, 2009). Interestingly, all of these regions were part of the nucleus (midline areas, superior frontal cortex, sensorimotor cortex, inferior and superior parietal cortex). Northoff et al. suggested that cortical midline structures, are essential components in generating a model of the self and can be referred to as the "core self"(Northoff and Bermpohl, 2004; Northoff *et al.*, 2006). According to this model a flow of information from the medial orbitofrontal and the frontal pole (both in the middle hierarchy) to cortical midline regions (the nucleus) enable the creation of a self-model, supporting a functional hierarchy as revealed by the k shell analysis. Furthermore, the default network, which has 81% overlap with the nucleus, has been shown to reflect internally focused thought that can occur in the form of mind wandering (Christoff *et al.*, 2009; Mason *et al.*, 2007; Raichle *et al.*, 2001; Smallwood and Schooler, 2006). Activation in the medial prefrontal part of the default network was specifically observed in association with subjective self-reports of mind wandering (Buckner *et al.*, 2008; Christoff *et al.*, 2009). Dominance of cingulate regions (found in the nucleus) were also associated with creative thinking (Christoff *et al.*, 2009; Kounios and Beeman, 2009; Kounios *et al.*, 2008; Kounios *et al.*, 2006; Spreng *et al.*, 2009), where executive regions such as the caudal anterior cingulate and the posterior cingulate cortex were activated before solving problems with insight.

Integrated information theory suggests that, "to generate consciousness, a physical system must have a large repertoire of available states (information) and it must be unified, i.e. it should not be decomposable into a collection of causally independent subsystems (integration)" (Balduzzi and Tononi, 2008; Tononi and Edelman, 1998). The nucleus can satisfy both of these requirements, receiving the most information from all other hierarchies, choosing relevant information from all different types of information and integrating it to a unified function using its global interconnected collective. According to the global work space theory (Baars, 1997; Dehaene and Changeux, 2003; Dehaene *et al.*, 1998; Dehaene and Naccache, 2001), consciousness should emerge in "a distributed neural system or `workspace' with long-distance connectivity that can potentially interconnect multiple specialized brain areas in a coordinated, though variable manner" (Dehaene and Naccache, 2001). Being an interconnected collective, the nucleus is a perfect candidate to be the region of global work space in which consciousness can emerge. According to Dehaene (Dehaene *et al.*, 1998), one requirement for the global work space is that neurons contributing to the workspace area should be distributed in at least five categories of circuits: high-level perceptual, motor, long-term memory, evaluative and attentional networks. Therefore, regions of the pre frontal, anterior cingulate, inferior parietal and speech production circuits in the left inferior frontal lobe (for the intentional guidance of actions), play a major role in the conscious workspace. In accordance, the nucleus revealed by k shell decomposition analysis integrates all of these categories: inferior parietal as high level perception, sensorimotor regions in the motor category, precuneus and posterior cingulate as integration of long term memory and anterior cingulate as evaluative and attentional regions (Northoff and Bermpohl, 2004; Northoff *et al.*, 2006). Since the nucleus contain more regions than required for the global workspace, according to Dehaene, using K shell decomposition might enable a better prediction of all the regions needed for conscious activity.

**Supplementary material 6: Cortical networks exhibit hierarchal structure**

Prior to applying k shell decomposition analysis, we examined basic properties of the cortical network. The degrees distribution of the cortical network best fitted a normal distribution (with a mean degree of 29.18±15). In accordance with previous work (Bassett, 2006; Bullmore and Sporns, 2009; Hagmann *et al.*, 2008; Hagmann *et al.*, 2007; He *et al.*, 2007; Reijneveld *et al.*, 2007; Rubinov and Sporns, 2010; Sporns *et al.*, 2004; Sporns *et al.*, 2000; Sporns and Zwi, 2004; Stam and Reijneveld, 2007; Van Den Heuvel and Pol, 2010; van den Heuvel *et al.*, 2008; Wang *et al.*, 2010), the cortical network exhibited "small world" organization, as well as a hierarchal structure (i.e. clustering coefficient distribution best fitted a power low distribution, $C \sim K^{-\beta}$ with $\beta=0.36$, meaning that as the node degree (K) increases the clustering coefficient (C) decreases). These results are depicted in figure S6a. Furthermore, as the degree of a node increased its average distance to the rest of the network decreased (see Fig. S7), suggesting that hubs in the cortex indeed add shortcuts to the network and connect nodes that are not directly connected. Contrary to the cortical network, no hierarchal structure was found in a random brain network (preserving the original network's degree distribution), i.e. clustering coefficient was constant and independent of the degree (see Fig. S6b).The cortical network also revealed positive average assortativity of 0.16, in contrast to a random network, in which average negative assortativity was -0.01. This finding suggests that hubs tend to connect other hubs more often in the cortical network compared to a random one.

**Supplementary References:**

Baars B J 1997 *In the theater of consciousness: The workspace of the mind*: Oxford University Press)

Balduzzi D and Tononi G 2008 Integrated information in discrete dynamical systems: motivation and theoretical framework *PLoS computational biology* **4** e1000091

Bassett D S 2006 Small World Brain Networks *The Neuroscientist* **12**

Buckner R L, Andrews-Hanna J R and Schacter D L 2008 The brain's default network *Annals of the New York Academy of Sciences* **1124** 1-38

Bullmore E and Sporns O 2009 Complex brain networks: graph theoretical analysis of structural and functional systems *Nature Reviews Neuroscience* **10** 186-98

Christoff K, Gordon A M, Smallwood J, Smith R and Schooler J W 2009 Experience sampling during fMRI reveals default network and executive system contributions to mind wandering *Proceedings of the National Academy of Sciences* **106** 8719-24

Dehaene S and Changeux J-P 2003 Neural mechanisms for access to consciousness *The cognitive neurosciences III*

Dehaene S, Kerszberg M and Changeux J-P 1998 A neuronal model of a global workspace in effortful cognitive tasks *Proceedings of the National Academy of Sciences* **95** 14529-34

Dehaene S and Naccache L 2001 Towards a cognitive neuroscience of consciousness: basic evidence and a workspace framework *Cognition* **79** 1-37

Fulton J F 1935 A note on the definition of the “motor” and “premotor” areas *Brain* **58** 311-6

Gazzaniga M S and Sperry R W 1967 Language after section of the cerebral commissures *Brain* **90** 131-48

Goodale M A and Milner A D 1992 Separate visual pathways for perception and action *Trends in neurosciences* **15** 20-5

Hagmann P, Cammoun L, Gigandet X, Meuli R, Honey C J, Wedeen V J and Sporns O 2008 Mapping the structural core of human cerebral cortex *PLoS Biol* **6** e159

Hagmann P, Kurant M, Gigandet X, Thiran P, Wedeen V J, Meuli R and Thiran J P 2007 Mapping human whole-brain structural networks with diffusion MRI *PLoS ONE* **2** e597

He Y, Chen Z J and Evans A C 2007 Small-world anatomical networks in the human brain revealed by cortical thickness from MRI *Cerebral cortex* **17** 2407-19

Kounios J and Beeman M 2009 The Aha! Moment the cognitive neuroscience of insight *Current directions in psychological science* **18** 210-6

Kounios J, Fleck J I, Green D L, Payne L, Stevenson J L, Bowden E M and Jung-Beeman M 2008 The origins of insight in resting-state brain activity *Neuropsychologia* **46** 281-91

Kounios J, Frymiare J L, Bowden E M, Fleck J I, Subramaniam K, Parrish T B and Jung-Beeman M 2006 The prepared mind neural activity prior to problem presentation predicts subsequent solution by sudden insight *Psychological Science* **17** 882-90

Mason M F, Norton M I, Van Horn J D, Wegner D M, Grafton S T and Macrae C N 2007 Wandering minds: the default network and stimulus-independent thought *Science* **315** 393-5

Newman M E J 2003 The structure and function of complex networks *SIAM review* **45** 167-256
Northoff G and Bermpohl F 2004 Cortical midline structures and the self *Trends in cognitive sciences* **8** 102-7
Northoff G, Heinzel A, de Greck M, Bermpohl F, Dobrowolny H and Panksepp J 2006 Self-referential processing in our brain—a meta-analysis of imaging studies on the self *Neuroimage* **31** 440-57
Owen A M, McMillan K M, Laird A R and Bullmore E 2005 N-back working memory paradigm: A meta-analysis of normative functional neuroimaging studies *Human brain mapping* **25** 46-59
Raichle M E, MacLeod A M, Snyder A Z, Powers W J, Gusnard D A and Shulman G L 2001 A default mode of brain function *Proceedings of the National Academy of Sciences* **98** 676-82
Reijneveld J C, Ponten S C, Berendse H W and Stam C J 2007 The application of graph theoretical analysis to complex networks in the brain *Clin Neurophysiol* **118** 2317-31
Rubinov M and Sporns O 2010 Complex network measures of brain connectivity: uses and interpretations *Neuroimage* **52** 1059-69
Seeley W W, Menon V, Schatzberg A F, Keller J, Glover G H, Kenna H, Reiss A L and Greicius M D 2007 Dissociable intrinsic connectivity networks for salience processing and executive control *The Journal of neuroscience* **27** 2349-56
Smallwood J and Schooler J W 2006 The restless mind *Psychological bulletin* **132** 946
Sporns O, Chialvo D R, Kaiser M and Hilgetag C C 2004 Organization, development and function of complex brain networks *Trends Cogn Sci* **8** 418-25
Sporns O, Tononi G and Edelman G M 2000 Theoretical neuroanatomy: relating anatomical and functional connectivity in graphs and cortical connection matrices *Cereb Cortex* **10** 127-41
Sporns O and Zwi J D 2004 The small world of the cerebral cortex *Neuroinformatics* **2** 145-62
Spreng R N, Mar R A and Kim A S 2009 The common neural basis of autobiographical memory, prospection, navigation, theory of mind, and the default mode: a quantitative meta-analysis *Journal of cognitive neuroscience* **21** 489-510
Stam C J and Reijneveld J C 2007 Graph theoretical analysis of complex networks in the brain *Nonlinear Biomed Phys* **1** 3
Tononi G and Edelman G M 1998 Consciousness and complexity *science* **282** 1846-51
Van Den Heuvel M P and Pol H E H 2010 Exploring the brain network: a review on resting-state fMRI functional connectivity *European Neuropsychopharmacology* **20** 519-34
van den Heuvel M P, Stam C J, Boersma M and Pol H H 2008 Small-world and scale-free organization of voxel-based resting-state functional connectivity in the human brain *Neuroimage* **43** 528-39
Vanhaudenhuyse A, Noirhomme Q, Tshibanda L J-F, Bruno M-A, Boveroux P, Schnakers C, Soddu A, Perlbarg V, Ledoux D and Brichant J-F 2009 Default network connectivity reflects the level of consciousness in non-communicative brain-damaged patients *Brain* awp313
Wang J, Zuo X and He Y 2010 Graph-based network analysis of resting-state functional MRI *Frontiers in systems neuroscience* **4**

## Supplementary Figures:

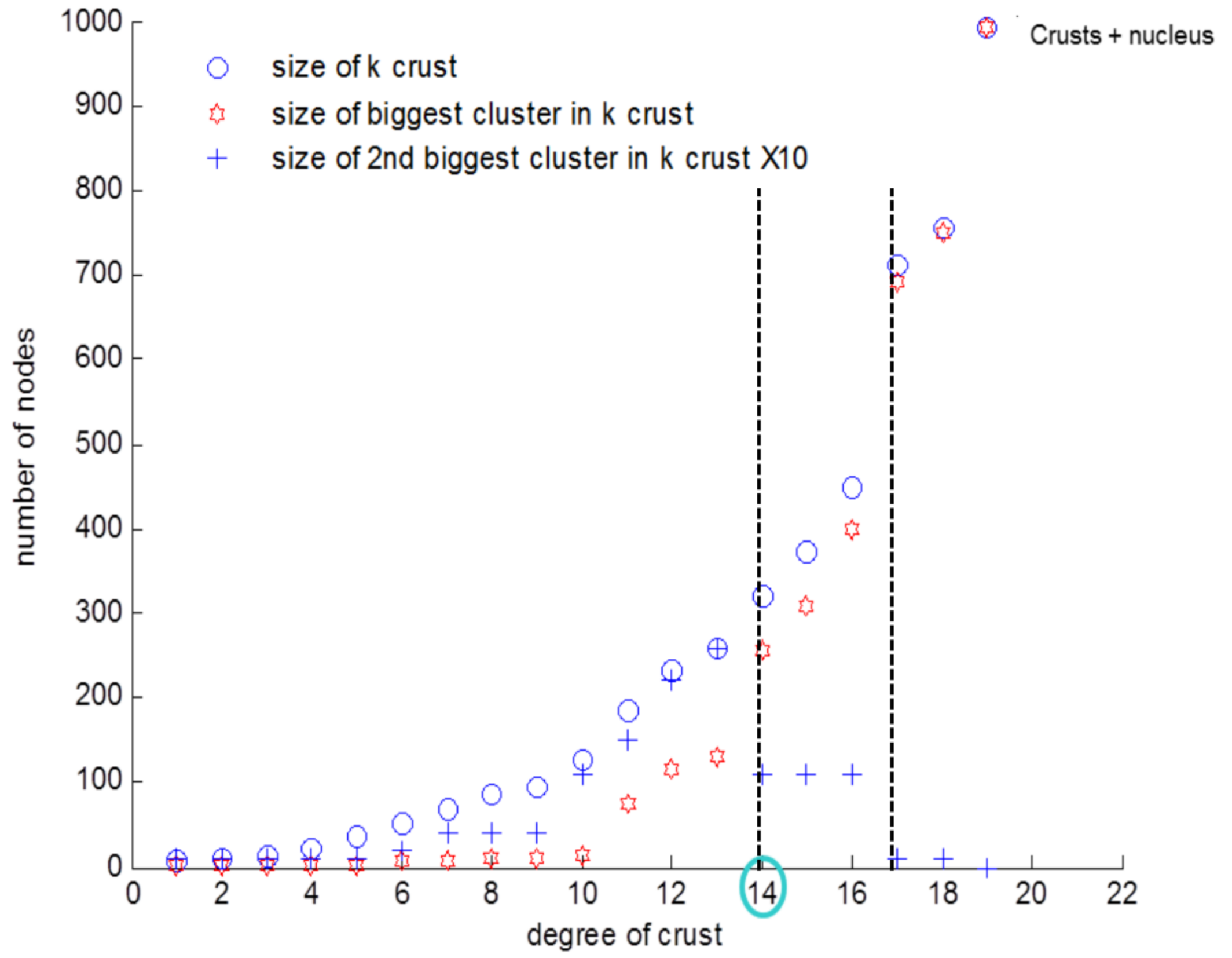

**Figure S1. The formation of the giant component.**
K-crust- includes the nodes that have been removed until step k of the process (K- the degree of the crust). The giant component was formed in a process similar to a first order phase transition. In the beginning of the process islands of removed nodes were forming and growing, but in k=14 (first critical point, left dashed line), most of these islands connect together to form the giant component and there is sharp decline in the size of the second biggest cluster in k-crust (i.e. its nodes were connected to the rest of the of the giant component and not to the second biggest cluster in k-crust). Later, in k=17, 261 nodes were removed at once to the giant component and the size of the second biggest cluster were reduced (second critical point, right dashed line). In k=19, the nucleus was revealed. X- axis: degree of the k-crust; Y-axis: number of removed Nodes; Blue cycles: size of k-crust; Red star: size of the biggest cluster in the k-crust (biggest island of the removed nodes); Blue plus: second biggest cluster in k-crust -second biggest island of the removed nodes (multiplied by 10 for viewing purposes). Last k denotes the whole network (crust + nucleus). Example taken from cortical network 1.

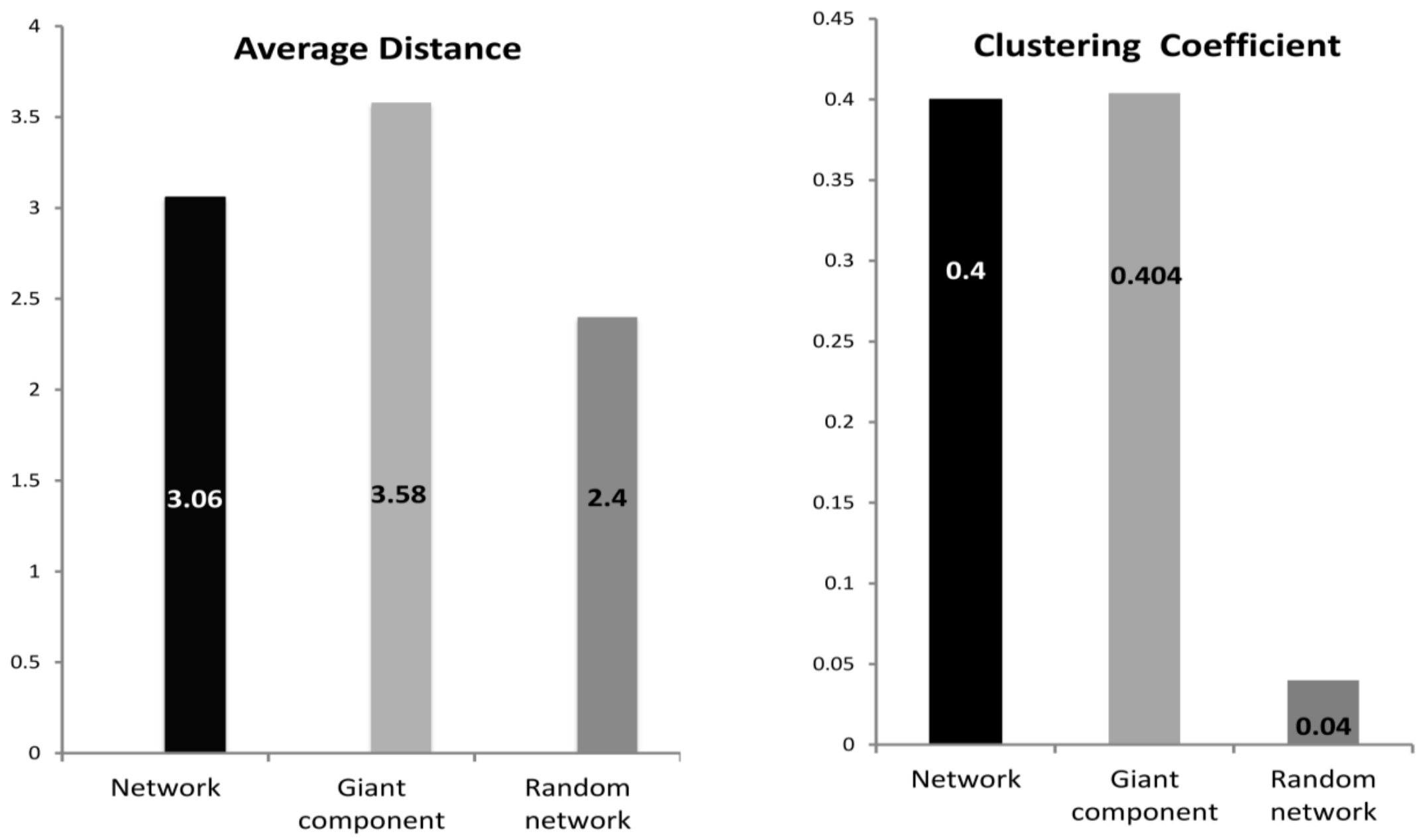

**Figure S2: Comparison between properties of the whole network and the giant component.**

Comparison between the average distance (left panel) and clustering coefficient (right panel) of the entire cortical network (red), the giant component (green) and the random cortical network (blue). Both the whole network (giant component + nucleus + isolated nodes) and the giant component have small world characteristics. Note that the entire network and the giant component have almost the same clustering coefficient while the entire network has smaller average distance compared to the giant component. These findings suggest that the nucleus mainly adds shortcuts and global (rather than local) structures to the whole network.

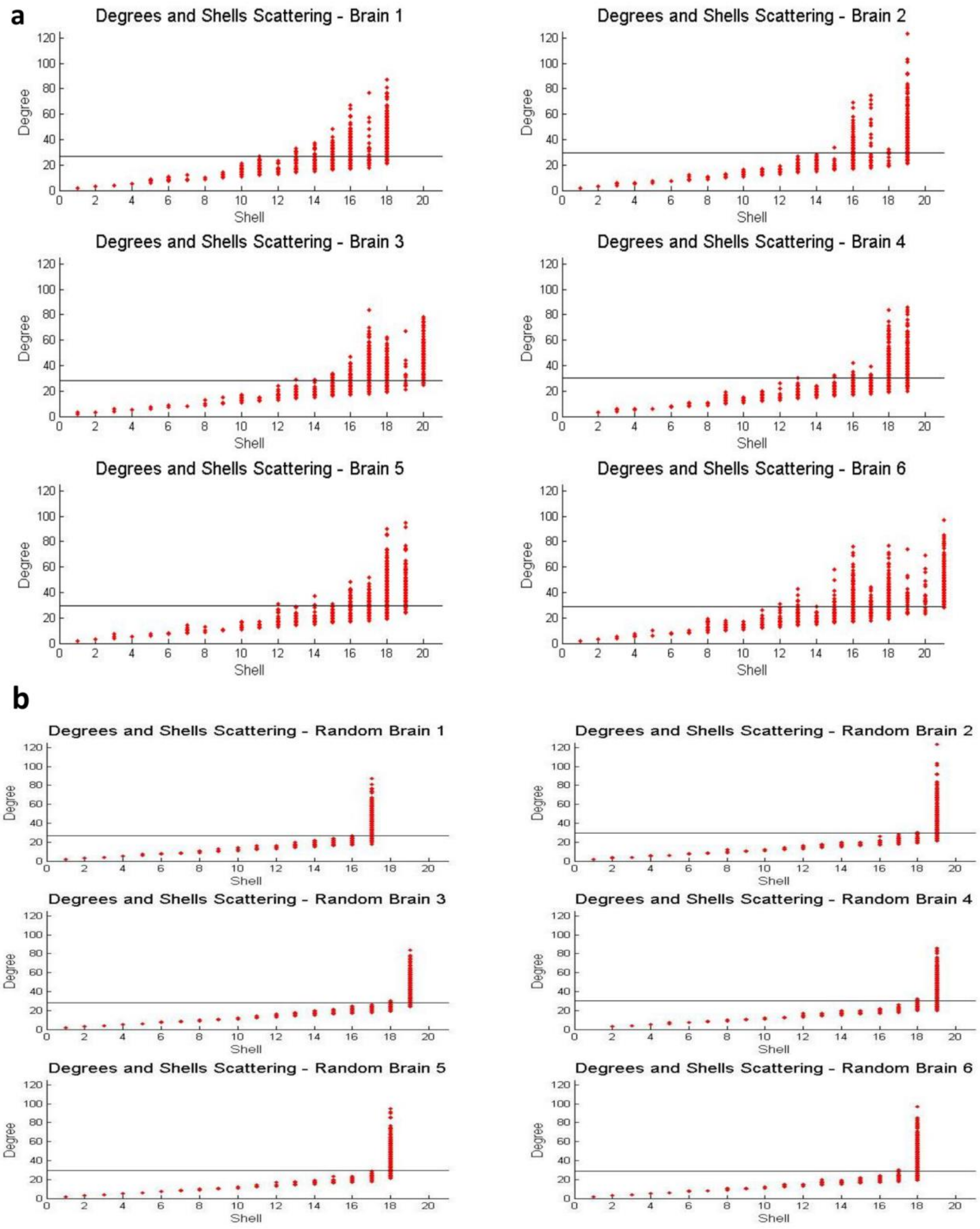

**Figure S3. Degree scattering across the different shells** for all networks. In the real cortices the hubs are distributed from shell 14 and above (A) while in the random cortices (B) all the hubs reached the nucleus. X-axis: k-shell; Y-axis: degree; the last k is the nucleus; Black line indicates average degree of the network.

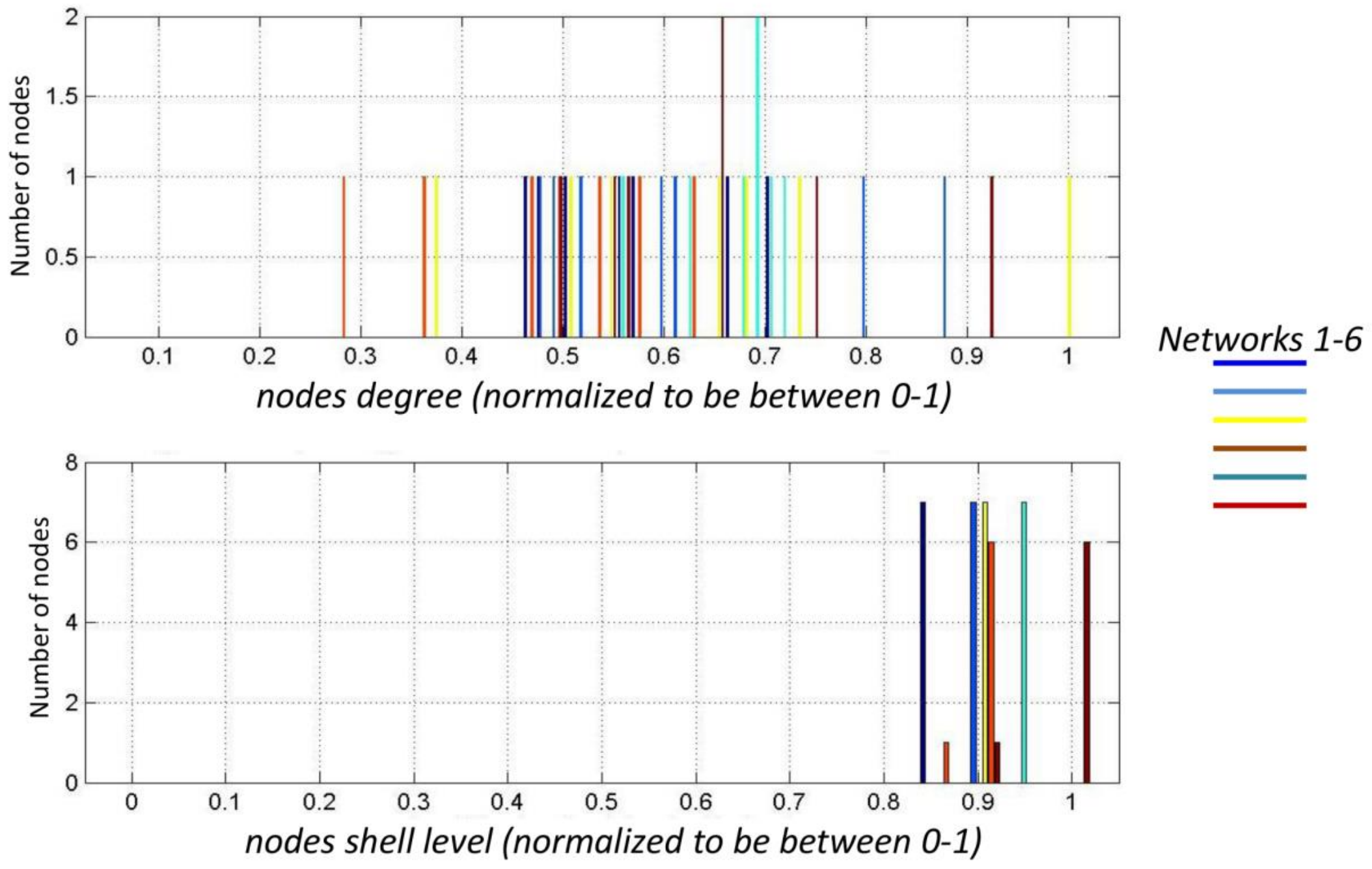

**Figure S4: Comparison between degree distribution and shell distribution** of right posterior cingulate. Different colors denote different cortical networks. Upper graph- X axis: nodes degree (normalized to be between 0-1); Y axis: number of nodes. Bottom graph- X axis: nodes shell level (normalized to be between 0-1); Y axis: number of nodes. The variance of the shells distribution is much smaller than the variance of the degree distribution. The same effect is shown in the variance between networks (see supplementary data 2).

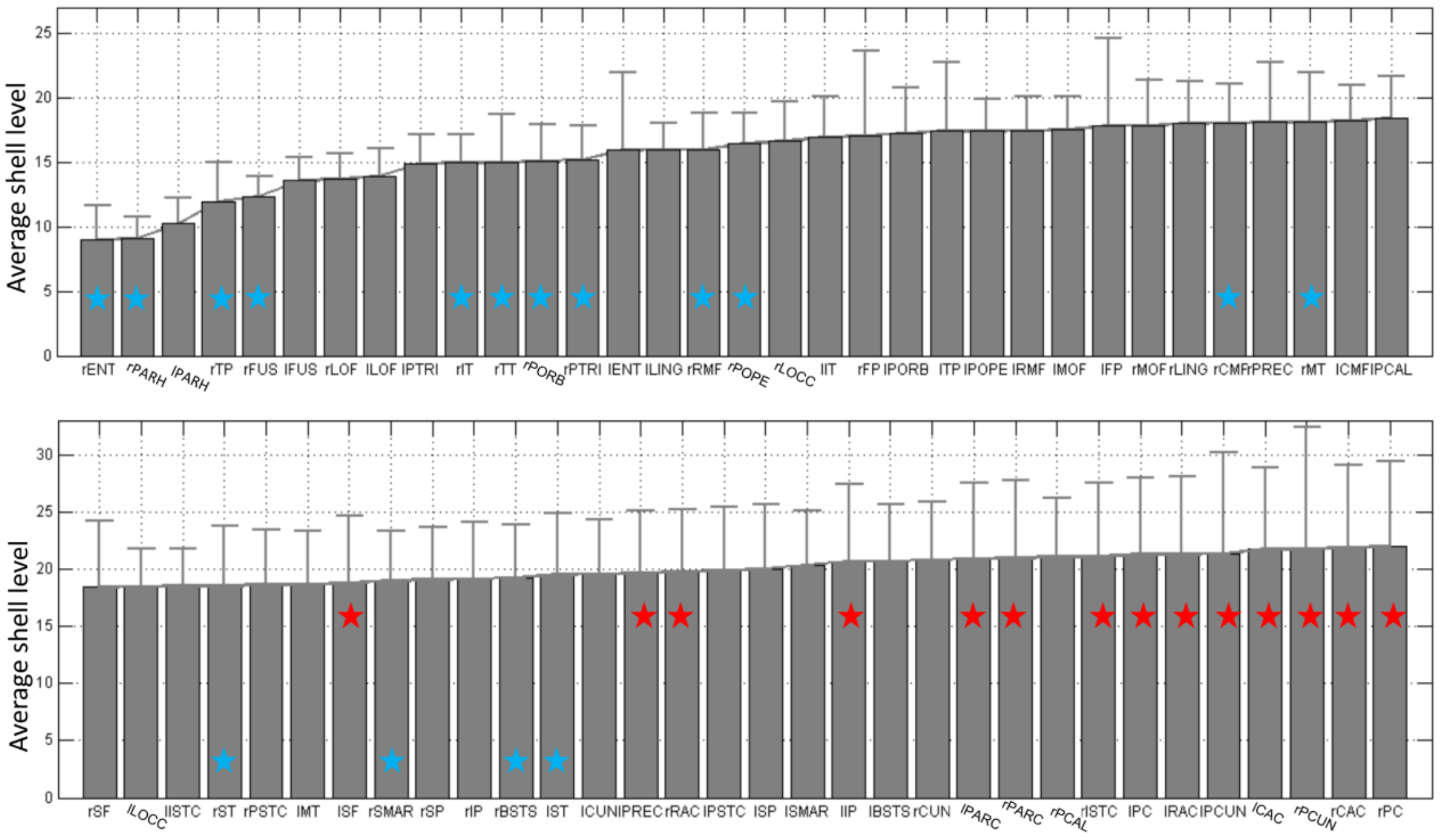

**Figure S5: Average shell level of each anatomical region.**

Bar graphs depicting the mean shell level of each anatomical region. Red asterisks denote regions that are always in the nucleus while blue asterisks denote regions that are never in the nucleus (across all six networks). The 66 cortical regions were labeled as follows (following Hagmann(Hagmann *et al.*, 2008)): each label consists of two parts, a prefix for the cortical hemisphere (r=right hemisphere, l=left hemisphere) and one of 33 designators: BSTS=bank of the superior temporal sulcus, CAC = caudal anterior cingulate cortex, CMF = caudal middle frontal cortex, CUN = cuneus, ENT = entorhinal cortex, FP frontal pole, FUS = fusiform gyrus, IP = inferior parietal cortex, IT = inferior temporal cortex, ISTC = isthmus of the cingulate cortex, LOCC = lateral occipital cortex, LOF = lateral orbitofrontal cortex, LING =lingual gyrus, MOF = medial orbitofrontal cortex, MT = middle temporal cortex, PARC = paracentral lobule, PARH=parahippocampal cortex, POPE=pars opercularis, PORB=pars orbitalis, PTRI=pars triangularis, PCAL=pericalcarine cortex, PSTS = postcentral gyrus, PC = posterior cingulate cortex, PREC = precentral gyrus, PCUN = precuneus, RAC = rostral anterior cingulate cortex, RMF =rostral middle frontal cortex, SF=superior frontal cortex, SP=superior parietal cortex, ST=superior temporal cortex, SMAR=supramarginal gyrus, TP=temporal pole, and TT = transverse temporal cortex.

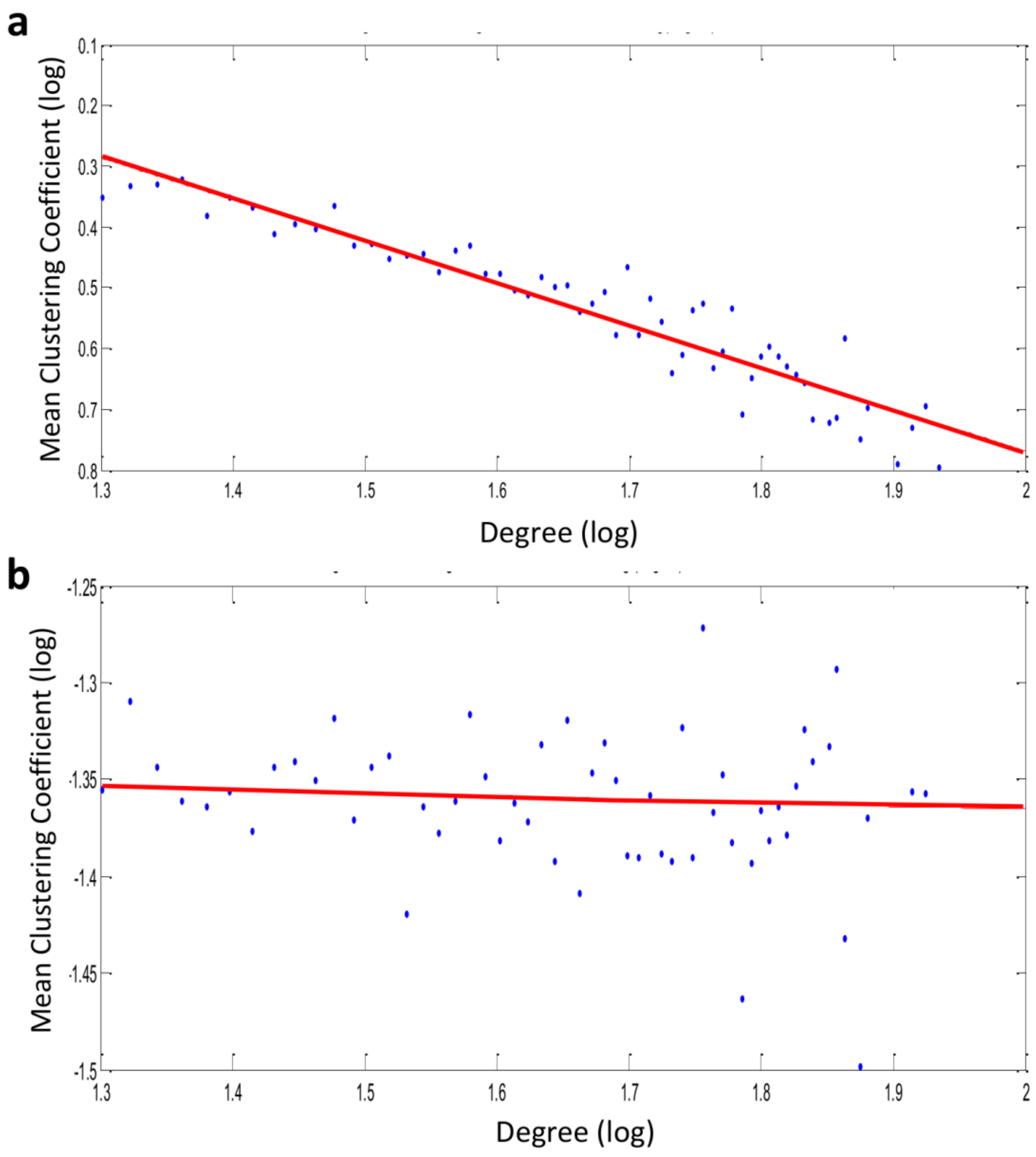

**Figure S6. Clustering Coefficient properties.**

A) Clustering coefficient as function of degree. Example from network 4. X- axis: log of degree, Y-axise: log of clustering coefficients. Fit to the form C~K-β with β=0.36 (red line). B) Clustering coefficient as function of degree in a random network. X- axise: log of degree, Y-axise: log of clustering coefficients. Clustering coefficient was constant and independent of the degree.

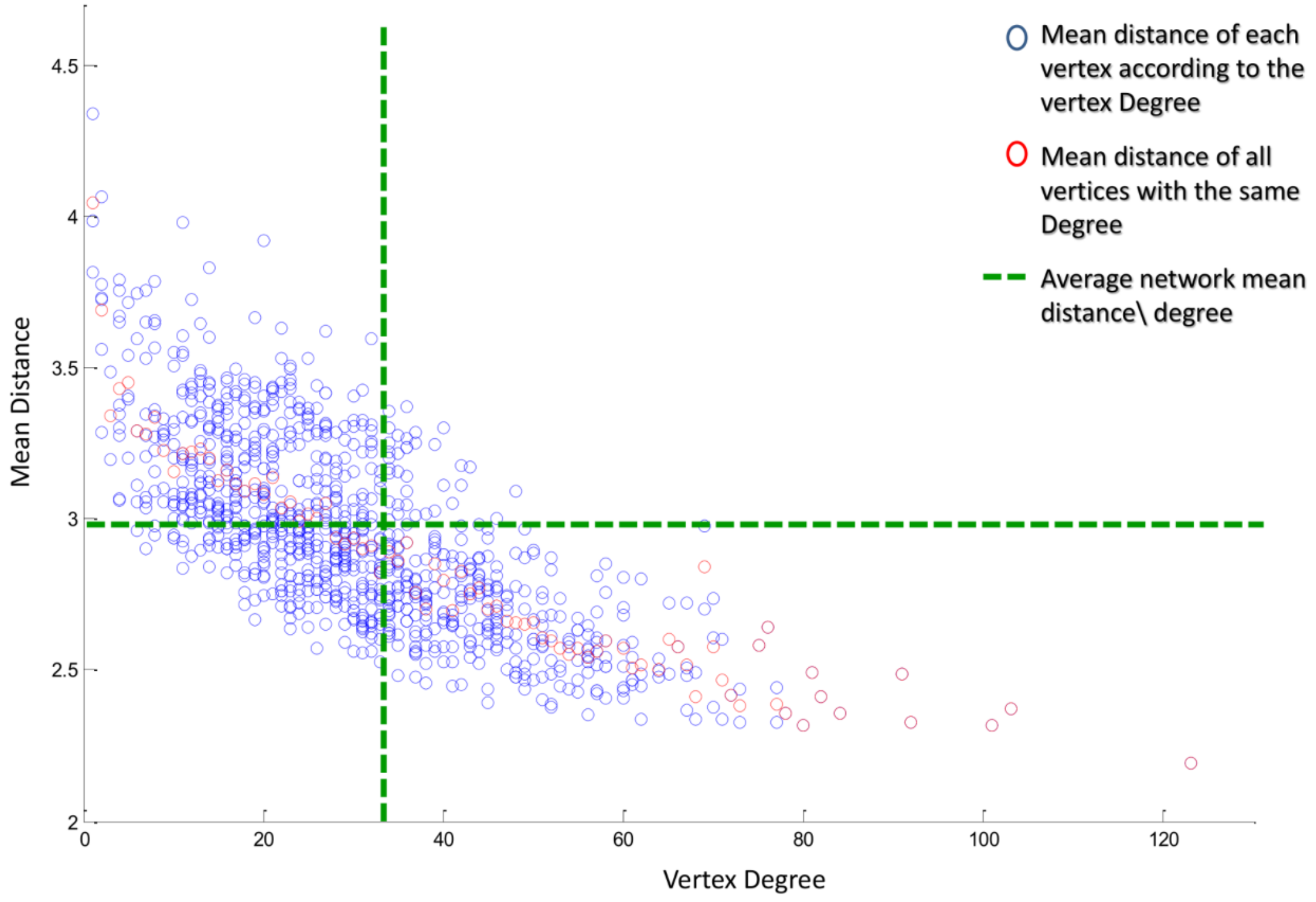

**Figure S7. Mean distance according to degree.**
Mean distance of each node according to its degree (blue circles). Mean distance of all nodes with the same degree is denoted by red circles. X- axis: degree, Y-axise: mean distance. Dashed green lines denote overall mean degree and mean distance of the whole network . Example taken from network 2.

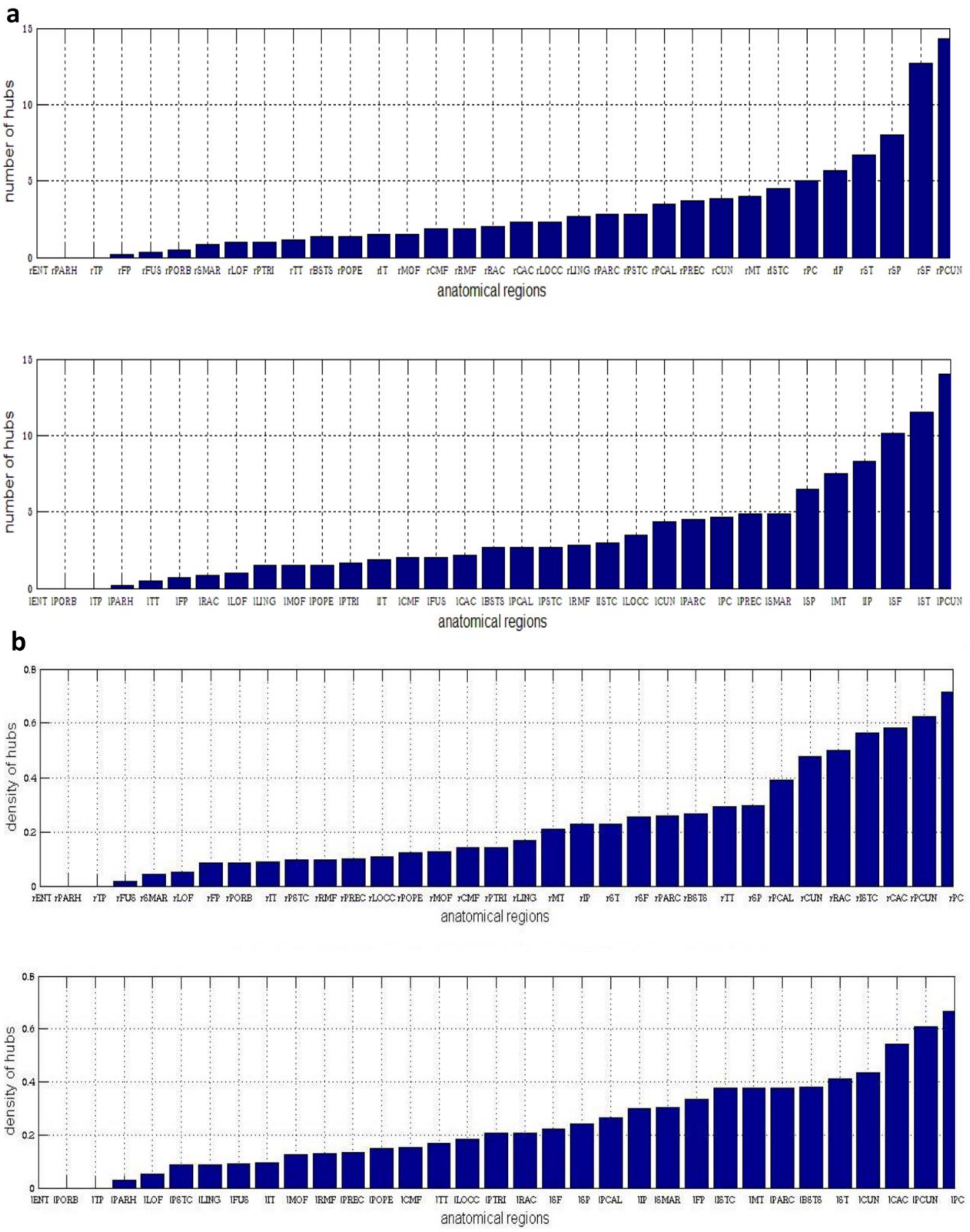

**Figure S8: Hub Distribution and Density.**

Distribution of hubs in the cortical network (A) and distribution according hub density (B). The 66 cortical regions were labeled as in Fig. S5

**Table S1 - Average connections between hierarchies**

| | Low hierarchy | Middle hierarchy | High hierarchy |
|---|---|---|---|
| **Low hierarchy** | **22% ±6.33%** | 30.2% ±3.36% | 47.7% ±4.24% |
| **Middle hierarchy** | 6.8% ±0.77% | **51.7% ±2.6%** | 41.4% ±2.6% |
| **High hierarchy** | 5.8% ±0.6% | 22.4% ±1.5% | **71.7% ±1.6%** |

-.Percent connections were normalized by the total amount of hierarchy connections. Bold results are significant at q<0.0005 (FDR corrected).

**Table S2 – Average shell level of functional networks**

| **Functional Network** | **Average shell level (mean ± SD)** |
|---|---|
| Default mode network(Raichle *et al.*, 2001) | 18.1±1.5 |
| Sensorimotor network(Fulton, 1935) | 17.5±1 |
| Salience network(Seeley *et al.*, 2007) | 17.34±2.37 |
| Visual network(Goodale and Milner, 1992) | |
| -Primary Visual areas | 16.93±1.8 |
| -Dorsal stream | 17.7±0.8 |
| -Ventral stream | 12±3.4 |
| Executive control network(Owen *et al.*, 2005) | 16.85±1.74 |

*Supplementary data files:*

1. **Set 2 Figures 66.ZIP** – 66 figures of cortical anatomical regions according to their shells distribution. X-axis: k-shell number; Y-axis: number of nodes. Different colors represent the 6 different cortical networks. STD= standard deviation across the shells, BRAINS STD= standard deviation across all 6 networks. The 66 cortical regions were labeled as in Fig. S5.
2. **Plot_var All.ZIP** - Comparison between degree distribution and shell distribution of all anatomical regions. Different colors denote different cortical networks. Upper graph- X axis: nodes degree (normalized to be between 0-1); Y axis: number of nodes. Bottom graph- X axis: nodes shell level (normalized to be between 0-1); Y axis: number of nodes (see Figure S4).VAR= variance across shells\degrees, BRAINS VAR= variance across networks. The 66 cortical regions were labeled as in Fig. S5.

# 3.2 Synchronization of chaotic systems: a microscopic description

Nir Lahav*[1], Irene Sendi˜na-Nadal[2, 3], Chittaranjan Hens[4], Baruch, Ksherim[4], Baruch Barzel[4], Reuven Cohen[4] and Stefano Boccaletti[5, 6]

1.Department of Physics, Bar-Ilan University, 52900 Ramat Gan, Israel
2.Complex Systems Group & GISC, Universidad Rey Juan Carlos, 28933 Mostoles, Madrid, Spain
3.Center for Biomedical Technology, Universidad Politecnica de Madrid, 28223 Pozuelo de Alarcon, Madrid, Spain
4.Department of Mathematics, Bar-Ilan University, 52900 Ramat Gan, Israel
5.CNR-Institute of complex systems, Via Madonna del Piano 10, 50019 Sesto Fiorentino, Italy
6.Unmanned Systems Research Institute, Northwestern Polytechnical University, Xi'an 710072, China

# 3.3. Topological synchronization of chaotic systems

Nir Lahav,[1,*] Irene Sendiña-Nadal,[2,3] Chittaranjan Hens,[4,5] Baruch Ksherim,[4] Baruch Barzel,[4] Reuven Cohen,[4] and Stefano Boccaletti[6,7]

[1.] *Department of Physics, Bar-Ilan University, 52900 Ramat Gan, Israel*
[2.] *Complex Systems Group & GISC, Universidad Rey Juan Carlos, 28933 Móstoles, Madrid, Spain*
[3.] *Center for Biomedical Technology, Universidad Politécnica de Madrid, 28223 Pozuelo de Alarcón, Madrid, Spain*
[4.] *Department of Mathematics, Bar-Ilan University, 52900 Ramat Gan, Israel*
[5.] *Physics and Applied Mathematics Unit, Indian Statistical Institute, 203 BT Road, Kolkata 700108, India*
[6.] *CNR-Institute of Complex Systems, Via Madonna del Piano 10, 50019 Sesto Fiorentino, Italy*
[7.] *Unmanned Systems Research Institute, Northwestern Polytechnical University, Xian 710072, China*

**Under preparation**

# Topological synchronization of chaotic systems

Nir Lahav,[1,*] Irene Sendiña-Nadal,[2,3] Chittaranjan Hens,[4] Baruch Ksherim,[4] Baruch Barzel,[4] Reuven Cohen,[4] and Stefano Boccaletti[5,6]

[1]*Department of Physics, Bar-Ilan University, 52900 Ramat Gan, Israel*
[2]*Complex Systems Group & GISC, Universidad Rey Juan Carlos, 28933 Móstoles, Madrid, Spain*
[3]*Center for Biomedical Technology, Universidad Politécnica de Madrid, 28223 Pozuelo de Alarcón, Madrid, Spain*
[4]*Department of Mathematics, Bar-Ilan University, 52900 Ramat Gan, Israel*
[5]*CNR-Institute of complex systems, Via Madonna del Piano 10, 50019 Sesto Fiorentino, Italy*
[6]*Unmanned Systems Research Institute, Northwestern Polytechnical University, Xi'an 710072, China*

Chaotic dynamics present two fundamental and unique emergence phenomena, strange attractors with their multi fractal structure and Chaotic synchronization, a distinctive emergent of self-organization in nature. Chaotic synchronization was classically characterized in terms of macroscopic parameters, such as Lyapunov exponents. In our previous paper we showed a microscopic description of this fundamental behavior. By presenting a new kind of synchronization - *topological synchronization* We showed that chaotic synchronization is a continuous process that starts in low density areas of the attractor. In this paper we analyze the relationship between the two emergent phenomena by shifting the descriptive levels and describing synchronization in the emergent multi fractal level. To capture the multi-fractal structure, we measured the general dimension of the system, and measured how it changed while increasing the coupling strength. We show that during the gradual process of topological adjustment in phase space the multi fractal structures of each strange attractor of the two coupled oscillators continuously converge, taking similar form, until complete topological synchronization ensues. Furthermore, according to our results chaotic synchronization has a universal property. Both in continuous systems and discrete maps, with the right coupling, synchronization initiates from the sparse areas of the attractor and creates zipper effect - a distinctive pattern in the multi-fractal structure of the system. Topological synchronization offers a new perspective to chaotic synchronization and allows us to find new universal properties and expand our understanding of the synchronization process.

Complex systems present us an immense challenge as we try to explain their behavior. One key element in their description is to show how do synchronization and self-organization emerge from systems that didn't have these properties to begin with. Especially if the systems have chaotic behavior. Synchronization underlies numerous collective phenomena observed in nature [1], providing a scaffold for emergent behaviors, ranging from the acoustic unison of cricket choruses and the coordinated choreography of starling flocks [2] to human cognition, perception, memory and consciousness phenomena [3–7]. surprisingly, although chaotic systems have high sensitivity to initial conditions and thus defy synchrony, in the 1980's it has been shown that even chaotic systems can be synchronized [8–11]. Understanding how such a process can happen and characterizing the transition from completely different activities to synchrony in chaotic systems is of fundamental importance in order to understand the emergence of synchronization and self-organization in nature.

Chaotic dynamics present two fundamental and unique emergence phenomena, strange attractors which, in most cases, will have multi fractal structure [12, 13] and Chaotic synchronization. Understanding how these two phenomena occur and relate to each other is essential in order to shed more light on the process of emergence in nature. Usually, chaotic synchronization is investigated by analyzing the time series of the system. often it observed by tracking the coordinated behavior of two slightly mismatched coupled chaotic systems, namely two systems featuring a minor shift in one of their parameters. As the coupling strength increases, a sequence of transitions occurs, beginning with no synchronization, advancing to phase synchronization [15], lag synchronization [16], and eventually, under sufficiently strong coupling, reaching complete synchronization. The process is typically characterized at the macroscopic level through the Lyapunov spectrum[15] and at the mesoscopic level through the nonlocalized unstable periodic orbits [17–21].

In our previous paper [22] we presented a new approach that revealed the microscopic level of the synchronization process. By presenting new kind of synchronization, a *topological synchronization*, we shifted descriptive levels of the synchronization process to the emergence level of the topology domain of the synced attractors. We discovered that at the microscopic level synchronization is a continuous process that starts from local synchronizations in different areas of the attractor. These local topological synchronizations start from the sparse areas of the attractor, where there are lower expansion rates, and accumulate until the system reaches complete synchronization. In this paper we investigate the relationship between the two emergent phenomena of chaos, the multi fractal structure and the synchronization process of strange attractors. In order to do so, we analyze the new phenomenon of topological synchronization. We show that indeed, Topological synchronization of strange attractors is a gradual process in the emergent multi-fractal level. In which, the multi fractal structures of each strange attractor of the two coupled oscillators continuously converge, taking similar form, until complete topological synchronization ensues. Topological synchronization unveils new detailed information about the synchronization process that never been shown before. For example, details about changes of the fractal dimensions along the synchronization process as well as details about the probability of each scaling law to appear on the synchronized attractor and the probability of the trajectory to remain on a scaling law along the synchronization process. In addition, we show evidence that chaotic synchronization process has universal properties. Both in our examined continuous system and discrete map, with the right coupling, synchronization initiates from the sparse areas of the attractor and creates zipper effect - a distinctive pattern in the multi-fractal structure of the system.

The emergence of strange attractors is typically characterized by multi-fractal structure [12, 13], which means that there are infinite number of scaling laws in their structure, each captured by different fractal dimension. Furthermore, every scaling law has different probability of the trajectory to follow it. [14]. *Hausdorff dimension*, that typically captured by *box count dimension*, is only one of these scaling laws. In order to demonstrate topological synchronization, we need to use more general definition of dimension to capture this multi fractality. To this end, we used *Rényi Generalized dimension* [12, 23] which fully describes the structure of a multi-fractal with respect to the different probabilities of each fractal:

$$D_q = \lim_{l\to 0}\left[\frac{1}{q-1}\frac{\ln(\sum_i {p_i}^q)}{\ln(\frac{1}{l})}\right], \tag{1}$$

Where $P_i$ is the Probability of a point (in state space) to be in sphere i, $l$ is the radius of spheres and $q$ is a parameter that can be any real number. Parameter q captures different fractal dimensions $D_q$ in the multi-fractal that have different probabilities for the trajectory to follow them. Thus, general dimension is not one value but a curve of values that depends on the parameter $q$ and represents the multi fractal structure of a strange attractor. The dominate dimension is the box counting dimension, which is a mixture of all the scaling laws that will appear the most in the attractor. On the curve of the general dimension it will have the value of $D_0$. $D_1$ is the information dimension and $D_2$ is the correlation dimension [24]. $D_{-\infty}$ represents a very rare scaling law that appears only once in the strange attractor with small probability of states obeying this law and $D_\infty$ represents yet another very rare scaling rule that also appears once, but this time with high probability of states obeying this law [12, 25].

Equipped with Eq. (1) we can fully describe topological synchronization. Let's demonstrate it on one of the most fundamental examples in the context of synchronization, capturing two slightly mismatched chaotic Rössler oscillators [26] coupled in a master-slave configuration. The equations of motion driving these oscillators take the form:

$$\begin{aligned}\dot{\mathbf{x}}_1 &= f_1(\mathbf{x}_1)\\ \dot{\mathbf{x}}_2 &= f_2(\mathbf{x}_2) + \sigma(\mathbf{x}_1 - \mathbf{x}_2),\end{aligned} \tag{2}$$

where $\mathbf{x}_1 \equiv (x_1, y_1, z_1)$ and $\mathbf{x}_2 \equiv (x_2, y_2, z_2)$ are the vector states of the master and slave oscillators respectively, σ is the coupling strength and $f_{1,2}(\mathbf{x}) = (-y - z, x + ay, b + z(x - c_{1,2}))$. Without loss of generality we set the parameters to $a = 0.1$ and $b = 0.1$ identically across the two oscillators, and express the slight mismatch between the master and the salve through the parameters $c_1 = 18.0$ vs. $c_2 = 18.5$. System (2) describes a unidirectional master ($\mathbf{x}_1$) slave ($\mathbf{x}_2$) form of coupling, uniformly applied to all coordinates $x, y$ and $z$. Under this directional coupling scheme we can track and quantify the process of synchronization in a controlled fashion, as the slave gradually emulates the behavior of the master, while the master continues its undisturbed oscillations.

In our previous paper we showed the microscopic build-up of synchronization in system (2) [22]. Local synchronization initiates in the sparse areas of the attractor and as the local synchronizations accumulate, Phase synchronization occurs for $\sigma_{ps}\geq 0.1$ and complete synchronization obtained for $\sigma_{cs} \geq 2.0$. In Fig. 1 we show the general dimension curves, $D_q$ of system (2). The Master (black) has a fixed curve while the slave starts with completely different $D_q$ than the master in low coupling σ = 0.07 (blue) and converge with the master $D_q$ at higher coupling σ = 0.12 (red dashed) showing process of topological synchronization between the master and the slave. Moreover, at the transition point to phase synchronization (red dashed) blow ups for the master and slave curves show that in the slave case, $D_q$ for the negative part of q (q < 0) is much closer to the master then $D_q$ for the positive part (q > 0. Compare the two zooms on the positive and negative parts of the curve, and take into account the difference in the vertical axis ranges: for negative $q$ it is 0.02, and for positive $q$ it is 0.15, a difference of almost one order of magnitude). This result corresponds to the fact that Local synchronization initiates in the sparse areas of the attractor where the probability of points is low. As we step to the negative part of the parameter q and approach $D_{-\infty}$ , we examine the sparse areas of the attractor with low probability scaling laws, and indeed they reached topological synchronization before the dense areas of the attractor (positive part of the parameter $q$).

The previous example demonstrates that synchronization process between different strange attractors can be understood as topological synchronization between the multi-fractal structures of the attractors. Topological synchronization means that the multi-fractal structure of one attractor predicts the multifractal structure of the second attractor and when complete topological synchronization occurs the multi-fractal structure of one attractor fully predicts the other. Therefore, Topological synchronization is characterized by the boundedness of the difference between the $D_q$ curves of the first oscillator and the second oscillator, over the whole dynamical evolution of the system. Consequently, the condition for complete topological synchronization between oscillator 1 and 2 is:

$$\Delta Dq = |Dq_1 - Dq_2| \to 0. \tag{3}$$

In order to farther analyze the properties of *topological synchronization* we chose a simple 1D discrete system from the Logistic map family, coupled in a master-slave configuration. The equations of motion driving these oscillators take the form [27]:

$$\begin{aligned} x_{n+1} &= c_1\left(1 - 2x_n^2\right) \\ y_{n+1} &= (1-k)c_2\left(1 - 2y_n^2\right) + c_1 k(1 - 2x_n^2), \end{aligned} \tag{4}$$

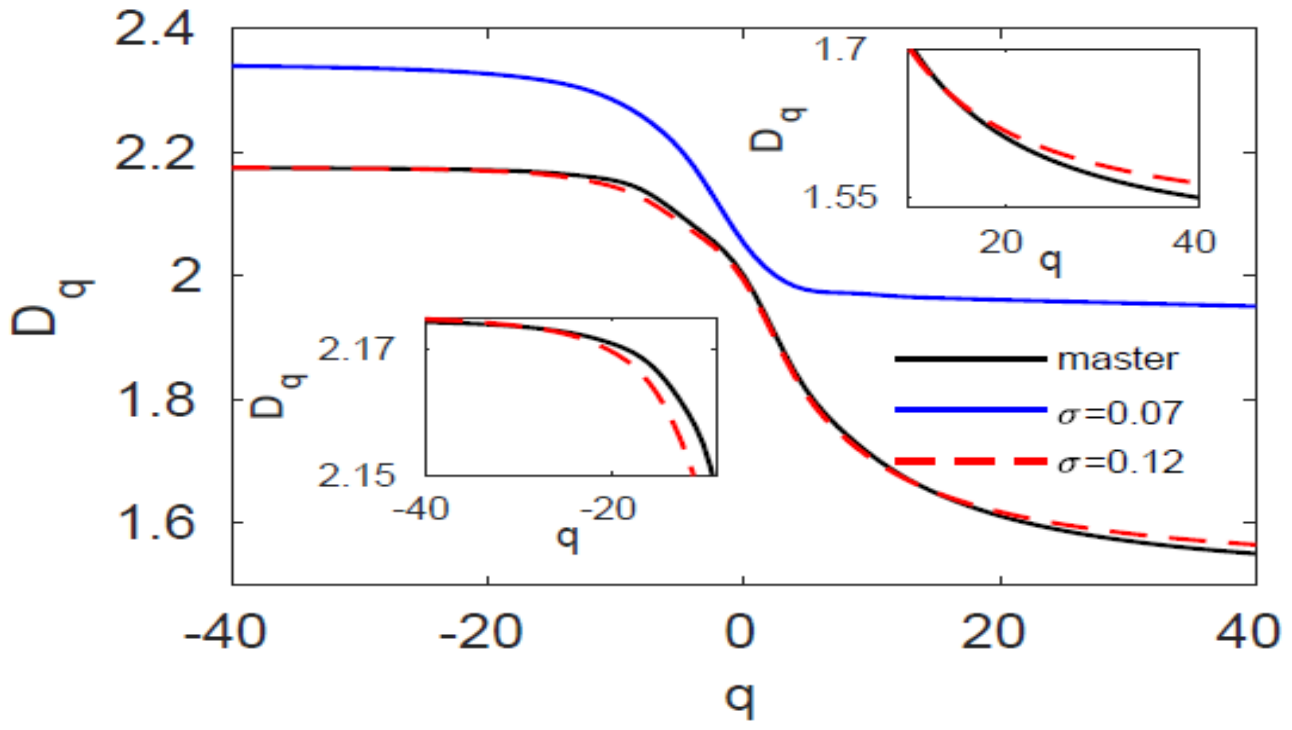

Figure 1: (Color online). **Generalized fractal dimension for slightly mismatched Rössler system**. General dimension $D_q$ as a function of parameter $q$ for the master (black) and slave when coupled to the master with $\sigma = 0.07$ (blue) and $\sigma = 0.12$ (red dashed). In sets are blow ups for the master and slave curves for $\sigma = 0.12$ in the q $<$ 0 (bottom left) and q $>$ 0 (top right) regions. Topological synchronization occurs as the $D_q$ curve of the slave converge into the $Dq$ curve of the master.

Where $k$ is the coupling strength. Without loss of generality we express the mismatch between the master and the salve through the parameters $c_1 = 0.89$ versus $c_2 = 0.8373351$ (onset of chaos). The slave oscillator ($y_n$) is on the onset of chaos with sparse strange attractor whereas the master oscillator ($x_n$) has a dense strange attractor. In Fig. 2a we present the synchronization error parameter $E$ versus $k$ [28]. As $E \to 0$ at $k_{CS} \sim 0.9$ complete synchronization emerges. Topological synchronization unveil the microscopic process underlying synchronization. This microscopic buildup is caused by a topological matching mechanism which eventually will lead to complete synchronization between the two attractors. Fig. 2b-e and Fig. 3 examine the general dimension of system (4) and reveals this topological synchronization process. Fig. 3 shows that gradual increase of $k$ causes a gradual decrease of the distance between the two $Dq$ curves to zero. Around $k = 0.21$ the distance of the negative part of the $Dq$ curves (q≤0) begins to decrease until it reaches zero around $k = 0.33$, whereas the distance of the positive part of the $Dq$ curves (q > 0) begins to decrease only at around $k = 0.3$. When the distance of the positive part also reaches zero around $k = 0.9$, the system reached complete topological synchronization with zero distance between the two $D_q$ curves.

Furthermore, in Fig. 2b-e we show that the changes of the slave $D_q$ curve versus $k$ revealed a **zipper effect** of the general dimension from the negative $q$ to the positive $q$. In low couplings there is a gradual synchronization of the negative part of the $D_q$ curve (q≤0). When the negative part of the $D_q$ curve is synchronized around, $k = 0.33$ (panel c), the positive part begins to gradually synchronize. More specifically, $D_1$ is synchronized at around $k = 0.36$, $D_2$ is synchronized at around $k = 0.42$, $D_3$ is synchronized at around $k = 0.51$ (panel d) and so on, "zipping" the topological synchronization process until at around $k = 0.9$ $D_{10}$ is synchronized and complete synchronization is achieved (panel e. For video of the whole zipper effect process, see supplementary video 1 where red curve is the slave and blue curve is the master, and supplementary video 2 Where left side is the slave and right side is the master attractors).

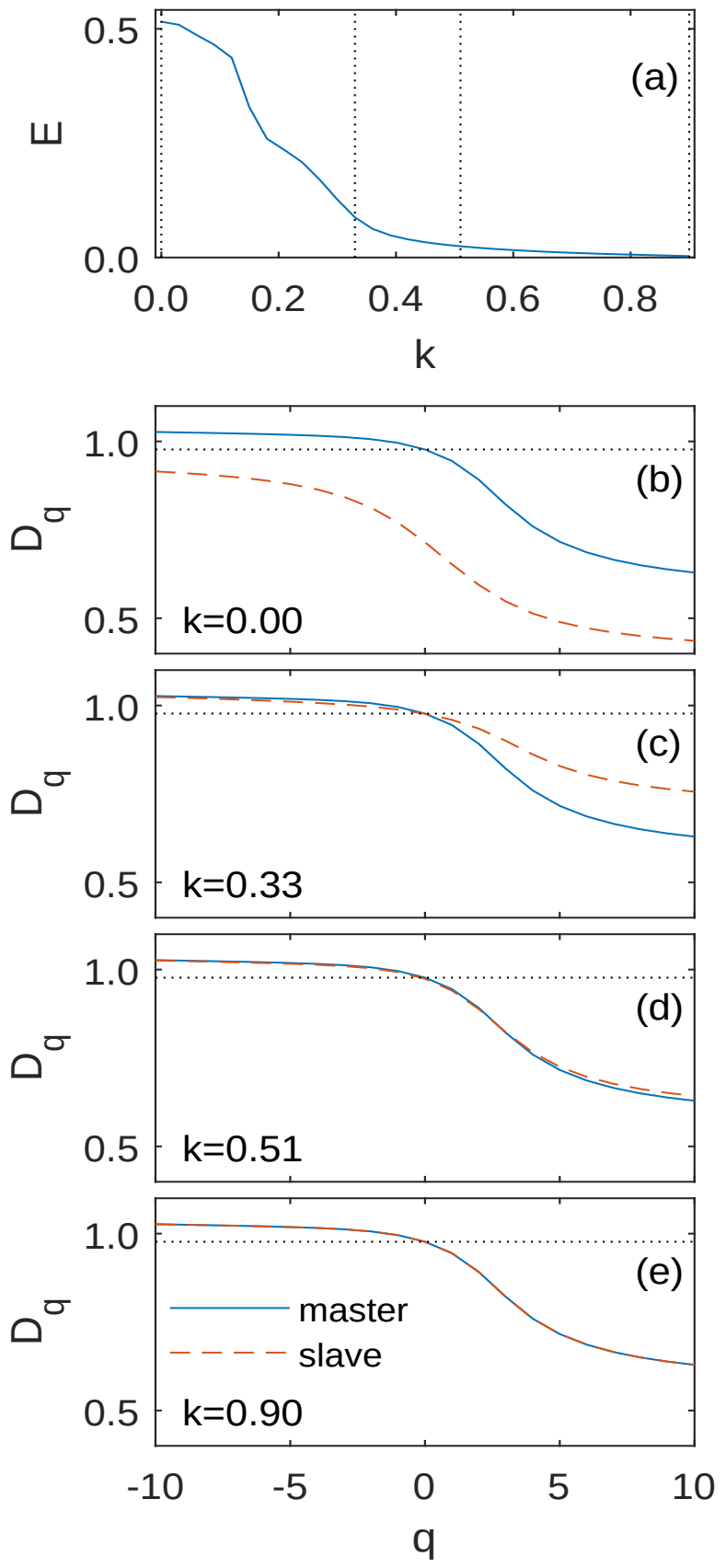

Figure 2: (Color online). **Microscopic build-up of synchronization for Logistic map**. (a) Synchronization error $E$ as function of coupling strength $k$. Under complete synchronization, $E \to 0$, obtained for $k \geq k_{CS} \sim 0.9$. (b)-(e) Topological synchronization and the zipper effect. General dimension $D_q$ as function of the parameter $q$ of master (blue) and slave (Red) attractors. As coupling $k$ increases a zipper effect from the negative ($q \leq 0$) to the positive ($q > 0$) part of $D_q$ can be seen.

The finding of negative to positive zipper effect in the $D_q$ curves concurs with the previous section on Rössler system. As, stepping from $D_{-\infty}$ to $D_{\infty}$ represent stepping from scaling laws with low occupation probability to scaling laws with high occupation probability. It implies that as in the Rössler case, also in logistic map, topological synchronization starts in low coupling strengths, with areas of the attractor that have low probability of points, and only when these areas complete their local synchronizations, at strong coupling strengths, the attractor will start to topologically sync also in areas with high probability of points.

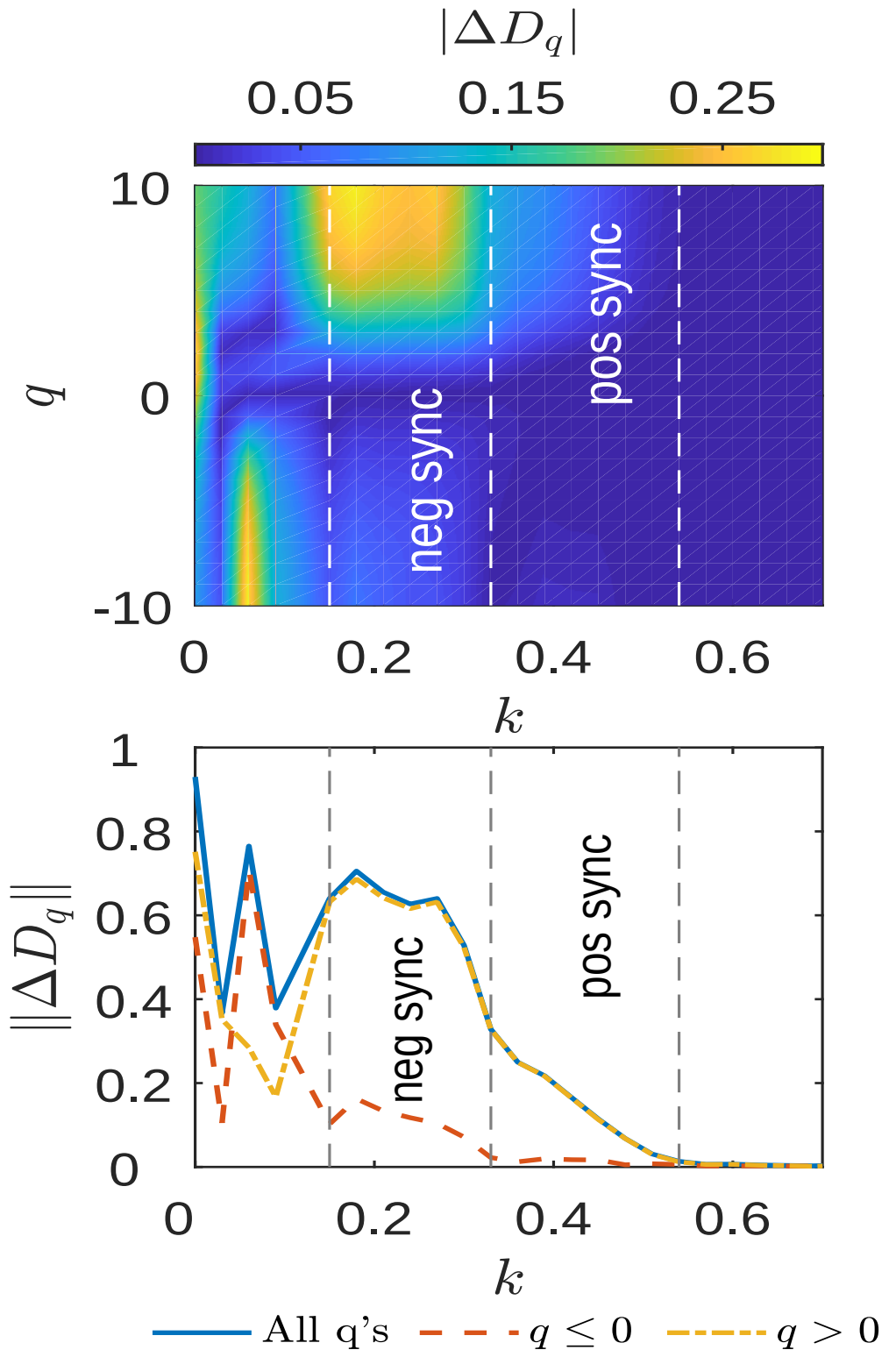

Figure 3: (Color online). **Distance between general dimension of master and slave for mismatched Logistic map system**. Upper panel, color map denoting the distance between the $D_q$ curves of the master and the slave, $|\Delta D_q|$ as function of the parameter q (y axis) and the coupling strength k (x axis). Dashed line shows the negative and positive zipper effect regions (as distance decreasing to zero). Bottom panel, distance between the $D_q$ curves of the master and the slave, $|\Delta D_q|$ as function of k. Red curve – distance for the negative part of $D_q$ curves ($q \leq 0$). Yellow curve - distance of the positive part of $D_q$ curves ($q > 0$). Blue curve – distance for the whole $D_q$ curves. In $k \sim 0.33$ the negative part of $D_q$ has completed its synchronization with the master and the positive part starts a gradual decrease of its distance to the master curve. The zipper effect is completed around $k \sim 0.9$ and the system reached complete topological synchronization. Dashed line shows the negative and positive zipper effect regions (as distance decreasing to zero).

In this paper we analyzed the relationship between the emergence phenomena of chaotic dynamics, the multi-fractal structure of a strange attractor and chaotic synchronization. We demonstrate this relationship by introducing topological synchronization, in which the multi-fractal structure of one strange attractor assimilate to the other until the multi-fractal structure of the attractors is the same. Topological synchronization shifts the descriptive levels of synchronization to the emergence level of the topology domain of the attractors. Topological synchronization is a powerful tool to investigate chaotic synchronization. It reveals that chaotic synchronization is a continuous process and hints for a universal zipper effect. The fact that in both discrete map and continuous system we see the same distinctive pattern in the multi-fractal structure, where topological synchronization starts from the sparse areas of the attarctor, suggests that chaotic synchronization has a universal property. If a system can reach chaotic synchronization, one can find a coupling form in which the synchronization initiates from the sparse areas of the attractor and creates zipper effect. The road to complete synchronization starts from the sparse areas in the attractor and continues with synchronizations of increasingly more crowded areas in the attractor until only with sufficient coupling strength a global complete topological synchronization can be achieved.

One application of these results is a way to determine how much synchronization a physical system has and where, in phase space, it occurred. For some real chaotic systems, complete synchronization will be detected whereas other systems may only sync until the point where their less crowded areas in the attractor will be synced. Topological synchronization can detect these differences and show which areas of the phase space have already synced. Since $D_0$ represents a mixture of all the scaling laws that will appear the most in the attractor [12], when $D_0$ is synced the structure of the slave attractor becomes very similar to the master (see supplementary video 1 and supplementary video 2 at around $k = 0.12$). Accordingly, we suggest that some biological systems may synchronize until a sufficient synchronization point, around the synchronization of $D_{0,}$ and will not reach complete synchronization.

**Acknowledgements**. The authors would like to thank Ashok Vaish and Ricardo Gutiérrez for their continuous support. C.H. wishes to thank the Planning and Budgeting Committee (PBC) of the Council for Higher Education, Israel, for support. This work was supported by the US National Science Foundation - CRISP Award Number: 1735505 and by the Ministerio de Economía y Competitividad of Spain (projects FIS2013-41057-P and FIS2017-84151-P).

* Corresponding author: freenl@gmail.com

[1] A. Pikvosky, M. Rosenblum, and J. Kurths, *Synchronization: a Universal Concept in Nonlinear Sciences* (Cambridge University Press, Cambridge, England, 2001); S. Boccaletti, J. Kurths, G. Osipov, D. L. Valladares, and C. S. Zhou, Phys. Rep. **366**, 1 (2002); S. Strogatz, *Sync: The emerging science of spontaneous order* (Hyperion, 2003); S. Boccaletti, A. Pisarchik, C.I. del Genio and A. Amann *Synchronization: from Coupled Systems to Complex Networks* (Cambridge University Press, Cambridge, England, 2018).

[2] L. Glass and M. C. Mackey, *From clocks to chaos: the rhythms of life* (Princeton University Press, 1988); A. T. Winfree, *The geometry of biological time, Vol. 12* (Springer Science & Business Media, 2001).

[3] P. J. Uhlhaas and W. Singer, *Neuron* **52**, 155 (2006); G. Buzsaki, *Rhythms of the Brain* (Oxford University Press, 2006); E. Bullmore and O. Sporns, Nat. Rev. Neurosc. **10**, 186 (2009).

[4] Varela, Francisco and Lachaux, Jean-Philippe and Rodriguez, Eugenio and Martinerie, Jacques, *Nature reviews neuroscience*

**2**, 4 (2001).

[5] E. Rodriguez, N. George, J.-P. Lachaux, J. Martinerie, B. Renault and F. J. Varela, *Nature* **397**, 430 (1999).

[6] W. Klimesch, *International Journal of Psychophysiology* **24**, 61 (1996).

[7] W. Singer, *Consciousness and neuronal synchronization (in: The neurology of consciousness)* (Academic Press, 2011).

[8] L.M. Pecora and T.L. Carroll, Phys. Rev. Lett. **64**, 821 (1990).

[9] L.M. Pecora and T.L. Carroll, Chaos **25**, 097611 (2015).

[10] A. Pikovskii, Z. Phys. B **55**(2), 149 (1984).

[11] L Huang, Q Chen, YC Lai, LM Pecora, Phys. Rev. E, **80**, 036204 (2009)

[12] Halsey, C. Thomas, et al, Physical Review A **33**.2, 1141 (1986).

[13] P Grassberger, I Procaccia, Physica D **9**, 189-208 1983a. P Grassberger, I Procaccia, Phys. Rev. Let. **50**, 346-349 1983b.

[14] Hentschel, HGE and Procaccia, Itamar, Physica D: Nonlinear Phenomena **8**, 435–444 1983.

[15] M.G. Rosenblum, A.S. Pikovsky, and J. Kurths, Phys. Rev. Lett. **76**, 1804 (1996).

[16] S. Boccaletti and D. L. Valladares, Phys. Rev. **E62**, 7497 (2000).

[17] A. Pikovsky, M. Zaks, M. Rosenblum, G. Osipov and J. Kurths, Chaos **7**, 680-687 (1997).

[18] D. Pazo, M. Zaks and J. Kurths, Chaos, **13**, 309-318 (2003).

[19] S. Yanchuk, Y. Maistrenko and E. Mosekilde, Chaos, **13**, 388-400 (2003).

[20] P. Cvitanović, Physica D, **51**, 138-151 (1991).

[21] Heagy, J. F., T. L. Carroll, and L. M. Pecora, Physical Review E **52**.2 (1995): R1253.

[22] N Lahav, et al., Physical Review E **98**.5, 052204 (2018).

[23] A. Renyi, Dimension, entropy and information, Transactions of the Second PragueConference on Information Theory, Academic Press, New York, 1960, 545-556.

[24] P Grassberger, I Procaccia, Physica D **13**.1-2, 34-54 (1984).

[25] Martinez, V.j, et al, Astrophysical Journal, **357**, 50 (1990).

[26] O. E. Rössler, Physics Letters A **57**, 397-398 (1976).

[27] Grassberger, Peter, Journal of Statistical Physics **26**, 173–179 (1981).

[28] In all our simulations, $E$ is not calculated as limits for $\tau$ going to infinity. In practice, we fixed a time step for integration of $h = 0.001$ time units (t.u.), we let the system evolve from random initial conditions from $t = 0$ until $t_0 = 50,000$ t.u., and then $E$ is calculated as $E = \tau^{-1} \int_{t_0}^{t_0+\tau} ||\mathbf{x}_1 - \mathbf{x}_2|| dt$ respectively, with $\tau = 1,000$ t.u.

[29] S. Boccaletti, L.M. Pecora and A. Pelaez, Phys. Rev. **E63**, 066219 (2001).

[30] L. Pastur, S. Boccaletti, and P. L. Ramazza, Phys. Rev. E **69**, 036201 (2004).

[31] L. M. Pecora, T. L. Carroll, and J. F. Heagy, Phys. Rev. E **52**, 3420 (1995).

[32] S. Boccaletti, V. Latora, Y. Moreno, M. Chavez, and D.-U. Hwang, Phys. Rep., **424**, 175-308 (2006)

המוכלל שלהם. המימד המוכלל לוקח בחשבון את כל הממדים הפרקטלים השונים שיש באותו מושך מוזר (מולטי-פרקטל). בעזרת המימד המוכלל הגדרנו מהו סנכרון טופולוגי וכימתנו כיצד המבנים הפרקטלים של מושך אחד הופכים לאורך תהליך הסנכרון למבנים הפרקטלים של המושך השני. את הסנכרון הטופולוגי בדקנו הן במערכת בדידה (המפה הלוגיסטית) והן במערכת רציפה (מערכת רוסלר). כך גילינו תופעה אוניברסלית של הסנכרון הכאוטי המתרחשת הן במערכות בדידות והן במערכות רציפות. כאשר יש צימוד מתאים, המערכות הכאוטיות יתחילו את הסנכרון שלהם מהאזורים הנדירים של המושך. מבחינה מבנית תופעה זו מתבטאת בתבנית ייחודית לה קראנו תופעת הריצ׳רץ׳ (Zipper effect).

האזורים בעלי הקישוריות הגבוהה ביותר ברשת להם היכולת הגבוהה ביותר לעשות אינטגרציה של מידע. עוד גילינו שאזורי הגרעין נמצאים בקורלציה עם פעילויות שקשורות למודעות.

כדי להבין את תופעת ההפצעה בטבע לא מספיק רק לחקור את מבנה הרשת. בנוסף צריך לחקור את הדינמיקה של קודקודי הרשת. תכונה מרכזית שיכולה להפציע במערכת היא הסנכרון. כאשר יש הפצעה של סנכרון ברשת, הכוונה היא שלקודקודים שונים תהיה אותה הדינמיקה (או לפחות חלק מהתכונות הדינמיות שלהם יהיו אותו הדבר). כך, בעזרת סנכרון, תבניות חדשות יכולות להפציע מתוך פעילות הרשת. יש דוגמאות רבות לסנכרון בטבע כמו למשל צרצור מסונכרן של צרצרים או איתות אורות מסונכרן של גחליליות. במאמר השני והשלישי חקרנו את תופעת הסנכרון הכאוטי מנקודת מבט חדשה. מערכות כאוטיות, למרות שאינן ניתנות לחזוי ארוך טווח, יכולות גם כן להסתנכרן. כך, כאוס מציע שתי הפצעות בעת ובעונה אחת. הפצעה של מושך מוזר בעל מבנה מולטי פרקטלי וכן הפצעה של סנכרון. אנו חקרנו, בפעם הראשונה, את הקשר בין שתי הפצעות אלו. חקרנו את התכונות המבניות של המושכים המוזרים וכיצד הן משתנות בעת סנכרון בין מושכים מוזרים שונים ברשת. כך גילינו סוג חדש של סנכרון, סנכרון טופולוגי, בו תופעת הסנכרון הכאוטי מתבטאת ברמת סנכרון המבנה הפרקטלי של המושכים המוזרים. במאמר השני הראנו שהסנכרון הטופולוגי שופך אור חדש על תהליך הסנכרון. אנו גילינו שסנכרון כאוטי הוא תהליך רציף שמתחיל באזורים הנדירים של המושך המוזר (אזורים על המושך אליהם כמעט לא מגיעים במרחב המצבים) ורק כאשר הצימוד חזק מספיק הסנכרון מגיע גם אל שאר האזורים במושך.

במאמר השלישי בחנו את הקשר בין המבנה המולטי פרקטלי של המושך לבין תהליך הסנכרון. כדי לבחון את המבנה הפנימי של המושכים המוזרים, מדדנו את המימד

# תקציר

כאשר אנו מסתכלים על העולם סביבנו, רוב הפעמים אנו נראה מערכות מורכבות בהן התרחשה הפצעה של תכונות חדשות שלא היו למרכיבים המקוריים שהרכיבו את המערכת. דוגמה לכך היא הפצעת החיים מתוך מערכת מורכבת של אינטראקציות כימיות. אחת המערכות המורכבות ביותר המוכרות לנו היא המוח האנושי. במוח בוגר ממוצע ישנם סביב מאה מיליארד תאי עצב וכמאה טריליון סינפסות שמחברות ביניהם. הרשת העצבית המורכבת והסבוכה הזו יוצרת הפצעה של תכונות חדשות מתוך האינטראקציות בין תאי העצב. תכונות כגון למידה וזיכרון, חישובים קוגניטיביים מורכבים, היכולת ליצור חוויות מודעות וליצור את חווית האני.

מחקר זה מנסה להתקדם לעבר הבנת תופעת ההפצעה בטבע ובמוח בעזרת שימוש בתורת הרשתות, תורת הכאוס וחקר תופעת הסנכרון. במאמר הראשון הפעלנו אנליזה בשם k-shell decomposition על רשתות קורטקס אמיתיות שנאספו בעזרת דימות מוחי (DSI,MRI). אנליזה זו בוחנת קישוריות לא רק של קודקודים בודדים אלא חושפת אוכלוסיות קישוריות שונות ברשת. מקליפת הקישוריות הנמוכה ביותר ועד לגרעין הרשת. הגרעין כולל את הקודקודים בעלי הקישוריות הגבוהה ביותר ברשת. בעזרת שיטה זו ניתן לצעוד אל מעבר לתפקיד של קודקודים ספציפיים ולבחון תפקידים של אזורי קישוריות שונים ברשת ולחשוף כיצד תכונות חדשות מפציעות בעקבות פעילות גלובלית של מספר גדול של קודקודים. כך יכולנו לחשוף את הטופולוגיה של הרשת ולהשוות בין אזורי הקישוריות השונים לבין התפקודים המוחיים שלהם. אנו גילינו שיש היררכיה של עיבוד מידע ואינטגרציית מידע המתרחשת בקורטקס. החל מהאזורים בעלי הקישוריות הנמוכה ויכולת אינטגרציית מידע נמוכה ועד לאזורי הגרעין –

# **תוכן עניינים**

עבודה זו נעשתה בהדרכתם של פרופסור שלמה הבלין ופרופסור ראובן כהן

מן המחלקה לפיסיקה והמחלקה למתמטיקה באוניברסיטת בר אילן.

# הפצעה וסנכרון במערכות כאוטיות וברשת המוחית של האדם

**חיבור לשם קבלת התואר "דוקטור לפילוסופיה"**

**מאת**

ניר להב

המחלקה לפיסיקה

**הוגש לסנט של אוניברסיטת בר-אילן**

רמת גן אדר תש"ף